\documentclass[a4paper,10pt]{article}

\usepackage{graphicx}
\usepackage{amsfonts}
\usepackage{amssymb}
\usepackage{amsmath}
\usepackage{fullpage}
\usepackage{url}
\usepackage[T1]{fontenc} % <-- Add this line to fix the error
\usepackage[utf8]{inputenc} % Ensures modern UTF-8 input handling
\usepackage{geometry}
\newenvironment{proof}%
{\begin{list}{ {\bf Proof}}%
		{\setlength{\leftmargin}{\labelwidth}} \item }%
	{\end{list}}

\newtheorem{definition}{Definition}
\newtheorem{theorem}[definition]{Theorem}

\newtheorem{prop}[definition]{Proposition}
\newtheorem{cor}[definition]{Corollary}

\newcommand{\qed}{\square}

\title{\sc Analysing Busy Beaver Machines \\ (Or How to Find Ossified Ocelots) }
\author{James Harland \\ 
School of Computing Technologies
RMIT University \\
GPO Box 2476 
Melbourne, 3001 \\ 
Australia \\ 
{\em james.harland@rmit.edu.au} }
\date{\today}
\begin{document}

\maketitle
\begin{abstract}
	The busy beaver problem involves working with machines which can produce very large numbers, both for the output of certain Turing machines and for the number of steps taken to execute them. Given that these numbers can involve thousands of digits, it is crucial to have a means of proving properties of the relevant machines which allow manipulation of such large numbers. In this paper we introduce the notion of \textit{shape computations} as a means of isolating key properties that allow efficient execution of the relevant Turing machines based on verifiable properties of state transitions. In particular, this notion allows us to provide an analytical result which demonstrates the quadratic relationship between the size of the output and the number of steps required to generate it. We have also developed an initial implementation of our approach, and we provide empirical evaluation of it on a suite of 240 example machines. 
\end{abstract}

\section{Introduction} 
\label{sec:intro}

The busy beaver problem was first defined in 1962 as a simple example of a non-computable function \cite{Rado62}. The classic proof of the non-computability of this function essentially shows that the busy beaver function ``grows faster'' than any computable function. This means that any attempt to find explicit values for the busy beaver must be able to deal with some relatively large numbers arising from computations on Turing machines of seemingly small size. For example, there is a well-known example of a 5-state Turing machine which terminates with 4,098 non-blank characters on the tape after 47,176,870 steps (which has recently been shown to be the largest such values for a 5-state machine \cite{BBChallenge26}). While these are relatively large numbers, it seems plausible that someone could check a claim like this by hand (possibly with some assistance from a calculator). However, it is not hard to find larger examples, such as the 2-state machine below, which terminates with string on the tape whose total of non-blank characters is a number with more than 100 digits in it,\footnote{The precise figure for this is given in the Appendix} and so it seems this is too large to allow for any such manual checking. We will refer to such machines as \textit{dreadful dragons} \cite{Harland16}. Given that there are examples known which involve numbers with over 18,000 digits (and a correspondingly larger number of computation steps), it should be clear that the verifiability of such results needs careful investigation. 

\begin{figure}[tb]
	\centering
	\includegraphics[width=0.48\textwidth]{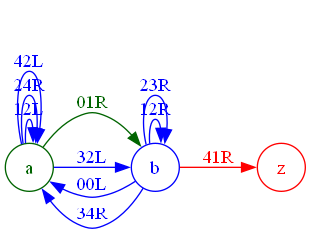} \\
	\caption{2-state machine of large productivity}
	\label{fig:dragon18}
\end{figure}

The main contribution of this paper is the introduction of \textit{shape computations}, which are a means of proving properties of sequences of configurations of a Turing machine. These computations are used as the basis for the derivation of \textit{ossified ocelot} transitions, ie transitions in the Turing machine which reflect verified patterns of operational behaviour. This work is based on the ``hypothetical engine'' of \cite{Harland07}; here it is defined in more detail and used as a basis for both theoretical results about dreadful dragons as well as verifiably correct deductions about the behavior of these machines, together empirical results showing their behaviour. This also builds on some earlier work \cite{Harland16} which established the practical importance of observing patterns in the execution of these machines (which we call \textit{observant otter} patterns). The shape computations derived from these patterns also allow use to derive analytical results about the relationship between the size of the configurations in these machines and the numebr of steps it takes to derive them. This has been observed empirically to be quadratic \cite{Quad,LP07}; we believe the results in this paper are the first analytical results to demonstrate this behaviour. Those same result show how we can exponentially increase execution speeds. 

We should note that the busy beaver problem has had a resurgence of interest with the emergence of the Busy Beaver Challenge Challenge \cite{BBCpage}, of which a key outcome is the setting of a decades-old conjecture about the value of the busy beaver function for $n = 5$ \cite{BBChallenge26}. This has also had the effect of renewing interest in how to evaluate the precise number of steps that such machines take to terminate, and providing new examples of machines with remarkably high such values. Hence we believe it is timely to consider what inference methods can be used in order to ``tame'' such unwieldly beasts. 

This paper is organised as follows. In Section~2 we review related work and in Section~3 we provide some basic definitions and concepts. Section~4 discusses execution methods for Turing machines, while Section~5 introduces the key concept of shape computations, and Section~6 providing some fundamental theoretical results based on these. In Section~7 we discuss some prgamatic issues related to finding shape computations, and in Section~8 we provide our empirical results based on 240 machines drawn from various sources. Finally in Section~9 we discuss our outcomes and in Section~10 our conclusions and ideas for further work. 

\section{Related Work}

It is now over 60 years since the busy beaver problem was originally conceived and defined by Rado \cite{Rado62}. The existence of non-computable functions had been known for some time by then \cite{Church36,Turing36}; Rado's motivation was to find a simple and specific example of such a function. The conception is based on a competition between players to propose champion machines, with the rules overseen by ``the International Busy Beaver Club''. Hence the idea of searching for increasingly large champion machines was inherent from the very beginning. 

This search has continued for decades since and continues today. Lin and Rado \cite{LR65} began the process with an analysis of machines with up to three states. Later work by Brady \cite{Brady83}, as well as Machlin and Stout \cite{MS90} confirmed and extended these result to four states. The work of Lafitte and Papazian \cite{LP07} extended this further still by broadening the outcomes to include cases with more than two symbols. The case for machines with 5 and 6 states were addressed by Marxen and Buntrock \cite{HM90}, particularly in the search for champion machines, and which produced some astronomically large figures, including the machine referred to above in which the number of non-blank characters is a number with over 18,000 digits. Analysis of the characteristics of these machines, particularly for the 5-state case, have been performed by Michel \cite{Michel93,Michel04,Michel13,Michel22}. There have also been discussion of various issues in this process by Aaronsen  \cite{Aaronsen20}, as well as variations of the busy beaver problem including the Beeping Busy Beaver and the placid platypus problem \cite{Harland06}. It is also interesting to note the recent suggestion by Nahin \cite{Nahin26} that the story of \textit{The Devil and Simon Flagg} from the 1950's \cite{Porges} (in which the devil is tasked with resolving whether or not Fermat's Last Theorem is true) should be updated to now have the devil determine an appropriately large busy beaver value (for which the present author's personal suggestion is 666!). 

The recent work of the Busy Beaver Challenge has not only had the remarkable result of finally verifying the long-standing conjecture that the busy beaver value for 5 states is indeed 4,098 
\cite{BBChallenge26}, but also of greatly expanding our knowledge of machines with 6 states (and more) as well as those with similar numbers of symbols. With the advent of such examples, it seems we have very much left behind any semblance of manual verification of results, simply due to the vast size of the numbers involved. This means we need to look well past the sheer size of the numbers to be able to understand the processes that lead to them. 

Two methodological comments seem appropriate here. One is that we are very much in agreement with de Mol \cite{deMol}, in that we believe that all work of this nature should include a description of the computational process, its output and the code used. This comment may date from a time when Github repositories of code, data and outcomes were less common, but it still seems worth restating today. The second is to pick up on an older theme elucidated by Milner in his lecture inaugurating the Laboratory for the Foundations of Computer Science at the University of Edinburgh \cite{Milner87}. While the aims of this paper are far more modest and in no way comparable to the grand program of research outlined therein, our approach is very much consistent with Milner's view that ``the design of computing systems can only properly succeed if it is well grounded in theory, and that the important concepts in a theory can only emerge through protracted exposure to application''. Here the theory and corresponding systems are very narrow and focused, but we believe there is an echo in this work of the same synergistic relationship between theory and application. 

\section{Preliminaries and Definitions}

There are many definitions of Turing machines \cite{Turing36}, with the differences being mostly inessential. Here we use a definition based on that of Sudkamp \cite{Sudkamp} but for which some slight modifications are made, following \cite{Harland16}. For our machines, the tape alphabet and the input alphabet are the same. 

\begin{definition}[Turing machine] \label{def:tm}
	A Turing machine is a quadruple $(Q \cup \{z\}, \Gamma, \delta, a)$ where
	
	\begin{itemize}
		\item $Q$ is a finite set of states
		\item $z$ is a distinguished state called the {\em halting state}
		\item $\Gamma$ is the tape alphabet
		\item $\delta$ is a partial function from $Q \times \Gamma$ to $Q \cup \{z\} \times \Gamma \times \{l,r\}$ called the {\em transition function}
		\item $a \in Q$ is a distinguished state called the {\em start state}
	\end{itemize} 
	
	An {\em $n$-state Turing machine} is one in which $|Q| = n$. 
	An {\em $m$-symbol Turing machine} is one in which $|\Gamma| = m$.  
\end{definition}

Note that due to the way that $\delta$ is defined, this is the so-called \textit{quintuple transition variation} of
Turing machines, in that a transition must specify for a given input state and input character, a new state, an output character and a
direction for the tape in which to move. Hence a transition can be specified by a quintuple of the form

\begin{center}
	$<State, Input, Output, Direction, NewState>$
\end{center}
where $State \in Q$, $NewState \in Q \cup \{z\}$, $Input, Output \in \Gamma$ and $Direction \in \{l,r\}$. 

We call a transition a \textit{halting transition} if $NewState = z$; otherwise it is a \textit{standard transition}. 

A Turing machine can be characterised by the tuples which make up the definition of $\delta$, assuming that there is some notational convention that is used for identifying the start state and halting state. 
In this paper machine states are labelled $a,b,c,d,e \ldots$ where $a$ is the initial state of the machine. 
The halting state is labelled $z$. Other states are referred to as \textit{standard states}. 
Symbols are labelled $0,1,2,3,\ldots $ where $0$ is the blank symbol.\footnote{Note that this means every tape alphabet must include the blank symbol.} This means that for any Turing machine we have $\Gamma = \{0,1,2,\ldots,m\}$ for some $m \geq 1$. 
%  We will also use $\_$ to indicate a variable whose values may be arbitrary but whose name is not significant (a la Prolog \cite{swi}), so that for example a tuple in which \textit{Input} is 1, \textit{Output} is 0 and all other values are arbitrary will be denoted as $(\_, 1, 0, \_, \_)$. 

As the study of the busy beaver problem involves limiting the size of Turing machines, we need to be specific about how we count states. 
As above, an $n$-state Turing machine has $n$ standard states and a halting state labelled $z$. Note that there are no transitions from state $z$, and that as $\delta$ is a partial function, there is at most one transition for any given pair of a standard state and a character in the tape alphabet. This means that none of our Turing machines are \textit{non-deterministic}\footnote{Our class of machines is sometimes called \textit{semi-deterministic}, i.e., there is \textit{at most} one transition for each possible combination of state and input symbol, rather than \textit{exactly} one transition for each such possible combination.}, so there is never a case when there are two possible transitions for a given state and input symbol. Note that it is possible for there to be no transition for a given state and symbol combination (i.e.\ for states other than just $z$). If such a combination is encountered during execution, the machine halts.  

A configuration of a Turing machine is the current state of execution, containing the current tape contents, the current machine state and the position of the tape head \cite{Sudkamp}. An informal definition like this is generally sufficient for most purposes. However, as we will be dealing later with methods for predicting tape configurations, and with more complex definitions of the corresponding configurations, we will make this definition formal. 

\begin{definition}[Turing machine tape] \label{def:tape}
	A {\em tape} is a total function from $\mathbb{Z}$ to the tape alphabet $\Gamma$.
	
	A {\em configuration} is a tuple $<S, T, i>$ where $S \in Q \cup \{z\}$, $T$ is a tape and $i \in \mathbb{Z}$. 
\end{definition}

Thus a configuration consists of the current state ($S$), the current tape contents ($T)$ and a specific tape cell which is currently under consideration ($i$). It is common for $i$ to be implicit, in that as long as the current cell is known (often by specifying a ``tape pointer''), the precise value of $i$ is often unimportant.
We distinguish one special tape $blank$, for which $blank(i) = 0 \; \forall i \in \mathbb{Z}$. 

Note that here we use $\mathbb{Z}$ as the ``index'' set for the tape. This reflects common practice in the busy beaver literature in which there is a preference for the particular variety of Turing machines which contain a single tape that extends infinitely in either direction. Of course in principle there are many choices for the index set, such as $\mathbb{N}$ for a semi-infinite tape, or $\mathbb{Z} \times \mathbb{Z}$ for two infinite tapes, or any countable set that suits. Whichever such set is chosen, we require that there be specific member of the index set which we use as the \textit{origin}, which is the initial value of $i$. When the index set is $\mathbb{Z}$, the origin is $0$. 

As we will introduce generalisations of both of these notions later, we will refer to these as \textit{standard tapes} and \textit{standard configurations} as necessary to avoid ambiguity. 

A computation consists of a sequence of configurations, starting from a given initial configuration, and with the machine specifying how one configuration leads to the next one. 
The initial configuration we denote as $config(0)$. For the busy beaver and related problems, $config(0)$ will always be $<a, blank, 0>$. 

The combination of an initial configuration $config(0)$ and a Turing machine $M$ is sufficient to define a unique potentially infinite sequence of configurations, as follows.
%\footnote{A non-deterministic machine may lead to several such sequences of configurations. We do not consider such machines in this paper.}

\begin{definition}[Turing machine configuration] \label{def:config}
	
	Let $M$ be a Turing machine and $config(n)$ = $<S,T,i>$ be a configuration. 
	
	If there is a tuple in $M$ of the form $<S, I, O, D, S'>$ where $I = T(i)$, then 
	$config(n+1)$ is the tuple $<S', T', k>$, where 
	
	\begin{itemize}
		\item $T'$ is the total function in which $\forall j \not = i \; T'(j) = T(j)$ and $T'(i) = O$
		\item $k = i+1$ if $D = r$ and $k = i-1$ if $D = l$.
	\end{itemize}
	
	Otherwise, $config(n+1)$ is undefined. 
\end{definition}

Note that $config(n+1)$ is only defined if there is a tuple in $M$ which matches the current state ($S$) and input symbol ($I$) in $config(n)$. Hence if the current state in $config(n)$ is $z$, there can be no such tuple, and so there is no successor configuration. It is also possible that there is no successor configuration in the case when the current state is not $z$, i.e., there is a combination of state and symbol which is not present in the transitions of $M$. 

We will refer to the sequence of configurations as a \textit{computation}, or the \textit{execution} of the machine. Note that a computation may be either finite or infinite. 
Note also that given our initial configuration, at any point during execution all but a finite number of the symbols on the tape will be blank. Hence it is often convenient to use the notation $111\{b\}011$ to denote a configuration in which the Turing machine is in state $b$ with the string 111011 on the tape and the tape head pointing at the 0, and in which all other tape cells are blank. 

A further point to note is that we will sometimes refer to this notion of execution as \textit{naive computation}, as we will shortly introduce some more intricate forms. We will also use the term \textit{hops} to refer to the number of transitions made in this form, so that the execution from $config(n)$ to $config(n+1)$ takes one hop. We will use the term \textit{steps} to refer to the number of transitions made in the more intricate computations to come, with the key point being that one step may correspond to a large number of hops, and the larger the number, the greater the efficiency of execution. 

The following definitions are the same as in \cite{Harland16,Harland22}.

\begin{definition}[Turing machine dimension, productivity and activity] \label{def:dim}
	Let $M$ be a Turing machine with $n$ states and $m$ symbols (where $n,m \geq 2$). 
	Then we say $M$ has \emph{dimension} $n \times m$.
	
	We say $M$ has \emph{activity} $k$ if the execution of $M$ on the blank input terminates in $k$ steps. If the execution of $M$ on the blank input does not terminate, we say the activity of $M$ is $\infty$. 
	
	The \emph{productivity} of a machine $M$ whose activity is finite is the number of non-blank characters on the tape after $M$ has halted. If the activity of $M$ is $\infty$, then the productivity of $M$ is not defined.
\end{definition}

\section{Executing Machines} 
\label{section:otter}

\subsection{Extended Transitions}

Turing machines are conceptually simple, particularly when it comes to executing them. However, as they are mostly used as small examples of particular computations, there is comparatively little work done on efficient execution of Turing machines for computations of significant length. One such technique is the \textit{macro machines} of Marxen and Buntrock \cite{HM90}. Essentially, the tape is considered to consist of blocks of characters of a given fixed length $k$, and the transitions for the machine are determined according to this structure. For example, if $k=3$, then when the machine encounters say the input string $001$ in state $c$, then it will execute naively until either the tape head moves outside the region of the tape containing these three specific symbols, halts, or repeats an earlier configuration.\footnote{Note that these are the only three possibilities, as the tape region is finite in length, and there are only a finite number of possible states and strings in this tape region.} In the last case, it is clear that the computation does not terminate. In the middle case, the computation halts. 
Otherwise, we note the state in which the machine exits from the area containing just the three given symbols, and the symbols that now occupy these places, and consider this a new `macro' transition for the machine. The only additional information that is necessary is the direction from which the tape head entered the current string, and the direction that it exited from the string. This means that for a given string, there are two transitions: one for when it enters from the left, and another for when it enters from the right (which of course coincide when the length is $1$). This also means that we need to keep track of not just the current (macro) symbol, but the current orientation of the tape head (left or right). Extending the notation for configurations above, we will use $111 \{b\} 001 \{l\} 110$ to denote a configuration in which the machine is in state $b$ with the string 111001110 on the tape, and the tape head pointing to the left hand end of the 001 string. We will use $111 \{b\} 001 \{r\} 110$ to denote the same configuration except that the tape head in this case is pointing to the right hand end of the 001 string. 

For example, consider a machine which encounters the string $001$ in state $c$. If the initial symbols $001$ are converted to $111$ and when the tape head leaves these three symbols, the machine is in the state $d$, then we may consider this a macro transition from state $c$ with input $001$ to new state $d$ with output $111$. The details are described by Marxen and Buntrock \cite{HM90}, and there are various extensions to this technique described on Marxen's web site \cite{Marxen} and in other places \cite{Holkner04}. 

This technique has been used to successfully execute some machines that take a very large number of steps to terminate \cite{HM90,Holkner04,Harland16}. However, one issue that arises is the correct choice of the length of the tape element, and the most efficient size to use varies from machine to machine. In addition, it is not obvious that having such a fixed size is necessarily the most appropriate way in which to execute all relevant machines. It seems natural to us to use a similar idea, but not to assume that the tape is uniformly divided into fixed-size blocks. Not only does this avoid the (minor but annoying) problem of having to find the most appropriate block size for a given machine, it also potentially allows for a more flexible analysis of the behaviour of machines. Accordingly, we do not assume that the tape is uniformly divided into fixed-size blocks, but rather we allow tape elements to be of any size, independently of the size of any other tape elements. This provides a more flexible tape representation, but which requires some more sophisticated execution techniques. 

A further issue is that macro machines per se can only increase efficiency by a constant factor, as they replace a fixed number of hops with a single step. While this is certainly a worthwhile increase, as shown in \cite{Harland16}, more significant increases are possible. However, these more sophisticated techniques require further analysis to ensure that they are verifiably correct. We will return to this issue in Section~\ref{section:shape}. 
%\textbf{Also need to emphasise the need for verification, and that macro machines alone only provide a linear speed up.}

We will now consider the appropriate modifications to our notion of tape, configuration and computation. 
Each unit of a tape is now not a single symbol, but a block of symbols.
As noted previously \cite{HM90, Harland16}, this naturally leads to considering ``compressed'' blocks of the form $(110)^3$ rather than the equivalent sequence of three blocks of $110$. Hence each block will consist of a string of tape symbols together with an exponent, which specifies the number of copies of the symbol.\footnote{We will often use some obvious simplifications in our notation, such as writing $0$ instead of $0^1$.} This simple technique alone can vastly reduce the size of the tape representation required. For example, a well-known example of a 6-state 2-symbol machine (Dragon 72 in \cite{Harland16}) terminates in $10^{36,536}$ steps leaving a pattern of the form $1(110)^X 11 $ on the tape, where $X$ is a number with $18,268$ digits. 

We now turn to the appropriate formal definitions. We will later have the need (see Section~\ref{section:shape}) to include not just natural numbers as exponents, but also variables and expressions based on them, such as $X + 2Y + 1$. This means that we make the definitions more general than strictly necessary at this point. 

\begin{definition}[Arithmetical Expressions]
	Let $V$ be a set of variables. We denote by $Exp(V)$ the expressions that can be formed from the binary operations $+$, $\times$, $-$, \verb!^!, $//$ and $\; \emph{mod}$ (where \verb!^! denotes exponentiation, and $//$ integer division) and elements of $V \cup \mathbb{N}$. % Do we need ``real'' division as well?
	We will refer to $Exp(V)$ as the set of \emph{arithmetical expressions} over $V$. 
	We denote by $vars(E)$ the variables that occur in the expression $E$. 
	
	If $vars(E) = \emptyset$ then we refer to $E$ as a \emph{numerical expression}.  
\end{definition}

We will normally require that in any expression of the form $X // Y$ or $X \; mod \; Y$ that $Y \in \mathbb{N} \backslash \{0\}$. % In other words, any divisor must be a specific positive integer. 

It should be noted that the variables in these expressions are over $\mathbb{N}$, and so these are expressions over integer arithmetic. In particular, we note the identity below; this will important in our later work (see Section~\ref{sec:moduli} for more details). 

\begin{equation}\label{equation:mod}
	m = n(m//n) + (m \; mod \; n) \;\; \forall n \in \mathbb{N} \backslash \{0\},  \forall m \in \mathbb{N}
\end{equation}

We now turn to the generalisation of a tape. 

\begin{definition}[Turing machine tape (generalised)]
	Let $V$ be a set of variables. 
	
	A {\em tape element} is an ordered pair $(I,E)$ where $I\in \Gamma^+$ and $E \in Exp(V)$. We refer to $I$ as the \emph{input string} and $E$ as the \emph{exponent}. 
	
	A {\em block tape} is a finite sequence of tape elements $(I_1,E_1), (I_2,E_2), \ldots , (I_k,E_k)$. 
	
	We denote this as $(I_1,E_1).(I_2,E_2). \cdots .(I_k,E_k)$. We say the {\em length} of the block tape is $k$. 
	
	We denote by {\em blank} the empty block tape (i.e.\ when the length of the block tape is 0). 
	
	A {\em block configuration} is a tuple $<Left, S, (I,E), Dir, Right>$ where $S \in Q \cup \{z\}$, $Left$ and $Right$ are block tapes, $(I,E)$ is a tape element and $Dir \in \{l,r\}$. 
	
	For a given block tape $T$, we define {\em next(T)} as follows.
	
	\begin{itemize}
		\item If the length of $T$ is 0, {\em next(T)} is $(0,1)$
		\item If the length of $T$ is greater than 0, {\em next(T)} is the first element of $T$
	\end{itemize}
	
	For a given block tape $T$, we define {\em rest(T)} as follows.
	
	\begin{itemize}
		\item If the length of $T$ is 0 or 1, {\em rest(T)} is $blank$
		\item If the length of $T$ is greater than 1, {\em rest(T)} is the tape containing all elements of $T$ apart from the first element
	\end{itemize}
	
	Hence in all cases we have $T = next(T).rest(T)$. 
	
	%For a given block tape $T$, we denote by {\em next(T)} the tape element $(0,1)$ when the length of $T$ is 0, and $(I,E)$ when the first element of $T$ is $(I,E)$. Similarly we denote by {\em rest(T)} the tape $blank$ when the length of $T$ is 0 or 1, and the tape containing all but the first element of $T$ otherwise.
	
	We say a tape element (respectively block tape, block configuration) is {\em numerical} if every exponent in the tape element (resp.\ block tape, block configuration) is a numerical expression. Otherwise we say the tape element (resp.\ block tape, block configuration) is {\em variable}. 
	
	The {\em size} of a tape element $(I,E)$ is $|I| \times E$. 
	
	The {\em size} of a block tape is the sum of the sizes of each tape element in it. 
	
	The {\em size} of a block configuration $<Left, S, (I,E), Dir, Right>$ is the sum of the sizes of $(I,E)$, $Left$, and $Right$. 
\end{definition}

As noted above, at any point in a computation, there are only finitely many non-blank characters on the tape, and so with block tapes we only give an explicit representation of these non-blank characters (i.e.\ the tape is blank apart from those elements explicitly listed in the block tape). Clearly for any numerical block tape, there is a corresponding standard tape, provided that there is a way to determine the origin of the standard tape from the block tape. We do so by the use of what we term the \textit{reference position;} see the remarks on this below. 

\begin{figure}[tb]
	\centering
	\includegraphics[width=0.7\textwidth]{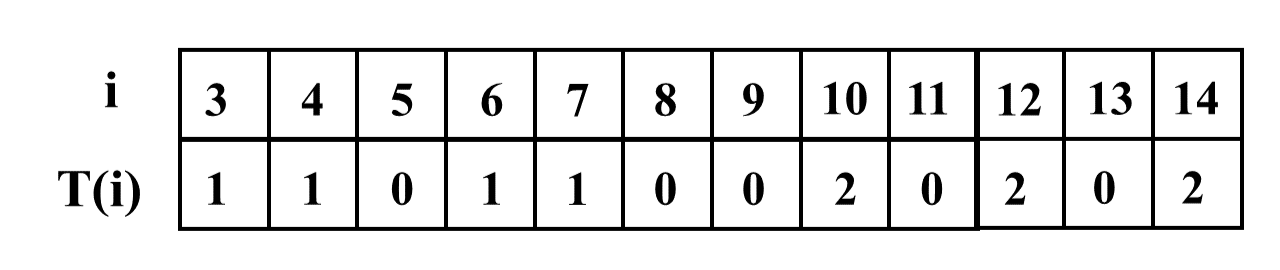} 
	\caption{Block tape example}
	\label{fig:tape}
\end{figure}

Note that the size of the tape element $(I,E)$ is the arithmetical expression $|I| \times E$. Hence in the standard tape $T$ corresponding to the block tape $(I_1,E_1)$, $(I_2,E_2), \ldots (I_k,E_k)$, if $T$ has the symbols from $(I_1,E_1)$ starting at $j$, the symbols from $(I_2,E_2)$ start at $j + |I_1| \times E_1$, and so on. For example, given the block tape $(110)^2(02)^3$, if the leftmost symbol is at position 3, in the corresponding standard tape $T$ we have $T(3) = T(4) = T(6) = T(7) = 1$, $T(5) = T(8) = T(9) = T(11) = T(13) = 0$, $T(10) = T(12) = T(14) = 2$, and all other tape cells are blank, as can be seen in Figure~\ref{fig:tape}.  

Note also that for a block tape $T = (I_1,E_1).(I_2,E_2).T'$, we have $next(T) = (I_1,E_1)$ and $rest(T) = (I_2,E_2).T'$. 

Configurations are a little more complex when block tapes are involved. This is due to some extra information that is needed, as well as the more complex way in which movement occurs. Firstly, as discussed above, there are two transitions for each tape element, one for each orientation of the tape head (left or right) \cite{Holkner04}. Secondly, there are more possible ways in which the movement may occur than simply one step to the left or right. For example, if the current tape element under consideration is $(110)^6$ from the left in state $b$, and the resulting transition exits to the right of $(110)$ also in state $b$, then the machine will move past all of the tape element $(110)^6$ \cite{Holkner04}. However, if the same occurs but with an output state different from $b$, then the next configuration will have the tape pointer on the left of the tape element $(110)^5$. 

We now define a new notion of transition for a machine. 

\begin{definition}[Extended transitions]
	An extended transition for a Turing machine $M$ is a tuple  
	
	$$<S, I, InDir, O, OutDir, S', Hops>$$ 
	
	where $S \in Q$, $S' \in Q \cup \{z, loops\}$, $I, O \in \Gamma^+$, $|I| = |O|$, $InDir, OutDir \in \{l,r\}$ and $Hops \in \mathbb{N} \backslash \{0\}$. 
\end{definition}

Note that $I$ and $O$ are now non-empty strings of symbols, not just single symbols. 

The transition $<S, I, InDir, O, OutDir, S', Hops>$ indicates that if $config(n)$ is in state $S$ with the tape index at the left or right end of $I$ as indicated by $InDir$, then $config(n + Hops)$ is in state $S'$, with $O$ on the tape in place of $I$, and the tape head moves in the direction indicated by $OutDir$. Note that we must have that $|I| = |O|$.  We include also the new distinguished state $loops$, for reasons that will become clear shortly.  

\begin{figure}[tb]
	\centering
	\begin{tabular}{cc}
		\includegraphics[width=0.48\textwidth]{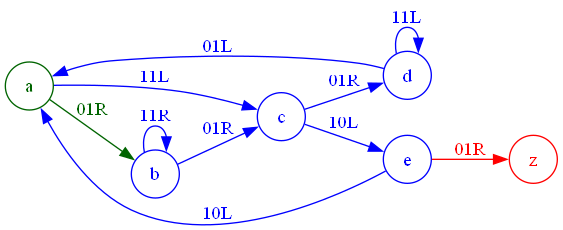} & \includegraphics[width=0.48\textwidth]{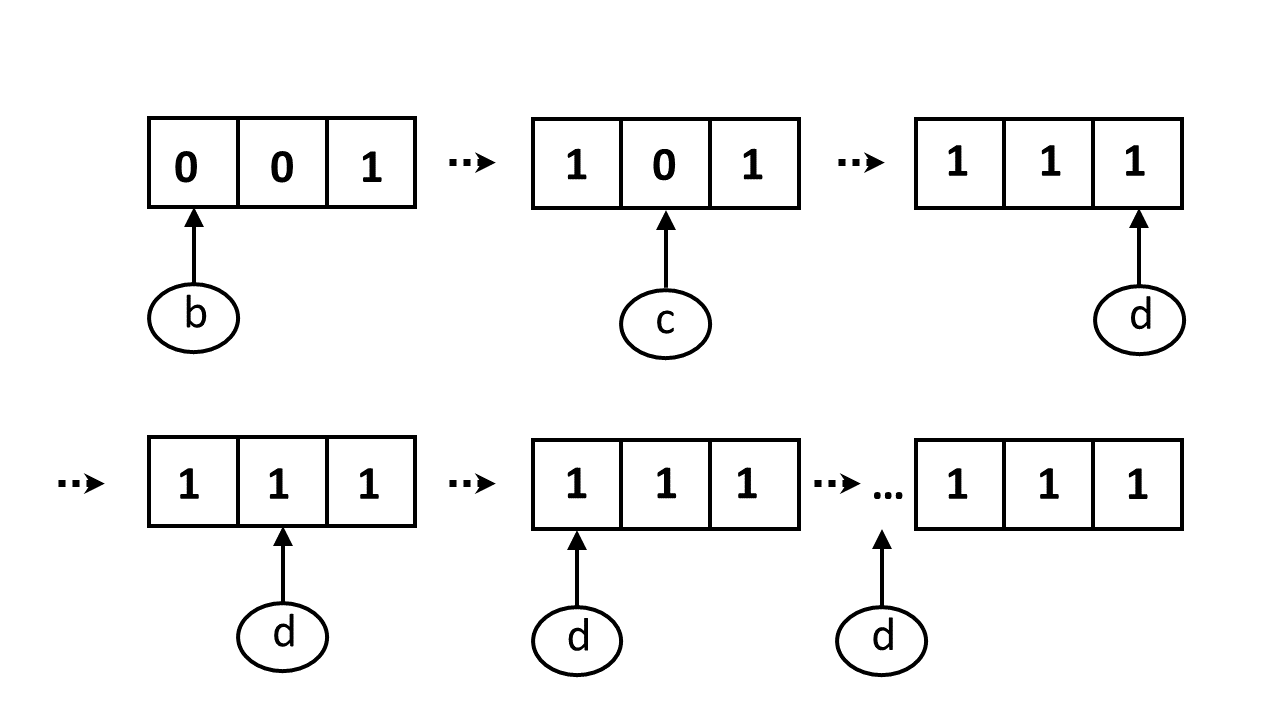} \\
	\end{tabular}
	\caption{Dragon 18 and extended transition}
	\label{fig:dragon18}
\end{figure}

For example, consider the machine in Figure~\ref{fig:dragon18}, which is the  5-state 2-symbol champion machine \cite{BBChallenge25,BBChallenge26} (Dragon 18 in \cite{Harland16}). 
From this machine, we can derive the transition $<b, 001, l, 111, l, d, 5>$ (see below for how this is done). 
This means that if $config(n)$ is say $1 \{b\} (001)^3 \{l\} \; 101$, applying this transition gives $config(n+5)$ as $\{d\}1\{l\} \; (111)^1 (001)^2 101$. If the transition was $<b, 001, l, 111, r, d, 5>$ instead, we would get $config(n+5)$ as $1 (111)^1 \{d\}(001)^2\{l\} \; 101$. This is shown in more detail in Figure~\ref{fig:dragon18}.  

Clearly the extended transitions of a given machine are determined by the original transitions of the machine. The way in which extended transitions are derived from the original ones is as follows. Given an $n$-state $m$-symbol Turing machine $M$, a tape element $(I,E)$ where $|I| = k$, a state $S$ and an initial direction $InDir$, we execute the machine (naively) on the string $I$, commencing in state $S$ and at the left or right end of $I$ as indicated by $InDir$. This execution proceeds until one of the following conditions occurs.

\begin{enumerate}
	\item The number of hops exceeds $n \times k \times m^k$
	\item The machine halts, with $O$ in place of $I$ on the tape element
	\item Execution moves the machine beyond the boundaries of $I$ in state $NS$, with $O$ in place of $I$ on the tape element
\end{enumerate}

We discuss each of these cases below. 

\begin{enumerate}
	\item The new transition is $<S, I, InDir, O, OutDir, loops, Hops>$, where $O$, $OutDir$ and $Hops$ are all undefined. This is because there can be at most $m^k$ possible tape elements of length $k$, and hence there are at most $n \times k \times m^k$ possible configurations (as there are $n$ states and $k$ different possible positions for the tape position in this element). Once the number of hops exceeds this number, we know the machine must be stuck in a loop, which we indicate by the special state $loops$. 
	
	\item The new transition is $<S, I, InDir, O, r, z, Hops>$, where $Hops$ is the number of hops taken since the start of the computation on this element. The machine may have halted either because it has encountered the halt state, or because there is no appropriate transition. As $OutDir$ is arbitrary in this case, we choose $OutDir = r$. It would be possible to include another additional special state, such as $abort$, to account for the second possibility (i.e.\ that no appropriate transition exists). However, distinguishing types of termination in this way is not important for us. If desired, it is always possible to add extra halting transitions to a machine which lacks a specific transition, in order to ensure that the only way a machine halts is when it encounters the halting state.\footnote{This increases by one the number of steps a machine takes to halt, which can become an annoying distraction when comparing results using slightly different conventions such as these.}
	
	\item The new transition is $<S, I, InDir, O, OutDir, S', Hops>$ where $S'$ is the new state, $Hops$ is the number of hops taken since the start of the computation on this element, and $OutDir$ is either $l$ or $r$, depending on the direction of exit from the element. 
\end{enumerate}

Hence it is relatively straightforward to compute an extended transition during the execution of a machine. Initially, there are no known extended transitions. When an ``unknown'' transition is encountered for the first time (i.e.\ the machine does not contain an extended transition for the current tape element), we determine the appropriate extended transition as above, and add it to the machine. For example, in Dragon 18 above, the first time the machine is in state $b$ with the input $001$ and pointing to the first 0, the extended transition $<b, 001, l, 111, l, d, 5>$ is calculated and added to the definition of the machine. The next time that this same state and input is encountered, the transition is already known, and hence need not be recomputed. In principle, it would be possible to compute many such extended transitions before commencing execution; in practice, comparatively few extended transitions tend to be needed, and so computing them as required seems appropriate. 

This leads us to an extended notion of a configuration. 

\begin{definition}[Block configurations]
	A {\em block configuration} is a tuple $<Left, S, (I,E), Dir, Right>$ 
	where $S \in Q \cup \{z\}$, 
	$(I,E)$ is a tape element, 
	% $Pos \in \mathbb{Z}$, 
	$Dir \in \{l,r\}$, and 
	$Left$ and $Right$ are block tapes. 
\end{definition}

The block configuration $<Left, S, (I,E), Dir, Right>$ indicates that the machine is currently in state $S$, with the current tape element $(I,E)$ and the two block tapes $Left$ and $Right$ on either side. 
If $Dir$ is $l$, then the tape pointer indicates the position of the leftmost  symbol of the tape element $(I,E)$. 
If $Dir$ is $r$, then the tape pointer indicates the position of the rightmost symbol of the tape element $(I,E)$. 
In either case, we refer to the position of the tape pointer as the {\em reference position} and the element $(I,E)$ as the {\em reference element}. 
It should be obvious that if we can map the reference position to a specific tape index, then it is possible to reconstruct a standard configuration (and hence a standard tape) from any numerical block configuration. 

We will some times refer to the block tape of a given configuration, by which we mean the tape consisting of the reverse of $Left$ followed by the tape element $(I,E)$ followed by $Right$. 

For example, the first configuration in Figure~\ref{fig:dragon18} is the block configuration $<L, b, (001,1), l, R>$, and the third such configuration is $<L, d, (111,1), r, R>$, where $L$ and $R$ are the contents of the tape to the left and right respectively of the reference elements $(001,1)$ and $(111,1)$ respectively. 

By default, the initial configuration for any computation is $<blank, a, (0,1), l, blank>$ which we denote as $config(0)$. As we shall see, we will have reason to use some other initial configurations as well. 

We are now in a position to define computations for block configurations. 

\begin{definition}[Turing machine execution (extended)]\label{def:extended}
	Let $M$ be a Turing machine and $config(n)$ = $<L, S, (I,E), InDir, R>$ be a block configuration. 
	
	If there is an extended transition $T$ for $M$ where $T = <S, I, InDir, O, OutDir, S', Hops>$, then the next configuration is as defined below.
	
	\begin{center}
		\begin{tabular}{llll}
			\textbf{Case} & \textbf{Condition} & \textbf{Configuration} & \textbf{Next} \\
			1 & $InDir = OutDir = l$ & $<rest(L), S', next(L), r,  (O,1).(I,E-1).R>$ & $n + Hops$ \\
			2 & $InDir = OutDir = r$ & $<(O,1).(I,E-1).L, S', next(R), l,  rest(R)>$ & $n + Hops$ \\
			3 & $InDir = l$, $OutDir = r$, $S \not = S'$ & $<(O,1).L, S', (I,E-1), l, R>$ & $n + Hops$ \\
			4 & $InDir = r$, $OutDir = l$, $S \not = S'$ & $<L, S', (I,E-1), r, (O, 1).R>$ & $n + Hops$ \\
			5 & $InDir = l$, $OutDir = r$, $S = S'$ & $<(O,E).L, S', next(R), l, rest(R)>$ & $n + (Hops \times E)$ \\
			6 & $InDir = r$, $OutDir = l$, $S = S'$ & $<rest(L), S', next(L), l, (O,E).R>$ & $n + (Hops \times E)$ \\
		\end{tabular}
	\end{center}
	
	In each case, $config(Next)$ is the configuration given.  
	
	We say $T$ is an \emph{acceleration transition} if either Case 5 or Case 6 holds. Otherwise it is a \emph{normal transition.} 
\end{definition}

Note that acceleration transitions are unaffected by the value of $E$, and in particular whether it is a numerical transition or otherwise. For normal transitions, we clearly must have that $E \geq 1$. 

\begin{figure}[tb]
	\centering
	\begin{tabular}{cc}
		\includegraphics[width=0.48\textwidth]{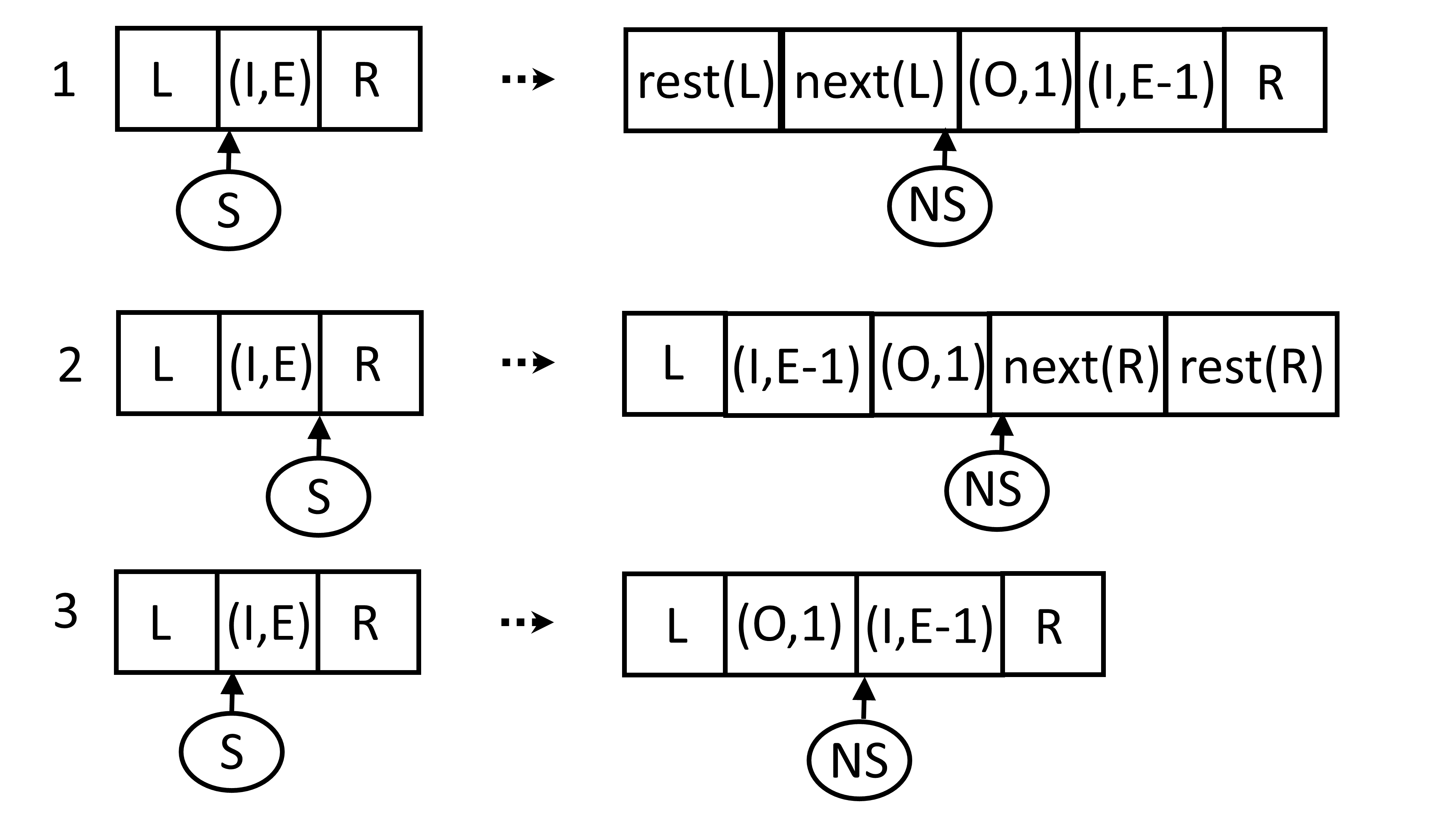} & \includegraphics[width=0.48\textwidth]{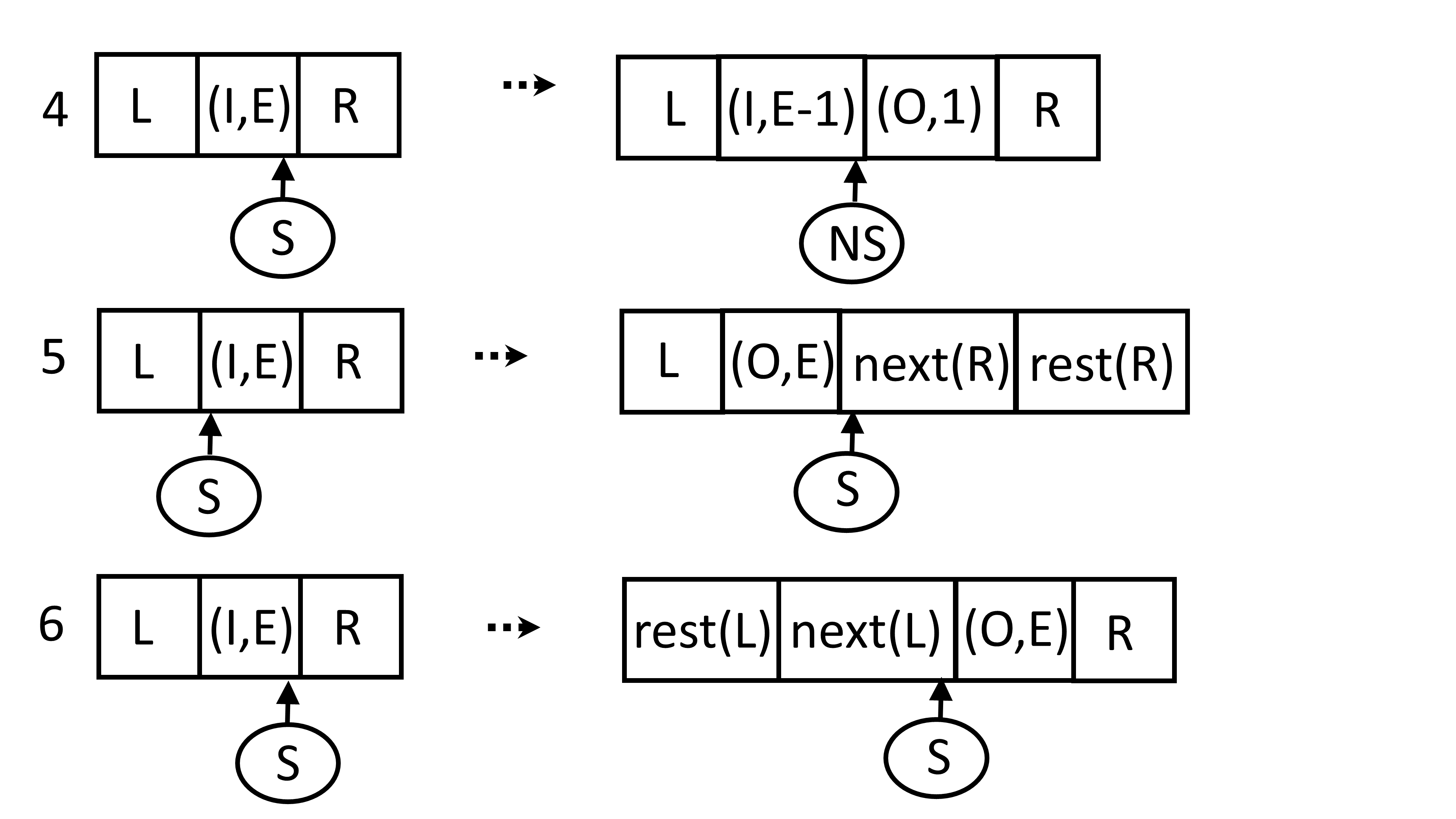} \\
	\end{tabular}
	\caption{New transitions}
	\label{fig:newtransitions}
\end{figure}

A diagrammatic illustration of each of these six cases is given in Figure~\ref{fig:newtransitions}. 

Note that in cases 1, 2, 3 and 4 above, if $E = 1$, then the tape element $(I,E-1)$ will be empty. Hence care needs to be taken when doing this to ensure that only non-blank tape elements are produced (which can be done by a simple test after the transition is performed). When $E$ is a variable expression, such as $X-2$, these same cases will require that $E-1 \geq 0$, and so these cases may generate extra constraints on the values of any variables present in the exponent. 

Cases 5 and 6 are known as the \textit{acceleration cases}, as when $E$ is a large number, these significantly improve performance \cite{Holkner04}. Note that these cases are appropriately defined for variable tape elements as well as numerical ones, with implication that $Hops$ may now also be an arithmetical expression rather than a natural number. 

As before, this defines a potentially infinite sequence of configurations. Again, a given configuration may have no successor, either due to being in the halting state, or because there is no extended transition for the current state and tape element. The main difference is that the successor block configuration to $config(n)$ need not be $config(n+1)$.

A pragmatic issue that arises during computations is the ability to compress block tapes appropriately. Some cases are obvious, such as the need to compress the block tape $(1,1).(001,1).(001,1).(001,1).(001,1)$. However, it is less clear which of the compressed tapes below is most appropriate. 

\begin{center}
	$(1,1).(001,4)$ \\
	$(10,1).(010,3).(01,1)$ \\
	$(100,4).(1,1)$ \\
\end{center}

The equivalence of these three representations is analogous to the equivalence of the regular expressions $1(001)^+$, $10(010)^*01$ and $(100)^+1$ \cite{Sudkamp}. We will refer to such equivalent representations as \emph{rotations}. For a tape element of size $k$, it should be clear that it is one of at most $k$ such rotations. This means that at key points in computations, we may need to consider alternative rotations. 

A further note is that the variable size blocks means that we may sometimes need to compress the current configuration. For example, when encountering a configuration such as $1\{b\}1\{l\}010102$, it seems natural and appropriate to use run-length compression to represent this configuration as $1\{b\}(10)^3\{l\}2$. Similarly a later step in execution may result in a configuration such as $1(11)^3\{c\}2\{l\}$ which we should compress in a likewise manner to $1^7\{c\}2\{l\}$.

% \textbf{\textit{More?}}

\subsection{Observant Otters}
\label{section:shape}

As discussed in \cite{Harland16}, many of the \textit{dreadful dragon} machines involve computations that have some predictable patterns. For example, consider the machine in Figure~\ref{fig:dragon18} again, whose computation includes the following sequence of configurations. 

\begin{center}
	\begin{tabular}{ll}
		\textbf{Hops} & \textbf{Configuration} \\
		131 & $\{a\}0\{r\} 1^3 (001)^6 1$ \\ 
		145 & $\{a\}0\{r\} 1^8 (001)^5 1$ \\
		169 & $\{a\}0\{r\} 1^{13} (001)^4 1$ \\
	\end{tabular}
\end{center}

Clearly there is a pattern at work here, and it is not hard to see that the following is the completion of this sequence. 

\begin{center}
	\begin{tabular}{ll}
		\textbf{Hops} & \textbf{Configuration} \\
		131 & $\{a\}0\{r\} 1^3 (001)^6 1$ \\
		145 & $\{a\}0\{r\} 1^8 (001)^5 1$ \\
		169 & $\{a\}0\{r\} 1^{13} (001)^4 1$ \\
		203 & $\{a\}0\{r\} 1^{18} (001)^3 1$ \\
		247 & $\{a\}0\{r\} 1^{23} (001)^2 1$ \\
		301 & $\{a\}0\{r\} 1^{28} (001)^1 1$ \\
		365 & $\{a\}0\{r\} 1^{33} 1$ \\
	\end{tabular}
\end{center}

Detecting this pattern and predicting the resulting configuration, without having to compute all of the intermediate ones, is the aim of the heuristic we call the \textit{observant otter} \cite{Harland16}. For the above example, once we have noticed the pattern for the first three configurations, we observe that all three have the same ``shape'', i.e.\ are of the form $a\{0\}r \; 1^X (001)^Y 1$, and that the above sequence can be summarised as 

$$\{a\}0\{r\} \; 1^X (001)^Y 1 \longmapsto \{a\}0\{r\} \; 1^{X+5} (001)^{Y-1} 1$$

with each such ``transition'' taking $2X+8$ hops, with the sequence commencing with $X=3$ and $Y=6$. We can determine this transition
by performing a slightly different computation, which commences with the variable configuration $00\{a\}0\{r\} \; 1^X (001)^Y 1$ and terminates when a second variable configuration with the same basic pattern is encountered, as below. 

\begin{center}
	\begin{tabular}{ll}
		\textbf{Hops} & \textbf{Configuration} \\
		0      & $00\{a\}0\{l\}1^{X}(001)^{Y}1$       \\
		1      & $00 1 \{b\}1^{X}\{l\} (001)^{Y}1$1   \\
		$X+1$  & $00 1^{X+1}\{b\}(001)^{Y}\{l\}1$      \\
		$X+6$  & $00 \{d\}1^{X+1}\{r\}1^3(001)^{Y-1}1$ \\
		$2X+7$ & $0\{d\}0\{l\}1^{X+4}(001)^{Y-1}1$ \\
		$2X+8$ & $\{a\}0\{l\}1^{X+5}(001)^{Y-1}1$ \\
		& \\
	\end{tabular}
\end{center}

Note that from the configuration at Hops $X+6$ in this sequence we derive the constraint that $Y \geq 1$. 
We can also see that by repeatedly applying this pattern, we can deduce that the configuration $\{a\}0\{r\} \; 1^3 (001)^6 1$ will lead to the configuration $\{a\}0\{r\} \; 1^{33} 1$ in 234 steps, and that more generally the configuration $\{a\}0\{r\} \; 1^X (001)^Y 1$ will lead to the configuration $\{a\}0\{r\} \; 1^{X+5Y} 1$ in $5Y^2 + 2XY + 3Y$ steps (see Section~\ref{sec:hopscalc} for details on how this is calculated). 

So to find such patterns, we first need to find a sequence of configurations of the same shape, such as those in Hops 131, 145 and 169 above. We then need to determine which parts of this shape are the same in each occurrence, and which vary. Finally we need to check that for the parts that vary, there is at least one ``descending'' variable (in the example above, this is $Y$) and that all of the others increase (in the example above,  $X$). The need for the descending variable $Y$ is to ensure that the process with terminate, and within a predictable number of steps. 
We can then predict the final configuration in this sequence. 

As we shall see, the precise increase in the ``increasing'' variable/s has a critical bearing on the magnitude of the sequence; in particular, a pattern involving a variable increasing from $X$ to say $X+5$ turns out to be much less significant in size than one for which the increase is from $X$ to say $2X+5$. This also requires more complex calculations to determine the final configuration in the sequence. 
The exact number of matching configurations that is required to determine a pattern is not obvious; however it seems reasonable to assume that 3 is a working minimum (as used in \cite{Harland16}), especially as some initial experiments have shown that reducing this minimum to 2 tends to generate far too many false positives for such patterns. In general it seems reasonable to expect this minimum to increase as the dimensions of the machines analysed increases. 

This means that our technical agenda is to find a way to formalise the process of finding and analysing patterns, in such a way that we can define transitions based on such patterns to execute in one step. In other words, we want to generalise the notion of (extended) transitions to include a new type, in which we can go from the configuration $\{a\}0\{r\} \; 1^X (001)^Y 1$ immediately to $\{a\}0\{r\} \; 1^{X+5Y} 1$ by applying an appropriate transition. This will be very much analogous to transitions in macro machines, in that when we find a section of the tape which matches a particular pattern, we apply the transition to this section of tape, independently of the rest of the tape. As we shall see, it sometimes takes a little care to define the appropriate section of tape exactly.

The first step in finding such a transition is to ensure that we commence computation with a particular type of variable configuration. 

\begin{definition}[Shapes]
	Let $V$ be a set of variables. 
	A {\em shape} is a block configuration $C$ for which each $E$ in each tape element $(I,E)$ we have either $E \in \mathbb{N}\backslash \{0\}$ or $E \in V$.  
\end{definition}

In this way a shape is a particularly simple type of block configuration, i.e.\ one in which the exponent in each tape element is either a positive integer or a single variable. For example, the shape for Dragon 18 discussed above is 

\begin{center}
	$\{a\}0\{r\} 1^X (001)^Y 1$
\end{center}

Note also that the block configuration $\{a\}0\{r\} 1^{Z+5} (001)^{W-1} 1$ is not a shape, but can be thought of as an instance of the shape $\{a\}0\{r\} 1^X (001)^Y 1$. In order to detect patterns of the above form, we need to be able to identify instances of the same shape. This leads us to the notion of substitution below. 

\begin{definition}[Substitutions]
	A {\em substitution} $\delta$ is a total function $\delta:V \rightarrow Exp(V)$ where $V$ is a set of variables. 
	
	The application of a substitution $\delta$ to a shape $S$ is denoted $S\delta$, which we call an {\em instance} of $S$. 
	
	We denote a substitution as $\delta = [X_1 \leftarrow E_1, X_2 \leftarrow E_2, \ldots , X_n \leftarrow E_n]$ 
	and the application of a substitution $\delta$ to a shape $S$ as $S\delta = S[X_1 \leftarrow E_1, X_2 \leftarrow E_2, \ldots , X_n \leftarrow E_n]$.
	
	We say $\delta$ is a {\em grounding substitution} if for each $E_i$ in $\delta$ we have $E_i \in \mathbb{N}$.
	
	We denote by {\em dom($\delta$)} the variables $\{X_1, X_2, \ldots , X_n\}$ and by {\em range(S)} the variables $\displaystyle \bigcup_{i=1}^n vars(E_i)$. 
\end{definition}

Note that unlike some similar definitions in the literature we do \textbf{not} require that $dom(\delta) \;\cap\; range(\delta) = \emptyset$; the reason for this will be discussed shortly. 

\begin{definition}[Shape instances]
	A block configuration $C$ is an {\em instance} of a shape $S$ if there is a substitution $\delta$ such that $C= S\delta$. 
	If $\delta$ is a grounding substitution, then $C$ is a {\em ground instance} of $S$. 
\end{definition}

For example, $\{a\}0\{r\} \; 1^8 (001)^5 \; 1$ is a ground instance of the shape $\{a\}0\{r\} \; 1^X (001)^Y \; 1$ due to the substitution $\{X \leftarrow 8, Y \leftarrow 5\}$.
The configuration $\{a\}0\{r\} 1^{Z+5} (001)^{W-1} 1$ is an instance of the same shape due to the substitution $\{X \leftarrow Z+5, Y \leftarrow W-1\}$.

The way we will identify an appropriate shape is from a pattern of instances in the execution of a machine. This leads us to the definition below. 

\begin{definition}[Minimal Shape]
	A subsequence $S_1, S_2, \ldots , S_k$ of the execution of a machine is a {\em k-instance} of a shape $S$ if there are grounding substitutions $\delta_1, \delta_2, \ldots , \delta_k$ such that $S_i = S\delta_i$. 
	
	Let $vars(S)$ be the set of variables in the shape $S$. We define a partial order on shapes such that $S_1 \leq S_2$ iff $vars(S_1) \subseteq vars(S_2)$. 
	
	We say $S$ is a {\em minimal shape} for a given $k$-instance $S_1, S_2, \ldots , S_k$ if $S$ is a minimal shape for this $k$-instance in this partial order. 
\end{definition}

For example, the above sequence of configurations of Dragon 18 from hop 131 to hop 365 is a $7$-instance of the shape $a \{0\} r 1^X (001)^Y 1$ where $\delta_1 = \{X \leftarrow 3, Y \leftarrow 6\}$, $\delta_2 = \{X \leftarrow 8, Y \leftarrow 5\}$ etc.

The reason use minimal shapes is to avoid shapes that are clearly too general. For example, for this same sequence of configurations, it should be intuitively clear that the most appropriate shape is $\{a\} 0 \{r\} \; 1^X (001)^Y 1$ rather than say $\{a\} 0^W \{r\} \; 1^X (001)^Y 1^Z$. Clearly both are shapes for this sequence, but there is no point in having $W$ and $Z$ as part of this shape if in every relevant substitution we have $W \leftarrow 1$ and $Z \leftarrow 1$. It should also be clear that this sequence is not the instance of any shape involving only one variable. 

We are now in a position to formally define the observant otter.

\begin{definition}[Observant Otter]
	Let $S$ be a minimal shape for a $k$-instance $S_1, S_2, \ldots , S_k$, let $\delta_1, \delta_2, \ldots , \delta_k$ be the corresponding grounding substitutions, and let $vars(S) = \{X_1, X_2, \ldots , X_n\}$. Let $x_{ij}$ be the value of $X_i$ under the application of $\delta_j$ (so $x_{ij} = X_i\delta_j$). Consider the sequence of instances $x_{i1}, x_{i2}, \ldots , x_{ik}$ of variable $X_i$.
	
	\begin{itemize}
		\item If $x_{i1} < x_{i2} < \ldots < x_{ik}$, we say $X_i$ is in a {\em progressive sequence}. 
		\item If $x_{i1} > x_{i2} > \ldots > x_{ik}$, we say $X_i$ is in a {\em regressive sequence}.
	\end{itemize}
	
	We say a $k$-instance $S_1, S_2, \ldots , S_k$ of a minimal shape $S$ is an {\em observant otter sequence} if both of the following conditions are satisfied.
	
	\begin{enumerate}
		\item Each variable in $vars(S)$ is either in a regressive sequence or in a progressive sequence.
		\item There is at least one variable in $vars(S)$ in a regressive sequence, and at least one variable in $vars(S)$ in a progressive sequence. 
	\end{enumerate}
	
	We say an observant otter sequence is {\em monoregressive} (resp. {\em monoprogressive}) if there there is exactly one variable in a regressive (resp. progressive) sequence. Otherwise it is {\em multiregressive} (resp. {\em multiprogressive}). 
	
\end{definition}

Note that as there is at most one variable in any tape element in a shape, in any observant otter sequence, each tape element can be identified with either a regressive or progressive sequence (if it contains a variable), or is unchanged (if it does not contain a variable). 

To return once more to the above-mentioned pattern in Dragon 18, it should be clear that the 3-instance below of the shape $\{a\}0\{r\} 1^X (001)^Y 1$ is an observant otter sequence. The sequence of values for $X$ is $3 < 8 < 13$, and for $Y$ it is $6 > 5 > 4$, and so $X$ is in a progressive sequence, and $Y$ is in a regressive sequence. Note also that the right most block 1 remains unchanged. 

\begin{center}
	\begin{tabular}{llll}
		\textbf{Hops} & \textbf{Configuration} & \textbf{Shape} & \textbf{Substitution} \\
		131 & $\{a\}0\{r\} \; 1^3 (001)^6 1$    & $\{a\}0\{r\} \; 1^X (001)^Y 1$ & $X \leftarrow 3,  Y \leftarrow 6$ \\ 
		145 & $\{a\}0\{r\} \; 1^8 (001)^5 1$    & $\{a\}0\{r\} \; 1^X (001)^Y 1$ & $X \leftarrow 8,  Y \leftarrow 5$\\
		169 & $\{a\}0\{r\} \; 1^{13} (001)^4 1$ & $\{a\}0\{r\} \; 1^X (001)^Y 1$ & $X \leftarrow 13, Y \leftarrow 4$\\
	\end{tabular}
\end{center}

In what follows, we will use the notational convention that $X$ is typically in a progressive sequence and $Y$ is typically in a regressive sequence.

\section{Shape Computations}

\subsection{Fundamentals}

Identifying an appropriate shape is a first step; what follows is how to use this to infer the relationship between the various instances, the final configuration in the sequence and the number of steps at which this will occur. This is relatively simple when the regressive and progressive sequences are both linear, as in the definitions below. 

\begin{definition}[Linear Observant Otter]
	Let $S_1, S_2, \ldots , S_k$ be an observant otter sequence,
	$\delta_1, \delta_2, \ldots , \delta_k$ be the corresponding grounding substitutions, 
	$vars(S) = \{X_1, X_2, \ldots , X_n\}$,
	and $x_{ij} = X_i\delta_j$. % be the value of $X_i$ under the application of $S_j$.
	
	The observant otter sequence $S_1, S_2, \ldots , S_k$ is {\em linear} if both of the following conditions are satisfied. 
	
	\begin{itemize}
		\item Every regressive sequence  $x_{i1}, x_{i2},  \ldots, x_{ik}$ is of the form $x_{i1}, x_{i1} - l, x_{i2} - l, \ldots, x_{i(k-1)} - l$ for some $l \in \mathbb{N} \backslash \{0\}$ (so $x_{i(j+1)} = x_{ij} - l$ for $1 \leq j \leq k-1$)
		\item Every progressive sequence $x_{i1}, x_{i2},  \ldots, x_{ik}$ is of the form $x_{i1}, m x_{i1} + n, m x_{i2} + n, \ldots, mx_{i(k-1)} + n$ for some $m,n \in \mathbb{N} \backslash \{0\}$ (so $x_{i(j+1)} = m x_{ij} + n$ for $1 \leq j \leq k-1$)
	\end{itemize}
	
	We refer to $l$ in a regressive sequence and the ordered pair $(m,n)$ in a progressive sequence as the {\em relative difference}. 
	
	A linear observant otter sequence is {\em unilinear} if in every progressive sequence we have $m = 1$. Otherwise it is \emph{multilinear}. 
\end{definition}

For a linear observant otter sequence, we can write the regressive sequences in the form $a_{j+1} = a_j - l$ where $a_0 = x{i1}$, and the progressive sequences in the form $a_{j+1} = m \times a_j + n$ where $a_0 = x{i1}$. We will also write the relative difference for a regressive sequence as the ordered pair $(1, l)$ when appropriate.  

Note that in the Dragon 18 example above, the relative difference for $X$ is $(1,5)$, while for $Y$ the relative difference is $1$. Hence this is a unilinear sequence. 

We will also find the notion of a \textit{difference block} useful. 

\begin{definition}[Difference Block]
	Let $Seq = S_1, S_2, \ldots , S_k$ be a linear observant otter sequence for a minimal shape $S$. For each tape element $(I,E)$ in $S$, we define the {\em difference block} as follows. 
	
	\begin{itemize}
		\item If $E$ is a variable in a regressive sequence with relative difference $l$, then the difference block is $(1,-l)$.
		\item If $E$ is a variable in a progressive sequence with relative difference $(m,n)$, then the difference block is $(m,n)$.
		\item If $E$ is a number, then the difference block is $(1,0)$.
		
	\end{itemize}
	
	We call a difference block $(m,n)$ in which $m = 1$ and $n = 0$ a {\em static} difference block. All other difference blocks are {\em dynamic}.
	
	The {\em application} of a difference block $(m,n)$ to the corresponding tape element $(I,E)$ in $S$ is the tape element $(I, m E + n)$. 
	
	The {\em difference tape} is the sequence of difference blocks in the corresponding order to $S$. 
	
	The {\em application} of the difference tape $D$ to $S$ is the block tape $T$ resulting from the application of each difference block in $D$ to the corresponding block in $S$. 
	
	We denote $S \longmapsto T$ as the \emph{otter pattern} corresponding to $Seq$, where $T$ is the block tape resulting from the application of the difference tape $D$ to $S$. 
	
\end{definition}

The intuition is that for any of the instances $S_1, S_2, \ldots , S_k$ in $Seq$, we can generate $S_{j+1}$ from $S_j$ and the difference tape. For any tape element $T = (I,E)$ in $S_j$, we find the corresponding difference block $B = (m,n)$, and then the block corresponding to $T$ in $S_{j+1}$ is $(I, m E + n)$. Note that this formulation accounts for both static and dynamic blocks. 

So for the example from Dragon 18 above, we have the block tape $(0,1).(1,3).(001,6).(1,1)$ with the corresponding difference tape being $(1,0).(1,5).(1,-1),(1,0)$. 

\begin{figure}[tb]
	\centering
	\includegraphics[width=0.5\textwidth]{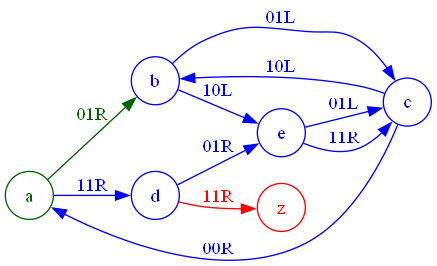}
	\caption{Dragon 25}
	\label{fig:dragon25}
\end{figure}

Note that the following sequence occurs in the execution of Dragon 25 from \cite{Harland16} (see Figure~\ref{fig:dragon25}).

\begin{center}
	\begin{tabular}{ll}
		\textbf{Hops} & \textbf{Configuration} \\
		9963 & $(10)^{3} \{a\}0\{l\} (10)^{71}1$ \\
		9967 & $(10)^{4} \{a\}0\{l\} (10)^{70}1$ \\
		9971 & $(10)^{5} \{a\}0\{l\} (10)^{69}1$ \\
		9975 & $(10)^{6} \{a\}0\{l\} (10)^{68}1$ \\
		... & \\
	\end{tabular}
\end{center}

The relevant shape here is $(10)^{X} \{a\}0\{l\} (10)^{Y} 1$, but clearly we require that $Y \geq 1$ in order for the pattern to continue. A similar issue arises from the following observant otter sequence also in Dragon 25. 

\begin{center}
	\begin{tabular}{ll}
		\textbf{Hops} & \textbf{Configuration} \\
		707 & $(1110)^3 \; \{a\} (10)^{13} \{l\} \; 1$ \\
		711 & $(1110)^4 \; \{a\} (10)^{11} \{l\} \; 1$ \\
		715 & $(1110)^5 \; \{a\} (10)^{9}  \{l\} \; 1$ \\
		719 & $(1110)^6 \; \{a\} (10)^{7}  \{l\} \; 1$ \\
		723 & $(1110)^7 \; \{a\} (10)^{5}  \{l\} \; 1$ \\
		727 & $(1110)^8 \; \{a\} (10)^{3}  \{l\} \; 1$ \\
		731 & $(1110)^9 \; \{a\} (10)^{1}  \{l\} \; 1$ \\
	\end{tabular}
\end{center}

The relevant shape here is $(1110)^{X} \{a\} (10)^Y \{l\} 1$, but clearly we require that $Y \geq 2$ in order for the pattern to continue. 
This means we need a method to determine such constraints. 

A further issue is that it is not entirely obvious is whether the observed pattern will \textbf{always} be appropriate. While this may seem intuitively natural, it is not a guarantee, and it is potentially sensitive to the chosen value of $k$ for the $k$-instances, in that choosing $k$ too low may result in misleading predictions. This leads us to introduce the notion of a \textit{shape computation}, which is inspired by the hypothetical engine of \cite{Harland07}.\footnote{In keeping with previous naming conventions, we shall often refer to shape computations as \textit{shapely shelducks}.} 

\begin{definition}[Shape Computation]
	Let $Seq$ be a linear observant otter sequence for some minimal shape $S$ in an execution of a machine $M$, and let $B$ be the corresponding difference tape. 
	
	For each tape element $T = (I,E)$ in $S$, let $(m,n)$ be the corresponding difference block in $B$. We say the {\em target block} for $T$ is the tape element $(I, m E + n)$. Let $D$ be the configuration resulting from $S$ by replacing each tape element with its corresponding target block. We say $T$ is the {\em target configuration} for $S$.
	
	If there is an execution of $M$ commencing with configuration $S$ that includes $T$, then we say this observant otter sequence has been {\em verified}.
	In such a case we refer to the section of the execution which commences with $S$ and concludes with $T$ as the {\em shape computation}. 
	
	We say the {\em initial position} of the shape computation is the reference position in $S$, and the {\em final position} is the reference position in $T$. In a given shape computation, the initial position is always assumed to be $0$. 
	
	The \emph{length} of a shape computation is the expression denoting the number of hops required to execute from $S$ to $T$. 
	
	We say the shape computation is \emph{unilinear} (resp.\ \emph{multilinear}) if the observant otter sequence on which it is based is linear (resp.\ multilinear).
	
\end{definition}

The idea is that $T$ is the configuration resulting from the application of a substitution of the form $X \leftarrow m X + n$ for each variable $X$ in $S$, where $(m,n)$ is the appropriate difference block.\footnote{This is the reason we do not insist that the domain and range of substitutions be distinct. This is more a matter of convenience rather than substance, though.}

Intuitively, the shape computation is a proof that any ground instance of $S$ will result in the corresponding ground instance of $T$, provided any constraints on the variables are met. We can also use the shape computation to determine these constraints. 

For example, consider the above example for Dragon 18, and the two above for Dragon 25. The corresponding shapes and target configurations are below. 

\begin{center}
	\begin{tabular}{lll}
		\textbf{Dragon} & \textbf{Shape} & \textbf{Target} \\
		18 & $\{a\} 0 \{r\} 1^X (001)^Y 1$ & $\{a\} 0 \{r\} 1^{X+5} (001)^{Y-1} 1$ \\
		25 & $(10)^{X} \{a\}0\{l\} (10)^{Y} 1$   & $(10)^{X+1} \{a\}0\{l\} (10)^{Y-1} 1$ \\
		25 & $(1110)^X \{a\} (10)^{Y} \{l\} 1$ & $(1110)^{X+1} \{a\} (10)^{Y-2} \{l\} 1$ \\
	\end{tabular} 
\end{center}

The shape computation of each of these is given below. We will refer to these as $SC_1$, $SC_2$ and $SC_3$ respectively. The initial and final position for each shape computation is given by $\downarrow$ and $\uparrow$ respectively.  

\begin{tabular}{lllllllll}
	& & & & & & & \\
	& & & $~~~~\downarrow$ & & \\
	0:      & & & $\{a\}0\{l\}$ & $1^{X}$ & $(001)^{Y}$ & & 1 & \\
	1:      & & & 1 & $\{b\}1^{X}\{l\}$   & $(001)^{Y}$ & & 1 & \\
	$X+1$:  & & & 1 & $1^{X}$ & $\{b\}(001)^{Y}\{l\}$   & & 1 & \\
	%$X+2$:  & & & 1 & $1^{X}$ & 1 $\{c\}0\{l\}$ 1 & $(001)^{Y-1}$ & 1 \\
	%$X+3$:  & & & 1 & $1^{X}$ & 1 1 $\{d\}1\{l\}$ & $(001)^{Y-1}$ & 1 \\
	$X+6$:  & & & 1 & $\{d\}1^{X}\{r\}$ & 1 1 1 & $(001)^{Y-1}$ & 1 \\
	$2X+7$: & & $\{d\}0\{l\}$ & 1 & $1^{X}$ & 1 1 1 & $(001)^{Y-1}$ & 1 \\
	$2X+8$: & $\{a\}0\{l\}$ & 1 & 1 & $1^{X}$ & 1 1 1 & $(001)^{Y-1}$ & 1 \\
	& $~~~~\uparrow$ & & & & & & \\
	& & & & & & & \\
\end{tabular}

\begin{tabular}{llllll}
	& & & & & \\
	& &              $~~~~\downarrow$ & & & \\
	$0$: & $(10)^{X}$ & $\{a\}0\{l\}$ & $(10)^{Y}$ & & $1$ \\ % Hops: 0 Ones: 1 Pos: 0 Transition: init 
	$1$: & $(10)^{X}$ & $1$ & $\{b\}(10)^{Y}\{l\}$ & & $1$ \\ % Hops: 1 Ones: 2 Pos: 1 Transition: complex(normal) 
	$2$: & $(10)^{X}$ & $\{e\}1\{r\}$ & $0 0$ & $(10)^{Y-1}$ & $1$ \\ % Hops: 2 Ones: 2 Pos: 0 Transition: complex(normal) 
	$3$: & $(10)^{X}$ & $1$ & $\{c\}0 0\{l\}$ & $(10)^{Y-1}$ & $1$ \\ % Hops: 3 Ones: 2 Pos: 1 Transition: complex(normal) 
	$4$: & $(10)^{X}$ & $1$ & $0 \{a\}0\{l\}$ & $(10)^{Y-1}$ & $1$ \\ % Hops: 4 Ones: 1 Pos: 2 Transition: complex(normal) 
	& &  &             $~~~~~\uparrow$ & & \\
	& & & & & \\
\end{tabular}

\begin{tabular}{llllll}
	& & & & & \\
	&              & $~~~~\downarrow$ & & & \\
	0: & $(1110)^{X}$ & $\{a\}(10)^{Y}\{l\}$ & 1 & & \\
	1: & $(1110)^{X}$ & 1 $\{d\}0\{l\}$ & $(10)^{Y-1}$ & 1 &  \\
	2: & $(1110)^{X}$ & 1 1 & $\{e\}(10)^{Y-1}\{l\}$ & 1 &  \\
	3: & $(1110)^{X}$ & 1 1 & 1 $\{c\}0\{l\}$ & $(10)^{Y-2}$ & 1 \\
	4: & $(1110)^{X}$ & 1 1 & 1 0 & $\{a\}(10)^{Y-2}\{l\}$ & 1 \\
	&              &     &     & $~~~~\uparrow$ & \\
	& & & & & \\
\end{tabular}

\begin{figure}[tb]
	\centering
	\includegraphics[width=0.6\textwidth]{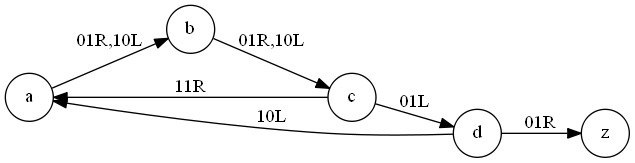}
	\caption{Otter machine example}
	\label{fig:problem}
\end{figure}

Note that in $SC_1$, the length depends on $X$, unlike $SC_2$ and $SC_3$. Note also that $SC_1$ ``travels'', in that it involves parts of the tape outside the initial configuration, i.e.\ two blank cells to the left. $SC_2$ and $SC_3$ do not involve any tape cells outside the initial configuration.

Note also the observant otter sequence below arising from the machine in Figure~\ref{fig:problem}.

\begin{center}
	\begin{tabular}{ll}
		\textbf{Hops} & \textbf{Configuration} \\
		17 & $1^5 \{d\} 1^2 \{l\}$ \\
		24 & $1^4 \{d\} 1^3 \{l\}$ \\
		31 & $1^3 \{d\} 1^4 \{l\}$ \\
		38 & $1^2 \{d\} 1^5 \{l\}$ \\
	\end{tabular}
\end{center}

This makes it seem that it is reasonable to apply the otter after hop 31, and so conclude that at hop 45 the configuration is $1 \{d\} 1^6 \{l\}$. However, this is not the case; the machine halts at hop 43. The reason that the otter application here is incorrect is that the otter requires at least three 1's on the left. This is made explicit by the shape computation, which is below. We will refer this shape computation as $SC_4$. 

\begin{tabular}{llllllll}
	& & & & & & \\
	& & & &         & $~~~~\downarrow$ & \\
	0: & & & & $1^{Y}$ & $\{d\} 1^{X} \{l\}$ & \\
	1: & & & & $\{a\}1^{Y}\{r\}$ & 0 & $1^{X-1}$ \\
	2: & & & $\{b\}1^{Y-1}\{r\}$ & 0 & 0 & $1^{X-1}$ \\
	3: & & $\{c\}1^{Y-2}\{r\}$ & 0 & 0 & 0 & $1^{X-1}$ \\
	4: & $1^{Y-3}$ & 1 & $\{a\}0\{l\}$ & 0 & 0 & $1^{X-1}$ \\
	5: & $1^{Y-3}$ & 1 & 1 & $\{b\}0\{l\}$ & 0 & $1^{X-1}$ \\
	6: & $1^{Y-3}$ & 1 & 1 & 1 & $\{c\}0$\{l\} & $1^{X-1}$ \\
	7: & $1^{Y-3}$ & 1 & 1 & $\{d\}1\{l\}$ & 1 & $1^{X-1}$ \\
	&           &   &   & $~~~~\uparrow$ & & \\
	& & & & & & \\
\end{tabular}

From this we see that reaching the target configuration of $1^{Y-1} \{d\} 1^{X+1} \{l\}$ requires that $Y \geq 3$, as we must ``visit'' at least three of the $1$'s in $1^Y$, and that $X \geq 1$ (as for each exponent $E$ we must have $E \geq 0$). As $X$ is in a progressive sequence, it is only necessary to check the constraint at the start of the shape computation. The constraint on $Y$ determines that the sequence ceases when $Y$ first becomes less than $3$. 

This means that the machine will move from the configuration $1^Y \{d\} 1^X \{l\}$ at step $N$ to the configuration $1^2 \{d\} 1^{Y+X-2} \{l\}$ at step $N+7(Y-2)$. When $X = 2, Y = 5$ and $N = 17$ as above, we arrive at configuration $1^2 \{d\} 1^{5} \{l\}$ at step $17+21=38$. 

Note also that for monoregressive sequences, it is simpler to predict the last occurrence than for multiply regressive sequences, for which we need to determine which of the regressive variables reaches its minimum first, which in turn depends on the initial values of the regressive variables and hence cannot be computed in advance. As it turns out, multiply regressive sequences are relatively rare, although not unknown. 

\subsection{Otter Machines}
\label{sec:otter}

Our technical goal is now to determine an appropriate transition, based on shape computations; we refer to the transition based on a particular observant otter pattern as an \textit{ossified ocelot.}
This means that we need to be able to find a way to isolate the changes to the tape to a particular region, so that we can apply the ocelot transition to this region independently of what is on the rest of the tape. In a similar manner to extended transitions for macro machines, this means that once we have found one otter pattern, we can repeatedly apply it without having to recompute it each time.

A key difference from macro machines is that there will be now three possibilities for an ocelot transition (apart from termination or looping): it can remain in the same region of the tape, or move to the next region to the left, or to the next region to the right. In a macro machine there is no possibility to remain in the same tape region, as this would indicate the machine is in a loop.  A further difference is that not only will be the relevant tape region be variable in size, we also need to calculate its boundaries precisely. For example, in each of $SC_1$, $SC_2$ and $SC_3$, there is a 1 on the right-hand side of the shape which is neither changed nor visited. Hence any transition defined from these shape computations should not include this 1, as the outcome of the transition will be independent of this block (and any blocks to the right of it), and so it inappropriate to require that this block be present before the transition can be applied. This means that we will need to be mindful of both including all necessary blocks in the transition, but also excluding any which are not necessary. 

In addition, as noted above, $SC_1$ requires two blank tape cells to the left of the initial position for each iteration of the shape computation. While in principle we can calculate the precise number of blanks needed for the corresponding transition, in practice it seems more appropriate to insist that the tape is blank to left of this position in order for the transition to be applied. This seems to reflect the observed practice that shapes do not decrease in size during a shape computation, and as we shall see, for some multilinear sequences, the increase can be very substantial. 

An ocelot transition (which we refer to as an \textit{ossified ocelot}) will consist of an input configuration (including any associated constraints), an output configuration, and an expression for the number of $Hops$ executed. All of these elements can be derived from the relevant shape computation, as below. 

\begin{definition}[Shape Computation Properties]
	Let $Seq$ be a linear observant otter sequence for some minimal shape $S$ in an execution of a machine $M$. Let $T$ be the relevant target configuration and let $SC$ be a shape computation from $S$ to $T$. 
	
	The {\em visited interval} of $SC$ is the interval $(Left, Right)$ where $Left$ is the leftmost position reached in $SC$, and $Right$ is the rightmost such position. The {\em width} of $SC$ is the width of the visited interval. 
	
	The {\em shape interval} of $SC$ is the interval denoting the extremities of $S$. 
	
	The {\em target interval} of $SC$ is the interval denoting the extremities of $T$.

	We say $SC$ is {\em increasing} (respectively {\em static}, {\em decreasing}) if the size of the shape interval is less than (resp.\ equal to, greater than) the size of the target interval.
	
\end{definition}

Returning to the four above \textit{shapely shelducks} $SC_2$, $SC_3$, and $SC_4$, we get the following table. 

\begin{center}
	\begin{tabular}{ccccc}
		\textbf{Id} & \textbf{Visited} & \textbf{Shape} & \textbf{Target} & \textbf{Status} \\
		$SC_1$ & $(-2,X+3)$ & $(0,X+3Y+1)$ & $(-2,X+3Y+1)$  & increasing   \\
		$SC_2$ & $(0,2)$   & $(-2X,2X+1)$   & $(-2X,2X+1)$  & static \\
		$SC_3$ & $(0,4)$    & $(-4X,2Y)$   & $(-4X,2Y)$     & static \\
		$SC_4$ & $(-3,0)$   & $(-Y,X)$     & $(-Y,X)$       & static \\
	\end{tabular}
\end{center}

The first step in the development of the relevant transition is to determine the {\em general instance} of the relevant shape. In particular, we want to find the general instance in terms of a parameter $k$, so that when $k = 0$ we have the input shape, and when $k = Count$ (ie the final instance of the shape) we have the output configuration. Once we have determined the general instance, we can then work out the value of $Count$ (i.e.\ how many iterations we can perform), and from this we can then determine both the output configuration of the transition and the relevant $Hops$ value. 

We first check whether the initial configuration and the target configuration in the shape computation are the same size. If so, that means there are no ``external'' constraints on the transition. In particular, an increasing shape computation will require the tape to the left or right of the shape internal to be blank to allow for this expansion. Precisely calculation of the number of blanks required is possible, but instead we simply annonate the shape as requiring blanks to the left or right as appropriate. We expect that shelducks will generally be static or increasing, and so we will not consider the decreasing case further.\footnote{unless and until we come across an example of such a shape computation!} 
%\textbf{\textit{Update this. multilinear otters require variable size padding, so it may be easier just have constraints of left-must-be-empty and right-must-be-empty, or both or neither. The last case is for ones in which no padding is needed at all.}}

For example, consider the increasing shape computation $SC_1$ above. As the target interval is $(-2,X+3Y+1)$ and the shape interval is $(0,X+3Y+1)$, for every iteration of the shape computation, we require two blanks to the left of the original shape. Note that $SC_1$ becomes static if we consider the shape as $00 \; \{a\} 0 \{r\} \; 1^X (001)^Y \; 1$ rather than $\{a\} 0 \{r\}  \; 1^X (001)^Y \; 1$. Hence for $i$ iterations of the pattern, we will require $i$ copies of $00$ to the left of the original shape, meaning that the original configuration of $(00)^i \; \{a\} 0 \{r\}  \; 1^X (001)^Y \; 1$ becomes $\{a\} 0 \{r\}  \; 1^{X+ 5\times i} (001)^{Y-i} \; 1$. However rather than be this precise, we simply write $(blank)$ to the left of the shape, as below. 

\begin{center}
	\textbf{Initial:} $(blank) \{a\} 0 \{r\} \; 1^X (001)^Y$ \textbf{Final:} $\{a\} 0 \{r\} \; 1^{X+5\times Count} (001)^{Y-Count}$
\end{center}

Note that from the shape computation we can deduce that the constraint on $Y$ is that $Y \geq 1$ (from the transition from $X+1$ to $X+2$) and so we have $Count = Y$ (as the first time that the constraint is violated is when $Y=0$). This means we can write the output configuration as $\{a\} 0 \{r\} \; 1^{X+5\times Y}$. 

% \textbf{\textit{As above stuff about multilinear and how we can approximate an exact outcome by left-is-blank or right-is-blank}}. 

In all the cases below, $k$ varies between 0 and $Count$, so the initial instance is when $k = 0$ and the final one is when $k = Count$. 
% The size of the general instance should be independent of $k$ (but not necessarily of $Count$) and the same as the length of the initial and final instances. 
The value of $Count$ will depend on the shape computation, which will determine what constraints need to be applied to calculate $Count$. 

To find the general case, we the relevant transition must include the reference tape element, and all blocks whose corresponding difference is dynamic (and any blocks between these). Then, if need be, we extend this to include any static blocks which have been visited in the shape computation. 

To return to our four example shelducks, we generate the respective general configurations as follows. 

\begin{description}
	\item[$SC_1$:] As above, we obtain $(blank) \; \{a\} 0 \{r\} \; 1^X (001)^Y \; 1$ for which the corresponding difference blocks are $m(1,0), m(1,5),m(1,-1),m(1,0)$ respectively\footnote{Note that we do not include any difference block for $blank$.}. As the visited interval is $(-2, X+3)$, the dynamic blocks and the blank include all visited blocks, and so the general instance in this case is $(blank) \{a\} 0 \{r\} \; 1^{X+5k} (001)^{Y-k}$. The input configuration is then  $(blank) \{a\} 0 \{r\} \; 1^{X} (001)^{Y}$ with the constraint $Y \geq 1$. The value of $Count$ is then $Y$, and the final configuration is $\{a\} 0 \{r\} \; 1^{X+5Y}$. As there are $2X+8$ steps in each instance of the shape computation, the value of $Hops$  is $Y \times (8 + 2 X + 2 \times 5 (Y-1)/2) = 8Y + 2XY + 10Y(Y-1)/2 = 5Y^2 + 3Y + 2XY$ (see Section~\ref{sec:hopscalc} for the general formula).
	
	\item[$SC_2$:] Here the initial and final configurations are the same size. The initial configuration includes the visited interval of $(0,2)$ provided that $Y \geq 1$ and so we have the general instance as $(10)^{X+k} \{a\}0\{l\} (10)^{Y-k}$. The input configuration is then $(10)^{X} \{a\}0\{l\} (10)^{Y}$ with the constraint $Y \geq 1$. The value of $Count$ is then $Y$, and the final configuration is $(10)^{X+Y}\{a\}0\{l\}$. As there are four steps in each instance of the shape computation, the value of $Hops$ is $4Y$. 
	
	\item[$SC_3$:] Here the initial and final configurations are the same size. The initial configuration includes the visited interval of $(0,4)$ provided that $Y \geq 2$ and so we have the general instance as $(1110)^{X+k} \; \{a\} (10)^{Y-2k} \{l\}$. The input configuration is then $(1110)^{X} \; \{a\} (10)^{Y} \{l\}$ with the constraint $Y \geq 2$. The value of $Count$ is then $Y // 2$, and the final configuration is $(1110)^{X+Y // 2} \; \{a\} (10)^{Y mod \;2} \{l\}$. As there are 4 steps in each instance of the shape computation, the value of $Hops$ is $(Y // 2) \times 4) = 2(Y - Y mod \;2)$. 
	
	\item[$SC_4$:] Here the initial and final configurations are the same size. The initial configuration includes the visited interval of $(-3,0)$ provided that $Y \geq 3$ and $X \geq 1$, and so we have the general instances as $1^{Y-k} \{d\}1^{X+k} \{l\}$. The input configuration is then $1^{Y} \{d\}1^{X} \{l\}$ with the constraints $Y \geq 3$ and $X \geq 1$. The value of $Count$ is then $Y-2$, and the final configuration is $1^{2} \{d\}1^{X+Y-2} \{l\}$. As there are 7 steps in each instance of the shape computation, the value of $Hops$ is $(Y-2) \times 7 = 7Y-14$. 
\end{description}

In summary we have

\begin{center}
	\begin{tabular}{cccc}
		\textbf{Id} & \textbf{Shape}                    & \textbf{Target}                          & \textbf{General}\\
		$SC_1$ & $(blank) \{a\} 0 \{r\} \; 1^X (001)^Y$ & $\{a\} 0 \{r\} \; 1^{X+5} (001)^{Y-1}$   & $(blank) \{a\} 0 \{r\} \; 1^{X+5k} (001)^{Y-k}$ \\
		$SC_2$ & $(10)^{X} \{a\}0\{l\} (10)^{Y}$        & $(10)^{X+Y}\{a\}0\{l\}$                  & $(10)^{X+k} \{a\}0\{l\} (10)^{Y-k}$     \\
		$SC_3$ & $(1110)^X \; \{a\} (10)^{Y} \{l\}$     & $(1110)^{X+1} \; \{a\} (10)^{Y-2} \{l\}$ & $(1110)^{X+k} \; \{a\} (10)^{Y-2k} \{l\}$   \\
		$SC_4$ & $1^Y \{d\}1^X \{l\}$                   & $1^{Y-1} \{d\}1^{X+1} \{l\}$             &  $1^{Y-k} \{d\}1^{X+k} \{l\}$       \\
	\end{tabular}
\end{center}

We now describe the process to determine the general instance (in terms of $Count$, $k$ and all variables in the shape).  

\begin{definition}[Difference Tape Applications]\label{def:application}
	Let $S$ be a shape and $D$ be a difference tape corresponding to $S$. We say the $k$th application of $D$ to $S$ results in the tape $T$ where each element $(I,E)$ in $S$ is transformed as follows. 
	
	\begin{center}
		\begin{tabular}{lll}
			\textbf{Difference block} & \textbf{Initial Element} & \textbf{Resulting Element} \\
			$(1,0)$ & $(I,E)$ &  $(I,E)$ \\
			$(1,n),n > 0$  & $(I,E)$ &  $(I, E + k n)$ \\
			$(1,-n),n > 0$ & $(I,E)$ &  $(I, E - k n)$ \\
			$(m,n), m > 1, n \geq 0$ & $(I,E)$ &  $(I, (m^k E) + n\frac{m^k-1}{m-1})$ \\
			$(m,-n), m > 1, n > 0$ & $(I,E)$ &  $(I, (m^k E) - n\frac{m^k-1}{m-1})$ \\
		\end{tabular}
	\end{center}
	
\end{definition}

We can now formally define the general instance of the otter pattern.

\begin{definition}[General Instance]
	
	% Let $n$ be the difference between the sizes of the target and initial configurations.  
	
	The \emph{padded initial configuration} is the initial configuration with the addition of the tape element $(blank)$ to the left or right tape of the initial configuration as appropriate.
	
	Let $GI$ be the largest contiguous sub-tape of the padded initial configuration which includes all dynamic blocks, the visited interval and the reference element. 
	
	Let $GD$ be the subtape of the difference tape which corresponds to $GI$. 
	
	The \emph{general instance} is then the $k$th application of $GD$ to $GI$. 
	
\end{definition}

Note that the largest contiguous sub-tape condition is required to exclude unused parts of the tape, such as the rightmost $1$ in $SC_1$, $SC_2$ and $SC_3$. Clearly all dynamic elements and the reference element must in the general instance. Any static elements which are in the visited interval must also be included in the general instance, but any elements outside the visited interval should not be included. 

We then use the $k=0$ case of the general instance as the initial configuration for the transition, and the $k=Count$ case of the general instance as the final configuration. 

Determining the relevant constraints on the variables from the shape computation takes a little care, as some transitions may impose constraints on these variables involved. In particular, the intuition is that an acceleration transition does not impose any constraints, but a normal transition does (see Definition~\ref{def:extended}). 

For example, consider the block configuration $1\{a\}(110)^X\{l\}01$ and the three transitions below.

\begin{center}
	\begin{tabular}{lll}
		$T_1$ & $<a,110,l,101,r,a,3>$ & acceleration \\
		$T_2$ & $<a,110,l,101,r,b,3>$ & normal \\
		$T_3$ & $<a,110,l,110,l,a,3>$ & normal \\
	\end{tabular}
\end{center}

Applying these transitions to $1\{a\}(110)^X\{l\}01$ gives us the following. 

\begin{center}
	\begin{tabular}{ll}
		$T_1$ & $1\{a\}(110)^X\{l\}01 \rightarrow 1(101)^X \{a\}0\{l\}1$ \\
		$T_2$ & $1\{a\}(110)^X\{l\}01 \rightarrow 1 101 \{b\}(110)^{X-1}\{l\}01$ \\
		$T_3$ & $1\{a\}(110)^X\{l\}01 \rightarrow \{a\}1\{l\}(110)^{X}01$ \\
	\end{tabular}
\end{center}

Applying $T_1$ does not impose any constraint on $X$. 
Applying $T_2$ imposes the constraint that $X \geq 1$. 
Applying $T_3$ imposes the constraint that $X \geq 1$, as otherwise we would be unable to apply this transition. Note that this transition is a special case of Case 1 in Definition~\ref{def:extended} above in which the input string is the same as the output string. So while the constraint is clear when we apply the definition strictly, it may not be apparent from a casual inspection of the shape computation. 

\begin{definition}[X-constraints]\label{def:Xtransition}
	
	A block configuration $<S,(I,E),InDir,L,R>$ is an \emph{X-configuration} if $E$ contains $X$, where $X$ is a variable. 
	
	We say an application of an extended transition $T$ to an $X$-configuration $C$ is $X$\emph{-constrained} if $T$ is a normal transition, and 
	the corresponding $X$\textit{-constraint} is $E \geq 1$. 
	Otherwise the application is not $X$-constrained.
	
	Let $SC$ be a shape computation with a linear observant otter sequence.  
	\begin{itemize}
		\item For each progressive variable $X$, the \emph{minimum value of $X$ in $SC$} is the minimal value satisfying $X \geq 0$ and all $X$-constraints in $SC$. 
		\item For each regressive variable $Y$,  the \emph{minimum value of $Y$ in $SC$} is the minimal value satisfying $Y \geq l$ and all $Y$-constraints in $SC$, where $l$ is the relative difference in the regressive sequence in $SC$.
	\end{itemize}
\end{definition}

We then use the minimum values of the variables as constraints on the application of the ossified ocelot. 

To work out the value of $Count$, we need to determine the minimum value for the regressive variables which will satisfy the constraints imposed by the shape computation. 
For a monoregressive linear observant otter with relative difference $l$, this means the value of $Count$ is 
%found by considering the first value of $k$ such that $Y - (k \times l) < Y_{min}$.  So we have $Y - (Count \times l) < Y_{min}$ and $Y - ((Count -1) \times l) \geq Y_{min}$. So we have $(Y-Y_{min})/l + 1 \geq Count > (Y-Y{min})/l$, and hence $Count$ is
$(Y-Y_{min})/l + 1$, where $Y_{min}$ is the minimum value of $Y$ required for the shape computation to be executed (see Section~\ref{sec:count} for how this formula is obtained). For multiply regressive linear observant otters, we perform a similar calculation for reach regressive variable and use the minimum of these as the value of $Count$.\footnote{The maximum number of regressive variables we have found so far is 2.} % , in which case $Count$ is $min((Y_1-Y_{1min})/l_1 + 1, (Y_2-Y_{2min})/l_2 + 1)$.}

We then use the general instance with $k = Count = (Y-Y_{min})/l + 1$ to determine the final configuration. We then also need to determine the number of steps taken from the initial configuration to the final configuration, whicn can be calculated from $Count$ and the value of $Hops$ in the shape computation; see Section~\ref{sec:hopscalc} for details of the calculations for both $Count$ and the number of steps in the transition. 

\section{Theoretical Results}

A key feature of shape computations is the ability to derive the corresponding transition from the computation itself. In particular, a crucial aspect of this is to be able to calculate the number of hops relevant to the application of the transition. There are two key attributes of the shape computation to be considered here: its length, and whether the otter pattern on which it is based is unilinear or multilinear. 
A shape computation corresponding to a linear observant otter sequence will be executed $(Y-Y_{min})/l + 1$ times for some regressive variable $Y$ and relative difference $l$. 
So if the shape computation has constant length, say $C$, then the number of hops in the transition will be $C \times ((Y-Y_{min})/l + 1)$, which is a linear function of $Y$.  However, more complex formulae emerge when the length of the shape computation is variable, as the number of steps will depend on the values of the progresive and regressive variables. For a shape computation whose length is linear in the progressive and regressive variables in it, and for which ottern pattern is unilinear, then the number of hops is quadratic in the regressive variables. In the multilinear case, though, a shape computation of linear length will result in the hops being exponential in the regressive variables. 

As the term linear is used in various contexts here, we find it appropriate to introduce the term \textit{affine}, as below. 

\begin{definition}[Affine and Quadratic Shape Computations]

The length of a shape computation is 
\begin{itemize}
	\item \emph{constant} if it is constant. 
	\item \emph{affine} if it is length is a linear function of the progressive and regressive variables in it.
	\item \emph{quadratic} if it is a quadratic function of the progressive and regressive variables in it.
\end{itemize}
\end{definition}

For example, if the regressive variables are $Y_1$ and $Y_2$ and the progressive variables are $X_1, X_2$ and $X_3$, then a shape computation of affine length would have length of the form $a_1X_1 + a_2X_2 + a_3X_3 + b_1Y_1 + b_2Y_2 + c$.  In particular, when there is exactly one regressive variable $Y$ and exactly one progressive variable $X$, an affine shape computation's length is of the form $aX + bY + c$ for some integers $a,b$ and $c$. This expression is determined in practice by the expression for the hops used when the target configuration is reached. 

A key point to note is that there is an important relationship between the length of the shape configuration, the hops for the resulting transition, and the activity of the resulting machine. 

\begin{theorem}\label{theorem:hops}
Let $S$ be a shape computation with a linear observant otter sequence which is monoprogressive (variable $X$) and monoregressive (variable $Y$). Then we have the following. 

\begin{enumerate}
	\item If $S$ has \textbf{constant} length, then the number of hops in the transition is \textbf{linear} in $Y$.
	\item If $S$ is \textbf{unilinear} and has \textbf{affine} length, then the number of hops in the transition is \textbf{quadratic} in $Y$.
	\item If $S$ is \textbf{multilinear} and has \textbf{affine} length, then the number of hops in the transition is \textbf{exponential with base $m$} in $Y$, where the relative difference is $(m,k)$.
	\item If $S$ is \textbf{multilinear} and has \textbf{quadratic} length over $X$ only, then the number of hops in the transition is \textbf{exponential with base $m^2$} in $Y$, where the relative difference is $(m,k)$.
\end{enumerate}

\end{theorem}

%\begin{proof}
\textbf{Proof: }\\

In all cases below, let $SC$ be the corresponding shape computation for $S$ with progressive variable $X$ and regressive variable $Y$ with relative differences $(m,n)$ and $d$ respectively, so $X_k = m X_{k-1} + n$ and $Y_k = Y_{k-1} - d$. Let $X_0$ and $Y_0$ be the initial values of $X$ and $Y$ respectively, and $y$ be the minimum value of $Y$ in $SC$. Let the number of times the shape computation is executed be $C$. 	

\begin{enumerate}
\item As the shape computation is executed $C$ times, then the number of hops is $cC$, where $c$ is the length of the shape computation. As $C = (Y_0 - y + d)/d$, the number of hops is linear in the regressive variable. 	

\item Let the length of $SC$ be $aX_0 + bY_0 + c$. As the sequence is unilinear, we have $m = 1$. 

Applying Proposition~\ref{prop:linearaffine} from Section~\ref{sec:hopscalc} with $p = q = 1$ and $C = (Y_0 - y + d)/d = (Y_0 - y)/d + 1$ gives the total hops as below. 

$$(Y_0 - y + d)/d \times \big( aX_0 + bY_0 + c + (Y_0 - y)/2d \times (an -bd) \big) $$

This is sufficient to show that the number of hops in the transition is dominated by the terms including $Y_0^2$ and $X_0Y_0$, and so the number of hops is quadratic in $Y$. 	

\item Let the length of $SC$ be $aX_0 + bY_0 + c$. 

Applying Proposition~\ref{prop:multiaffine} from Section~\ref{sec:hopscalc} with $p = q = 1$ and $Count = (Y_0 - y + d)/d = (Y_0 - y)/d + 1$ gives the total hops as below. 

$$\displaystyle \frac{aX_0(m^{C}-1)}{m-1} +\frac{an(m^{C}-1)}{(m-1)^2} + C \times bY_0 + c - \frac{an}{m-1} - \frac{C(C-1)bd}{2}$$

This is sufficient to show that the formula is dominated by the first two terms, which include subterms of the form $m^{(Y_0-k)/d}$ for some (small) fixed integers $k$ and $d$. These will clearly dominate the remaining quadratic, linear and constant terms. 	

\item Let the length of $SC$ be $aX_0^2 + bX_0 + c$. 

Applying Proposition~\ref{prop:multiquad} from Section~\ref{sec:hopscalc} with $p = q = 1$ and $C = (Y_0 - y + d)/d = (Y_0 - y)/d + 1$ gives the total hops as below. 

% $$\displaystyle \frac{aX_0(m^{Count}-1)}{m-1} +\frac{an(m^{Count}-1)}{(m-1)^2} + Count \times bY_0 + c - \frac{an}{m-1} - \frac{Count(Count-1)bd}{2}$$

$$\displaystyle \frac{  A + B + D }{(m-1)^3(m^2-1)}$$

where \\
$A = \left[ aX_0^2(m-1)^2 + 2anX_0(m-1) +  an^2 \right](m^{2C}-1)(m-1)$ \\
$B = \left[ bX_0(m-1)^2  - 2anX_0(m-1) +  bn(m-1)  - 2an^2 \right](m^C-1)(m^2-1)$ \\
$D = C \times \left[ an^2 -bn(m-1) + c(m-1)^2\right](m-1)(m^2-1) $ \\
$C = (Y_0 - y + d)/d$

This is sufficient to show that the number of hops in the transition is dominated by the terms $m^{2C}$ and $m^C$, with the former dominating the latter, and which can be rewritten to $(m^2)^C$. 	

\end{enumerate}
$\qed$
%\end{proof}

While this result is restricted to the monoprogressive monoregressive case, it can clearly be generalised to multiprogressive and/or multiregressive sequences as well, albeit at the cost of some simplicity and elegance, particularly for cases 3 and 4. 

As we shall see in Section~\ref{sec:results}, it seems that case 3 above does not occur in practice; in other words, it seems reasonable to conject that any multilinear shape computation will be of quadratic length over the progressive variables. This will require further investigation, and more familiarity with examples of multilinear shape computations. 

It should also be noted that that we can similarly analyse the size of the resulting configuration after the application of the ocelot transition. Note that the size of a minimal shape $S$ is unilinear in the progressive and regressive variables, in that if the progressive variables are $X_1, \ldots , X_p$ and the regressive variables are $Y_1, \ldots , Y_q$ then the size of $S$ is $\displaystyle \sum_{i=1}^p X_i + \sum_{i=1}^q Y_i + c$ for some constant $c$. Noting this means that we can also derive the results below. 

\begin{theorem}\label{theorem:size}
Let $S$ be a shape computation with a linear observant otter sequence which is monoprogressive (variable $X$) and monoregressive (variable $Y$). Then we have the following. 

\begin{enumerate}
	\item If $S$ is \textbf{unilinear}, then the size of the final configuration in the transition is \textbf{affine} in $X$ and $Y$.
	\item If $S$ is \textbf{multilinear}, then size of the final configuration in the transition is \textbf{exponential with base $m$} in $Y$, where the relative difference is $(m,k)$.
\end{enumerate}

\end{theorem}

% \begin{proof}
\textbf{Proof:} \\ 

Let the number of times the shape computation is executed be $C$. 

\begin{enumerate}
	\item As the sequence is unilinear, from Definition~\ref{def:application} we have that each tape element will be of the form $(I, E \pm nC)$ where $n$ the relevant tape difference and $C = (Y_0 - y + d)/d$, where $y$, $Y_0$ and $d$ are respectively the minimum value, the initial value and the relative difference for $Y$. As $E$ is either a constant or a variable, this establishes the result. 
	
	\item As the sequence is multilinear, from Definition~\ref{def:application} we have that there is at least one tape element of the form $(I, (m^CE) + n \frac{m^C-1}{m-1})$ where the relevant tape difference is $(m,n)$, $m > 1$ and $C = (Y_0 - y + d)/d$, where $y$, $Y_0$ and $d$ are respectively the minimum value, the initial value and the relative difference for $Y$. This means the size of the final configuration is dominated by the term $m^C = m^{(Y_0 - y + d)/d)}$, which establishes the result. 
\end{enumerate}

$\qed$	
%\end{proof}

Note that Theorems ~\ref{theorem:hops} and ~\ref{theorem:size} taken together allow us to show some general relationship between productivity and activity. In particular, the corollary below follows directly from these two results. 

\begin{cor}\label{corollary:prodact}
	Let $S$ be a shape computation with a linear observant otter sequence which is monoprogressive and monoregressive. 
	
	\begin{enumerate}
		\item If the sequence is\textbf{ unilinear} and the shape computation has \textbf{affine} length, then in the corresponding ocelot transition the increase in productivity is \textbf{affine} and the increase in activity is \textbf{quadratic}. 
		\item If the sequence is \textbf{multilinear} and the shape computation has \textbf{quadratic} length, then in the corresponding ocelot transition the increase in productivity is \textbf{exponential with base $m$} and the increase in activity is \textbf{exponential with base $m^2$.} 
	\end{enumerate} 
	
\end{cor} 

Hence any computation which is dominated by ocelot transitions of the above type will tend to have a quadratic relationship between productivity and activity, whether unilinear or multilinear. This means that shape computation analysis can provide some justification behind the observed relationship between productivity and activity, i.e.\ that in many machine of high productivity, the activity is around the square of the productivity. We believe this is the first analytical result (as distinct from empirical work) to show this relationship. 

\section{Finding and Managing Shape Computations}\label{sec:managing}
Defining shape computations is one thing; finding them once a pattern has been identified is another. Here we address some of the issues involved in managing the search for shapely shelducks. 

Once have observed a pattern and have formulated a shape $S$ and a target configuration $T$, we proceed by executing the machine with the initial configuration $S$ until either the current configuration is $T$ (in which case the search for the shape computation has been successful) or we conclude that no such shape computation exists (which is usually by exceeding a given length bound, but possibly also for more technical reasons explained below). A key point to note is that as our configurations now involve expressions including variables, there are some subtleties involved with do not occur with ground configurations.  

One issue that can arise is the interaction between the two operators $//$ and $mod$. For example, consider the sequence below which occurs in the execution of Dragon 7, and the corresponding shape computation. 

\begin{center}
	\begin{tabular}{ll}
		\textbf{Hops} & \textbf{Configuration} \\
		53 & $\{c\}0\{r\}2^{8}1$ \\
		66 & $\{c\}0\{r\}1^{5}2^{6}1$ \\
		99 & $\{c\}0\{r\}1^{10}2^{4}1$ \\
	\end{tabular}
\end{center}

We can see that this suggests the pattern $\{c\}0\{r\}1^{X}2^{Y}1 \longmapsto \{c\}0\{r\}1^{X+5}2^{Y-2}1$ which has the shape computation below. 

\begin{center}
	\begin{tabular}{ll}
		\textbf{Hops} & \textbf{Configuration} \\
		$0$     & $\{c\}0\{r\}1^{X}2^{Y}1$ \\ % Hops: 0 Ones: 1 Pos: 0 Transition: init 
		$1$     & $\{a\}0\{r\}1^{X+1}2^{Y}1$ \\ % Hops: 1 Ones: 1 Pos: -1 Transition: complex(normal) 
		$2$     & $1\{b\}1^{X+1}\{l\}2^{Y}1$ \\ % Hops: 2 Ones: 2 Pos: 0 Transition: complex(normal) 
		$X+3$   & $12^{X+1}\{b\}2^{Y}\{l\}1$ \\ % Hops: X+3 Ones: 2 Pos: X1+1 Transition: complex(acceleration) 
		$X+4$   & $1\{b\}2^{X+1}\{r\}12^{Y-1}1$ \\ % Hops: X+4 Ones: 3 Pos: X1 Transition: complex(normal) 
		$2X+5$  & $\{b\}1\{r\}1^{X+2}2^{Y-1}1$ \\ % Hops: 2*X+5 Ones: 2 Pos: -1 Transition: complex(acceleration) 
		$2X+6$  & $2\{b\}1^{X+2}\{l\}2^{Y-1}1$ \\ % Hops: 2*X+6 Ones: 2 Pos: 0 Transition: complex(normal) 
		$3X+8$  & $2^{X+3}\{b\}2^{Y-1}\{l\}1$ \\ % Hops: 3*X+8 Ones: 1 Pos: X1+2 Transition: complex(acceleration) 
		$3X+9$  & $\{b\}2^{X+3}\{r\}12^{Y-2}1$ \\ % Hops: 3*X+9 Ones: 2 Pos: X1+1 Transition: complex(normal) 
		$4X+12$ & $\{b\}0\{r\}1^{X+4}2^{Y-2}1$ \\ % Hops: 4*X+12 Ones: 1 Pos: -2 Transition: complex(acceleration) 
		$4X+13$ & $\{c\}0\{r\}1^{X+5}2^{Y-2}1$ \\ % Hops: 4*X+13 Ones: 1 Pos: -3 Transition: complex(normal) 
	\end{tabular}
\end{center}

This results in the ocelot transition below. 

\begin{center}
	$(blank)\{c\}0\{r\}1^{X}2^{Y} \leadsto \{c\}0\{r\}1^{X+5(Y//2)}2^{Y mod 2}$, $X \geq 0, Y \geq 2$ in $(Y//2)(4X+10(Y//2)+3)$ steps
\end{center}

As the relative difference for the regressive variable $Y$ here is $2$, applying the transition involves expressions which include both $//$ and $mod$. From a calculation perspective, applying such formula for ground values of $X$ and $Y$ is trivial. However when finding shape computations, which involves comparing the current and target configurations, we need to be able to appropriately manipulate and simplify algebraic expressions, including those containing variables. This will not only ensure that such expressions are compactly maintained, but also so that we can easily tell when a target configuration has been reached in a shape computation. These seems simple enough in principle, but in practice some care is needed, particularly when it comes to judicious use of Equation~\ref{equation:mod} above, 
%identity $n(m//n) = m - m \; mod \; n$ (see Section~\ref{sec:moduli} for more details), 
which establishes a useful relationship between $//$ and $mod$. This relationship is a key step used in the simplification of expressions which are progressively updated in the shape computation for exponents in the configuration as well as the number of hops used so far. This involves both relatively simple cases such as simplifying $2X + 4 + Y + X + 12$ to $3X+Y+16$ and more compex ones such as simplifying $4X+(Y//2)+3+(Y//2)+2$ to $4X+2(Y//2)+5$ and then to $4X+Y - Y \;mod \; 2 +5$. 

A related issue is the interaction between transitions which leads to observable patterns. 
Consider the otter pattern and shape computation below from Dragon 60. 

\begin{center}
	\begin{tabular}{ll}
		\textbf{Hops} & \textbf{Configuration} \\
		160 & $1^{9}\{d\}0^{2}\{l\}(10)^{9}01^{2}$ \\
		218 & $1^{19}\{d\}0^{2}\{l\}(10)^{7}01^{2}$ \\
		306 & $1^{29}\{d\}0^{2}\{l\}(10)^{5}01^{2}$ \\
	\end{tabular}
\end{center}

\begin{center}
	\begin{tabular}{ll}
		\textbf{Hops} & \textbf{Configuration} \\
		$0$ & $1^{X}\{d\}0^{2}\{l\}(10)^{Y}01^{2}$ \\ % Hops: 0 Ones: 2 Pos: 0 Transition: init 
		$1$ & $1^{X+1}\{a\}0\{l\}(10)^{Y}01^{2}$ \\ % Hops: 1 Ones: 2 Pos: 1 Transition: complex(normal) 
		$2$ & $1^{X+2}\{b\}(10)^{Y}\{l\}01^{2}$ \\ % Hops: 2 Ones: 2 Pos: 2 Transition: complex(normal) 
		$4$ & $1^{X+4}\{a\}(10)^{Y-1}\{l\}01^{2}$ \\ % Hops: 4 Ones: 2 Pos: 4 Transition: complex(normal) 
		$5$ & $\{c\}1^{X+4}\{r\}0^{2}(10)^{Y-2}01^{2}$ \\ % Hops: 5 Ones: 2 Pos: 3 Transition: complex(normal) 
		$X-X mod \;2+9$ & $\{c\}1^{X mod \;2}\{r\}(10)^{(X+4)//2}0^{2}(10)^{Y-2}01^{2}$ \\ % Hops: X-X mod 2+9 Ones: 2 Pos: -2*((X1+4)//2)+3 Transition: expansion(2) 
		$X+8$   & $\{c\}1\{r\}(10)^{(X+4)//2}0^{2}(10)^{Y-2}01^{2}$ \\ % Hops: X+8 Ones: 3 Pos: -X1 Transition: modulus 
		$X+9$   & $\{e\}0\{r\}0(10)^{(X+4)//2}0^{2}(10)^{Y-2}01^{2}$ \\ % Hops: X+9 Ones: 2 Pos: -X1-1 Transition: complex(normal) 
		$X+10$  & $\{f\}0\{r\}(10)^{(X+4)//2+1}0^{2}(10)^{Y-2}01^{2}$ \\ % Hops: X+10 Ones: 2 Pos: -X1-2 Transition: complex(normal) 
		$X+11$  & $\{a\}0\{r\}0(10)^{(X+4)//2+1}0^{2}(10)^{Y-2}01^{2}$ \\ % Hops: X+11 Ones: 2 Pos: -X1-3 Transition: complex(normal) 
		$X+12$  & $1\{b\}0\{l\}(10)^{(X+4)//2+1}0^{2}(10)^{Y-2}01^{2}$ \\ % Hops: X+12 Ones: 3 Pos: -X1-2 Transition: complex(normal) 
		$X+13$  & $\{a\}1\{r\}1(10)^{(X+4)//2+1}0^{2}(10)^{Y-2}01^{2}$ \\ % Hops: X+13 Ones: 4 Pos: -X1-3 Transition: complex(normal) 
		$X+14$  & $\{c\}0\{r\}01(10)^{(X+4)//2+1}0^{2}(10)^{Y-2}01^{2}$ \\ % Hops: X+14 Ones: 3 Pos: -X1-4 Transition: complex(normal) 
		$X+15$  & $\{b\}0\{r\}0^{2}1(10)^{(X+4)//2+1}0^{2}(10)^{Y-2}01^{2}$ \\ % Hops: X+15 Ones: 3 Pos: -X1-5 Transition: complex(normal) 
		$X+16$  & $\{a\}0\{r\}1001(10)^{(X+4)//2+1}0^{2}(10)^{Y-2}01^{2}$ \\ % Hops: X+16 Ones: 4 Pos: -X1-6 Transition: complex(normal) 
		$X+17$  & $1\{b\}1\{l\}001(10)^{(X+4)//2+1}0^{2}(10)^{Y-2}01^{2}$ \\ % Hops: X+17 Ones: 5 Pos: -X1-5 Transition: complex(normal) 
		$X+18$  & $11\{d\}0\{l\}01(10)^{(X+4)//2+1}0^{2}(10)^{Y-2}01^{2}$ \\ 
		$X+19$  & $1^{3}\{a\}0\{l\}1(10)^{(X+4)//2+1}0^{2}(10)^{Y-2}01^{2}$ \\ % Hops: X+19 Ones: 6 Pos: -X1-3 Transition: complex(normal) 
		$X+20$  & $1^{4}\{b\}1\{l\}(10)^{(X+4)//2+1}0^{2}(10)^{Y-2}01^{2}$ \\ % Hops: X+20 Ones: 7 Pos: -X1-2 Transition: complex(normal) 
		$X+21$  & $1^{5}\{d\}(10)^{(X+4)//2+1}\{l\}0^{2}(10)^{Y-2}01^{2}$ \\ % Hops: X+21 Ones: 7 Pos: -X1-1 Transition: complex(normal) 
		$3X+31$ & $1^{X+10}\{d\}0^{2}\{l\}(10)^{Y-2}01^{2}$ \\ % Hops: 3*X+31 Ones: 2 Pos: 4 Transition: expansion(1) 
	\end{tabular}
\end{center}

There are two keys points to note about this particular shape computation. One is the simplification of $X - X mod \;2 + 9$ to $X+8$; the other is the interaction between the two (extended) transitions below. 

\begin{center}
	\begin{tabular}{lllllll}
		\multicolumn{3}{c}{Initial} & \multicolumn{3}{c}{Final} & Steps \\
		c & 11 & r & c & 10 & l & 2 \\
		d & 10 & l & d & 11 & r & 4 \\ 
	\end{tabular}
\end{center}

Note that at step 5 above we have the configuration $\{c\}1^{X+4}\{r\}0^{2}(10)^{Y-2}01^{2}$. Now a naive execution at this point would progress from here in one step to $\{e\}1^{X+3}\{r\}0^{3}(10)^{Y-2}01^{2}$, and then to $\{c\}1^{X+2}\{r\}10^{3}(10)^{Y-2}01^{2}$ which then leads to another single step and so on. However, with a little more care we note that the sequence of transitions below means that here we have an ``expanded'' notion of an acceleration transition. 

\begin{center}
	$111\{c\}1\{r\} \rightarrow 11\{e\}1\{r\}0 \rightarrow 1\{c\}1\{r\}10 \rightarrow ...$
\end{center}

More specifically, noting the above the above extended transition for the string $11$ in the state $c$, we first rewrite $1^{X+4}$ to $1^{X mod \;2}(11)^{(X+4)//2}$ and hence apply the extended acceleration transition to obtain $\{c\}1^{X mod \;2}\{r\}(10)^{(X+4)//2}0^{2}(10)^{Y-2}01^{2}$. 

This means that whenever we encounter a tape element $(I,E)$ where $E$ is variable, we need to check whether we an acceleration transition like this can be found by finding an appropriate \emph{expansion} of the tape element. 

\begin{definition}
	Let $(I,E)$ be a tape element. An \emph{expansion} of $(I,E)$ is a block tape of the form $(I^n,E//n).(I,E \; mod \;n)$ or  $(I,E \; mod \;n).(I^n,E//n)$  for some $n \geq 2$.
\end{definition}

Clearly the choice which option for an expansion is appropriate will depend on whether the tape pointer is at the left or right end of $(I,E)$. 
In principle, there is an infinite number of expansions of a given tape element. In practice, as we are only looking for expansions that lead to extended acceleration transitions, there is no need to consider any cases when the length of $I^n$ exceeds the number of states in the machine. Hence we only need to consider a fairly small number of possible expansions. In the above case, for instance, we only need to consider the case $n=2$ before finding the relevant expansion. Note that there is a natural ``contraction'' in the shape computation that essentially counteracts this expansion at step $3X+31$. At the previous step we have 

\begin{center}
	$1^{5}\{d\}(10)^{(X+4)//2+1}\{l\}0^{2}(10)^{Y-2}01^{2}$ 
\end{center}

and applying the second extended (acceleration) transition from the table above, we get the unsimplified configuration 

\begin{center}
	$1^{5}(11)^{(X+4)//2+1}\{d\}0^{2}\{l\}(10)^{Y-2}01^{2}$
\end{center} 

Now as we can rewrite any tape block of the form $(11)^X$ as $1^{2X}$, we can rewrite this as 

\begin{center}
	$1^{5+2((X+4)//2+1)}\{d\}0^{2}\{l\}(10)^{Y-2}01^{2}$
\end{center}

and as we have 

\begin{center}
	$5+2((X+4)//2+1) = 5+2(X+4)//2 + 2 = 7+ (X+4) - (X+4) mod \;2 = X+ 11 - X mod \;2$
\end{center} 

we arrive at the configuration below. 

\begin{center}
	$1^{X+11 - X mod \;2}\{d\}0^{2}\{l\}(10)^{Y-2}01^{2}$
\end{center}

Now as above, this leaves us with the issue about how to deal with exponents such as $X-X mod \;2+9$ and $X+ 11 - X mod \;2$, and in particular what transition to apply to $\{c\}1^{X mod \;2}\{r\}$. Clearly $X mod \;2$ can only be either 0 or 1, but each case will be handled by different transitions in the machine, and hence behave differently. One possibility is to consider both alternatives in parallel, and recombine them later into one shape computation. Another is to consider a further constraint on $X$ as follows. We note that in the original configuration for this otter pattern (at hop 160 above), we have $X = 9$ and so $X mod \;2 = 1$. We note that as the relative difference here is $10$, and that $X+10 \; mod \;2 = X \; mod \;2$, meaning that not only is this status ``preserved'' in the shape computation, we may assume that $X mod \;2 = 1$ throughout, provided that this is the case initially. Hence as along as equivalence modulo 2 is preserved in the shape computation, we can simplify the exponents $X-X mod \;2+9$ and $X+ 11 - X mod \;2$ to $X+8$ and $X+10$ respectively. This will also require adding the additional constraint $X mod \;2 = 1$ to the transition, which can only be applied if this is satisfied (along with any other constraints).  
While this seems a little restrictive compared to the potential approach of considering all possible values of $X \; mod \; 2$, it seems that the alternative may lead to a significant number of alternative branches, especially for cases in which the base is more than $2$. 

This approach does not work in all cases, in that there are some in which the relative difference does not preserve the modulus status. Such a case occurs in Dragon 30, as below.   

\begin{center}
	\begin{tabular}{ll}
		\textbf{Hops} & \textbf{Configuration} \\
		158 & $\{a\}3^{2}\{r\}13^{9}$ \\
		178 & $\{a\}3^{7}\{r\}13^{5}$ \\
		222 & $\{a\}3^{12}\{r\}13$ \\
	\end{tabular}
\end{center}

The relevant shape computation with target $\{a\}3^{X+5}\{r\}13^{Y-4}$ proceeds as below. 

\begin{center}
	\begin{tabular}{ll}
		\textbf{Hops} & \textbf{Configuration} \\
		$0$ & $\{a\}3^{X}\{r\}13^{Y}$ \\ 
		$X+1$ & $1\{b\}1^{X+1}\{l\}3^{Y}$ \\
		$2X-(X+1)mod \;2+2$ & $13^{X-(X+1)mod \;2+1}\{b\}1^{(X+1)mod \;2}\{l\}3^{Y}$ \\
	\end{tabular}
\end{center}

Here our expansion is from $1\{b\}1^{X+1}\{l\}3^{Y}$ to $1\{b\}(11)^{(X+1)//2}\{l\}1^{(X+1)mod \;2}3^{Y}$ which becomes \\
$1(33)^{(X+1)//2}\{b\}1^{(X+1)mod \;2}\{l\}3^{Y}$. 
However, the relative difference here is $5$ and $X + 5 \; mod \;2 \not = X \; mod \;2$. Hence the constraint that $X = 0 \;mod \;2$ is not maintained throughout, and so this is an invalid shape computation. A way around this is to reconfigure the detected pattern in such a way as the modulus status is preserved; in particular we note that the slightly weaker pattern (and hence hypothesis) of $\{a\}3^{X}\{r\}13^{Y} \longmapsto \{a\}3^{X+10}\{r\}13^{Y-8}$ does indeed preserve equivalence modulo 2. Hence we need to be able to reformulate the shape computation in this way, i.e. when a modulus status is not preserved for $X = k \; mod \; n$ for a relative difference $m$, then we multiply all relative differences in the target configuration by the factor $lcm(n,m)/m$ where $lcm(n,m)$ is the lowest common multiple of $n$ and $m$. We then restart the shape computation with the reformulated target, which we known in advance will preserve the modulus status.  

A further issue that arises in finding shape computations is the nesting of patterns within shapely shelducks i.e.\ using ossified ocelots to verify observant otter patterns. 

Consider the observant otter pattern below, which is Otter 5 of Dragon 25. 

\begin{center}
	\begin{tabular}{ll}
		\textbf{Hops} & \textbf{Configuration} \\
		10559 & $1(1101)^{38}\{d\}0\{l\}$ \\
		10601 & $1(1101)^{36}(01)^{5}\{d\}0\{l\}$ \\
		10703 & $1(1101)^{34}(01)^{10}\{d\}0\{l\}$ \\
	\end{tabular}
\end{center}

The relevant shape computation proceeds as below. 

\begin{center}
	\begin{tabular}{ll}
		\textbf{Hops} & \textbf{Configuration} \\
		$0$     & $1(1101)^{Y}(01)^{X}\{d\}0\{l\}$ \\ % Hops: 0 Ones: 1 Pos: 0 Transition: init 
		$1$     & $1(1101)^{Y}(01)^{X}1\{e\}0\{l\}$ \\ % Hops: 1 Ones: 2 Pos: 1 Transition: complex(normal) 
		$2$     & $1(1101)^{Y}(01)^{X}\{c\}1\{r\}1$ \\ % Hops: 2 Ones: 3 Pos: 0 Transition: complex(normal) 
		$3$     & $1(1101)^{Y}\{b\}(01)^{X}\{r\}01$ \\ % Hops: 3 Ones: 2 Pos: -1 Transition: complex(normal) 
		$5$     & $1(1101)^{Y}\{c\}(01)^{X-1}\{r\}1001$ \\ % Hops: 5 Ones: 3 Pos: -3 Transition: complex(normal) 
		$2X+3$  & $1\{c\}(1101)^{Y}\{r\}(10)^{X}01$ \\ % Hops: 2*X+3 Ones: 2 Pos: -2*X1-1 Transition: complex(acceleration) 
		$2X+7$  & $1\{e\}(1101)^{Y-1}\{r\}0010(10)^{X}01$ \\ % Hops: 2*X+7 Ones: 3 Pos: -2*X1-5 Transition: complex(normal) 
		$2X+8$  & $1(1101)^{Y-1}\{c\}0010\{l\}(10)^{X}01$ \\ % Hops: 2*X+8 Ones: 3 Pos: -2*X1-4 Transition: complex(normal) 
		$2X+14$ & $1(1101)^{Y-1}(01)^{2}\{b\}(10)^{X}\{l\}01$ \\ % Hops: 2*X+14 Ones: 4 Pos: -2*X1 Transition: complex(normal) 
		$2X+15$ & $1(1101)^{Y-1}\{e\}(01)^{2}\{r\}0^{2}(10)^{X-1}01$ \\ % Hops: 2*X+15 Ones: 4 Pos: -2*X1-1 Transition: complex(normal) 
		$2X+16$ & $1(1101)^{Y-1}(01)^{2}\{c\}0^{2}\{l\}(10)^{X-1}01$ \\ % Hops: 2*X+16 Ones: 4 Pos: -2*X1 Transition: complex(normal) 
		$2X+17$ & $1(1101)^{Y-1}(01)^{2}0\{a\}0\{l\}(10)^{X-1}01$ \\ 
		$2X+18$ & $1(1101)^{Y-1}(01)^{3}\{b\}(10)^{X-1}\{l\}01$ \\
	\end{tabular}
\end{center}

At this point, the execution of the shape computation will endlessly repeat steps similar to the steps from $2X+14$ to $2X+18$, resulting in a cycle of ever-decreasing values of $X$ (ie, $X$, $X-1$, $X-2$, ...). 

However, we note that Otter 4 for this same dragon results in the following ocelot transition. 

\begin{center}
	$(10)^{X}\{a\}0\{l\}(10)^{Y} \leadsto (10)^{X+Y}\{a\}0\{l\}$ in $4Y$ steps with constraints $X \geq 0, Y \geq 1$.
\end{center} 

By rotating the configuration at step $2X+17$ to $1(1101)^{Y-1}0(10)^2\{a\}0\{l\}(10)^{X-1}01$ we can apply Otter 4 at this point to obtain the rest of the shape computation as below. 

\begin{center}
	\begin{tabular}{ll}
		\textbf{Hops} & \textbf{Configuration} \\
		$6X+13$  & $1(1101)^{Y-1}(01)^{2}0(10)^{X-1}\{a\}0\{l\}01$ Otter 4 \\ % Hops: 6*X+13 Ones: 4 Pos: -1 Transition: ocelot(4) 
		$6X+14$  & $1(1101)^{Y-1}0(10)^{X+1}1\{b\}0\{l\}1$ \\ % Hops: 6*X+14 Ones: 3 Pos: 0 Transition: complex(normal) 
		$6X+15$  & $1(1101)^{Y-1}0(10)^{X+1}\{c\}1\{r\}1^{2}$ \\ % Hops: 6*X+15 Ones: 4 Pos: -1 Transition: complex(normal) 
		$8X+19$  & $1\{c\}(1101)^{Y-1}\{r\}(10)^{X+2}1^{2}$ \\ % Hops: 8*X+19 Ones: 3 Pos: -2*X1-5 Transition: expansion(1) 
		$8X+23$  & $1\{e\}(1101)^{Y-2}\{r\}0010(10)^{X+2}1^{2}$ \\ % Hops: 8*X+23 Ones: 4 Pos: -2*X1-9 Transition: complex(normal) 
		$8X+24$  & $1(1101)^{Y-2}\{c\}0010\{l\}(10)^{X+2}1^{2}$ \\ % Hops: 8*X+24 Ones: 4 Pos: -2*X1-8 Transition: complex(normal) 
		$8X+30$  & $1(1101)^{Y-2}(01)^{2}\{b\}(10)^{X+2}\{l\}1^{2}$ \\ % Hops: 8*X+30 Ones: 5 Pos: -2*X1-4 Transition: complex(normal) 
		$8X+31$  & $1(1101)^{Y-2}\{e\}(01)^{2}\{r\}0^{2}(10)^{X+1}1^{2}$ \\ % Hops: 8*X+31 Ones: 5 Pos: -2*X1-5 Transition: complex(normal) 
		$8X+32$  & $1(1101)^{Y-2}(01)^{2}\{c\}0^{2}\{l\}(10)^{X+1}1^{2}$ \\ % Hops: 8*X+32 Ones: 5 Pos: -2*X1-4 Transition: complex(normal) 
		$8X+33$  & $1(1101)^{Y-2}(01)^{2}0\{a\}0\{l\}(10)^{X+1}1^{2}$ \\ % Hops: 8*X+33 Ones: 5 Pos: -2*X1-3 Transition: complex(normal) 
		$12X+37$ & $1(1101)^{Y-2}(01)^{2}0(10)^{X+1}\{a\}0\{l\}1^{2}$ Otter 4\\ % Hops: 12*X+37 Ones: 5 Pos: -1 Transition: ocelot(4) 
		$12X+38$ & $1(1101)^{Y-2}0(10)^{X+3}1\{b\}1^{2}\{l\}$ \\ % Hops: 12*X+38 Ones: 4 Pos: 0 Transition: complex(normal) 
		$12X+39$ & $1(1101)^{Y-2}0(10)^{X+3}\{e\}1\{r\}01$ \\ % Hops: 12*X+39 Ones: 3 Pos: -1 Transition: complex(normal) 
		$12X+40$ & $1(1101)^{Y-2}0(10)^{X+3}1\{c\}0\{l\}1$ \\ % Hops: 12*X+40 Ones: 3 Pos: 0 Transition: complex(normal) 
		$12X+41$ & $1(1101)^{Y-2}0(10)^{X+4}\{a\}1\{l\}$ \\ % Hops: 12*X+41 Ones: 2 Pos: 1 Transition: complex(normal) 
		$12X+42$ & $1(1101)^{Y-2}0(10)^{X+4}1\{d\}0\{l\}$ \\ % Hops: 12*X+42 Ones: 2 Pos: 2 Transition: complex(normal) 
	\end{tabular}
\end{center}

Note that there is a second application of Otter 4 at step $8X+33$, and this allows us to find the target configuration of $1(1101)^{Y-2}0(10)^{X+4}1\{d\}0\{l\}$ as desired. This means that just as in execution of ground configurations, the executions of shape computations seeking to verify otter patterns should also make use of previously determined ocelot transitions. 

This means that when searching for shape computations, we have some extra issues that we need to account for compared to ``normal'' execution. In particular the use of expansions and ocelot transitions during execution may mean that the target configuration is not explicitly presented, and hence can be ``missed''. For example, the application of the ocelot transition at step $8X+33$ in the above shape computation means that several intermediate configurations, which would be present in a naive execution from the same configuration, are omitted. In normal execution, when we are simply interested to find out when (or even whether) a given machine's computation terminates, applying ocelot transitions whenever possible seems appropriate. For shape computations, it is less clear cut, as we are looking for a specific target configuration. In order to allow for both the use of more powerful methods as well as ensuring that critical configurations are not skipped, our strategy is to search with a variety of modes, with the first search for the target configuration being naive (i.e.\ without expansion or ocelot transitions), and if that fails (usually by exceeding a length bound) then we try again by successively adding expansion and ocelot transitions to the strategy used. We also allow the possibility of ``breaking up'' transitions like the one for $c$ with input $11$ in Dragon 60 above (which we refer to as the \textit{modify} option) to ensure that a target configuration is not missed by skipping over the configuration involving the state $\{e\}$. In our outputs we annotate each search with the relevant options, giving 4 variations overall (naive, expansion only, expansion + ocelot, and expansion + ocelot + modify). Naturally other combinations or more sophisticated heuristics can be used as well, but this seems an appropriate starting point, which we expect will be refined over time as we gather more experience with shape computations. 

\section{Empirical Results}\label{sec:results}

We have developed an implementation of an interpreter for the Turing machine extensions described above, including a detector for otter patterns, a procedure for verifying shapely shelducks, deriving ocelot transitions from shape computations and applying these transitions in execution, as well as auxiliary functions such as simplifying expressions, compressing and expanding configurations as appropriate, and calculations involving various formulae. There is undoubtedly significant scope for improvement, particularly in terms of execution speed. However, it should be noted that while efficiency is an important aspect of this work, it is not all-important; our intention is to be able to provide a clear justification for each transition derived, hence ensuring accuracy of our outcomes, whilst also being only too aware of the possibility of overly-naive techniques producing impractical execution times. 

This implementation consists of around 5,600 lines of SWI-Prolog \cite{swi} version 10.1.15 written by the author and is available at \url{https://github.com/JamesHarland/Ossified-Ocelot}. 

In order to determine an appropriate dataset for evaluation, we have two main sources. The first is a historical collection of 100 machines of high activity and productivity reported by Harland \cite{Harland16}, supplemented by a further 15 such machines with 5 states and 2 symbols due to the work of Michel \cite{Michel22}. 
However this collection has been dwarfed by the recent work of the Busy Beaver Challenge \cite{BBCpage}, which has greatly enhanced and expanded our knowledge of such dreadful dragons. Given the large number of example machines, we have chosen to take the top 20 known machines in each dimension together with some others (such as a machine with 15 states), which (allowing for some duplicates) has more than doubled the number of machines in the dataset to 240. 
In order to simplify comparison, the 100 machines of \cite{Harland16} have retained numbers 1-100, with the extra 15 of Michel \cite{Michel22} numbered 101-115, and the machines from the Busy Beaver 
Challenge \cite{BBCpage} numbered from 116 upwards. 
The author would very pleased to hear of further examples of dreadful dragons.

A summary of the outcomes is given in Table~\ref{table:overall}. Note that we denote an $n$-state $m$-symbol machine as having dimension $n \times m$ and class $nm$. 
Machines listed as \textbf{Terminates} are those which we have successfully run to completion (which we call \textit{terminating termites}). \textbf{Unproven} lists the machines for which we have not (yet!) achieved this (which we call \textit{unproven urchins}), while \textbf{Overflow} lists those for which computation exceeded our memory limits (which we call \textit{clattering clancies}).  All of our results were obtained on a PC running Windows 11 with an Intel i7-11700 2.5 GHz processor with 8 cores, 32 GB of RAM and a disk capacity of 930 GB. This particular combination of technologies (i.e.\ the above hardware and  SWI Prolog 10.1.15) has an upper bound of $2^{2,147,479,551}$ (or just under $2^{2^{31}}$) on the size of the integers that can be used; execution of the \textbf{Overflow} machines exceeded this limit, which meant we were unable to obtain any meaningful results for these machines (apart from the generation of some ocelot transitions).
This point will be discussed in more detail a little later. More detailed results are given in Tables~\ref{table:detail1},~\ref{table:detail2} and~\ref{table:detail3}. All data supporting these results, including executions for all 240 machines classified into termites, urchins and clancies, can be found at \url{https://github.com/JamesHarland/Ossified-Ocelot}. 

\begin{table}
	\centering
	\begin{tabular}{lcccccc}
		\textbf{Range} & \textbf{Dimension} & \textbf{Class} & \textbf{Total} & \textbf{Terminates}  & \textbf{Unproven} & \textbf{Overflow}  \\
		1-4                     & 2x4 &  8  & 4   & 4   & 0  & 0  \\
		5-17,  \textit{116-126} & 3x3 &  9  & 24  & 24  & 0  & 0  \\
		18-26, \textit{101-115} & 5x2 & 10  & 24  & 24  & 0  & 0  \\
		27-50, \textit{127-137} & 2x5 & 10  & 35  & 24  & 11 & 0  \\
		51-72, \textit{200-219} & 6x2 & 12  & 42  & 22  & 10 & 10 \\
		73-81, \textit{158-177} & 3x4 & 12  & 29  & 12  & 1  & 16 \\
		82-91, \textit{178-199} & 4x3 & 12  & 32  & 11  & 3  & 18 \\
		92-98, \textit{138-157} & 2x6 & 12  & 27  & 7   & 3  & 17 \\
		99,    \textit{220-239} & 7x2 & 14  & 21  & 1   & 2  & 18 \\
		\textit{240}            & 3x5 & 15  & 1   & 0   & 1  & 0  \\
		100                     & 8x2 & 16  & 1   & 1   & 0  & 0  \\
		\textbf{1-240}          &     &     & \textbf{240} & \textbf{130} & \textbf{31} & \textbf{79} \\
	\end{tabular}
	\caption{Overall results}
	\label{table:overall}
\end{table}

\begin{table}
	{\footnotesize
		\begin{tabular}{lllllllll}
			\textbf{No.} & \textbf{Dim.} & \textbf{Time} & \textbf{Otters} & \textbf{Ocelots} & \textbf{Ocelot Steps} & \textbf{Productivity} & \textbf{Activity} & \textbf{Predicted} \\
			1  &	2x4 &	0.05 &	3 &	6  &	5400 & 	84 &	6445&	88.5 \\
			2  &	2x4 &	0.03 &	1 &	4  &	6310 & 	90 &	7195&	89.5 \\
			3  &	2x4 &	0.03 &	1 &	4  &	6310 & 	90 &	7195&	89.5 \\
			4  &	2x4 &	1.36 &	1 &	10 &	3913480 & 	2050 &	3932964 & 99.5 \\
			5  &	3x3 &	9985.18 & 2 & 2801 & 36953 &  31 &	2315619 & 1.6 \\
			6  &	3x3 &	0.86 &	2 &	13 &	29340870 & 	5600 &	29403894 &	99.8 \\
			7  &	3x3 &	0.04 &	1 &	8  &	92586026 & 	13949 &	92649163 &	99.9 \\
			8  &	3x3 &	1.3  &	1 &	10 &	3913480 & 	2050 &	3932964 &	99.5 \\
			9  &	3x3 &	0.14 &	2 &	22 &	310280672 & 	36089 &	310341163 &	100.0 \\
			10 &	3x3 &	0.47 &	1 &	9  &	544607710 & 	32213 &	544884219 &	99.9 \\
			11 &	3x3 &	1.2  &	1 &	16 &	1808265960 & 	43925 &	1808669046 &	100.0 \\
			12 &	3x3 &	0.12 &	4 &	18 &	4938512275 & 	107900 &	4939345068 &	100.0 \\
			13 &	3x3 &	1.42 &	1 &	16 &	1808265960 & 	43925 &	1808669066 &	100.0 \\
			14 &	3x3 &	0.09 &	1 &	20 &	987508487207 & 	1525688 &	987522842126 &	100.0 \\
			15 &	3x3 &	0.06 &	1 &	14 &	4144451519458 & 	2950149 &	4144465135614 &	100.0 \\
			16 &	3x3 &	0.19 &	1 &	17 &	{16 digits} &	95524079 &	{16 digits} &	100.0 \\
			17 &	3x3 &	0.3  &	3 &	149 &	{18 digits} &	374676383 &	{18 digits} & 	100.0 \\
			18 &	5x2 &	0.05 &	1 &	13 &	47115534 & 	4098 &	47176870 &	99.9 \\
			19 &	5x2 &	0.04 &	1 &	12 &	11768114 & 	4098 &	11798826 &	99.7 \\
			20 &	5x2 &	0.35 &	2 &	20 &	23523766 & 	4097 &	23554764 &	99.9 \\
			21 &	5x2 &	0.06 &	2 &	11 &	11762577 & 	4097 &	11798796 &	99.7 \\
			22 &	5x2 &	0.07 &	2 &	11 &	11753274 & 	4096 &	11804910 &	99.6 \\
			23 &	5x2 &	0.07 &	2 &	11 &	11753274 & 	4096 &	11804896 &	99.6 \\
			24 &	5x2 &	0.08 &	1 &	6  &	2124788 & 	1915 &	2133492 &	99.6 \\
			25 &	5x2 &	3.98 &	11 & 32 &	2323796 & 	1471 &	2358064 &	99.0 \\
			26 &	5x2 &	0.22 &	4 &	8  &	105356 & 	501 &	134467 &	94.3 \\
			27 &	2x5 &	0.21 &	1 &	10 &	8617433920 & 	90604 &	8619024596 &	100.0 \\
			28 &	2x5 &	2.88 &	2 &	12 &	4560983494 & 	64665 &	4561507071 &	100.0 \\
			29 &	2x5 &	0.23 &	2 &	19 &	7543194926 & 	97104 &	7543673517 &	100.0 \\
			30 &	2x5 &	0.68 &	3 &	102 &	233418621414 & 	458357 &	233431192481 &	100.0 \\
			31 &	2x5 &	2.74 &	1 &	13 &	469111851960 & 	668420 &	469121946086 &	100.0 \\
			32 &	2x5 &	0.08 &	1 &	13 &	912589873782 & 	1957771 &	912594733606 &	100.0 \\
			33 &	2x5 &	0.22 &	2 &	36 &	924176893940 & 	1137477 &	924180005181 &	100.0 \\
			34 &	2x5 &	0.19 &	1 &	41 &	3793233420583 & 	2576467 &	3793261759791 &	100.0 \\
			35 &	2x5 &	0.14 &	3 &	35 &	14103229301517 & 	4848239 &	14103258269249 &	100.0 \\
			36 &	2x5 &	Unproven & - & - & - & - & - & - \\
			37 &	2x5 &	0.07 &	3 &	14 &	15715206 & 	4099 &	15754273&	99.8 \\ 
			38 &	2x5 &	0.11 &	1 &	13 &	16204576 & 	3685 &	16268767&	99.6 \\
			39 &	2x5 &	0.15 &	3 &	18 &	148108638 & 	11120 &	148304214&	99.9 \\
			40 &	2x5 &	0.16 &	1 &	29 &	417310640627872 & 	36543045 &	417310842648366&	100.0 \\
			41 & 	2x5 & 	0.09 &	1 &	18 & 	{17 digits}  &	114668733 & 	{17 digits} & 	100.0 \\
			42 &	2x5 &	0.06 &	1 & 19 & 	{18 digits} &   398005342 & 	{18 digits} & 	100.0 \\
			43 & 	2x5 & 	0.67 & 	1 & 20 & 	{18 digits} & 	620906587 & 	{18 digits} & 	100.0 \\
			44 & 	2x5 & 	0.1  & 	1 & 20 & 	{18 digits} & 	1194050967 & 	{18 digits} & 	100.0 \\
			45 & 	2x5 & 	0.08 & 	1 & 20 & 	{18 digits} & 	1194050967 & 	{18 digits} & 	100.0 \\
			46 & 	2x5 & 	0.08 & 	1 & 26 & 	{23 digits} & 	172312766455 & 	{23 digits} & 	100.0 \\
			47 & 	2x5 & 	0.37 & 	1 & 189 & 	{63 digits} & 	{32 digits}  & 	{63  digits} &	100.0 \\
			48 & 	2x5 &	0.69 &	1 & 595 & 	{212 digits} &	{107 digits} &	{212 digits} & 	100.0 \\
			49 & 	2x5 &	0.7  &	1 & 595 & 	{212 digits} &	{107 digits} &	{212 digits} & 	100.0 \\
			50 & 	2x5 &	2.33 &	1 & 1995 & 	{705 digits} &	{353 digits} &	{705 digits} & 	100.0 \\
			51 &	6x2 &	0.14 &	2 &	18 &	13121030712 & 	136612 &	13122572797 &100.0 \\
			52 & 	6x2 &	0.07 &	1 & 18 & 	{17 digits} &	95524079 & 	{17  digits} & 	100.0 \\
			53 &	6x2 &	0.66 &	4 &	31 &	95547148585246 & 	17485734 &	95547257425490&	100.0 \\
			54 & 	6x2 & 	0.33 & 	2 & 31 & 	{16 digits} & 	36109969 & 	{16 digits} & 	100.0 \\
			55 &	6x2 &	0.15 &	1 &	15 &	758649893236585 & 	36109970 &	758650111948072&	100.0 \\
			56 & 	6x2 & 	2.51 & 	3 & 116 & 	{21 digits} & 	5234513991 & 	{21 digits} & 	100.0 \\
			57 & 	6x2 & 	0.33 & 	1 & 23 & 	{22 digits} & 	33299939444 & 	{22 digits} & 	100.0 \\
			58 & 	6x2 & 	0.37 & 	1 & 29 & 	{27 digits} & 	11974457230330 & 	{27 digits} & 	100.0 \\
			59 & 	6x2 & 	0.64 & 	2 & 164 & 	{29 digits} & 	40740206640846 & 	{29 digits} & 	100.0 \\
			60 & 	6x2 &	1.69 &	1 &	45 & 	{39 digits} &	{20 digits} &	{39 digits} & 	100.0 \\
			61 & 	6x2 &	3.69 &	3 & 209 & 	{43 digits} &	{22 digits} &	{43 digits} & 	100.0 \\
			62 & 	6x2 &	4.15 &	3 & 264 & 	{96 digits} &	{49 digits} &	{96 digits} & 	100.0 \\
			63 & 	6x2 &	3.04 &	4 & 544 & 	{96 digits} &	{49 digits} &	{96 digits} & 	100.0 \\
			64 & 	6x2 &	7.46 &	3 & 383 & 	{99 digits} &	{50 digits} &	{99 digits} & 	100.0 \\
			65 & 	6x2 &	2.11 &	3 &	400 & 	{121 digits} &	{61 digits} &	{121 digits} & 	100.0 \\
			66 & 	6x2 &	13.01 &	5 & 424 & 	{101 digits} &	{51 digits} &	{101 digits} & 	100.0 \\
			67 & 	6x2 &	6.29  &	4 & 44 & 	{927 digits} &	{464 digits} &	{927 digits} & 	100.0 \\
			68 & 	6x2 &	22.68 &	3 & 9812 & 	{1731 digits} &	{866 digits} &	{1731 digits} & 	100.0 \\
			69 & 	6x2 &	25.65 &	3 & 2684 & 	{1764 digits} &	{882 digits} &	{1764 digits} & 	100.0 \\
			70 & 	6x2 &	268.15 & 5 & 16211 & 	{2880 digits} &	{1441 digits} &	{2880 digits} & 	100.0 \\
			71 & 	6x2 &	0.32 &	2 & 9 & 	{21133 digits} &	{10567 digits} &	{21133 digits} & 	100.0 \\
			72 & 	6x2 &	3.14 &	5 & 12 & 	{36536 digits} &	{18268 digits} &	{36536 digits} & 	100.0 \\
		\end{tabular}
	}
	\caption{Detailed results (1)}
	\label{table:detail1}
\end{table}

\begin{table}
	{\footnotesize
		\begin{tabular}{lllllllll}
			\textbf{No.} & \textbf{Dim.} & \textbf{Time} & \textbf{Otters} & \textbf{Ocelots} & \textbf{Ocelot Steps} & \textbf{Productivity} & \textbf{Activity} & \textbf{Predicted} \\
			73 &	3x4 &	6157.42  &  4  & 49001  &   123590123      & 17323         & 262759288 &	47.0 \\
			74 & 	3x4 &	0.51     &	3  &  174   & 	{54 digits}    & {27 digits}   &	{54 digits} & 	100.0 \\
			75 & 	3x4 &	2.71     &	1  &  808   & 	{282 digits}   & {142 digits}  &	{282 digits} & 	100.0 \\
			76 & 	3x4 &	66.57    &	3  &  1141  &   {870 digits}   & {435 digits}  &	{870 digits} & 	100.0 \\
			77 & 	3x4 &	2.21     &	13 & 43     & 	{1257 digits}  & {629 digits}  &	{1257 digits} & 	100.0 \\
			78 & 	3x4 &	0.27     &	2  & 6      & 	{2603 digits}  & {1302 digits} &	{2603 digits} & 	100.0 \\
			79 & 	3x4 &	42.99    &	4  & 71774  &   {4712 digits}  & {2356 digits} &	{4712 digits} & 	100.0 \\
			80 & 	3x4 &	365.58   &	9  & 13485  & 	{4746 digits}  & {2373 digits} &	{4746 digits} & 	100.0 \\
			81 & 	3x4 &	0.29     &	2  & 11     & 	{13038 digits} & {6520 digits} &	{13038 digits} & 	100.0 \\
			82 &	4x3 &	0.22     &	7  & 88     &   249851670      & 15008 &	250096775&	99.9 \\
			83 & 	4x3 &	9.29     &	2  & 11029  & 	{1427 digits}  & {714 digits} &	{1427 digits} & 	100.0 \\
			84 & 	4x3 &	151.52   &	14 & 5911   & 	{1620 digits}  & {810 digits} &	{1620 digits} & 	100.0 \\
			85 & 	4x3 &	47.91    &	1  & 7889   & 	{1975 digits}  & {988 digits} &	{1975 digits} & 	100.0 \\
			86 & 	4x3 &	558.91   &	4  & 24416  & 	{7722 digits}  & {3861 digits} &	{7722 digits} & 	100.0 \\
			87 & 	4x3 &	170.18   &	2  & 39990  & 	{9123 digits}  & {4562 digits} &	{9123 digits} & 	100.0 \\
			88 & 	4x3 &	0.8      &  2  & 12     & 	{9865 digits}  & {4933 digits} &	{9865 digits} & 	100.0 \\
			89 & 	4x3 &	215.17   &	4  & 34264  & 	{12070 digits} & {6035 digits} &	{12070 digits} & 	100.0 \\
			90 & 	4x3 &	1280.84  &	6  & 39597  & 	{14073 digits} & {7037 digits} &	{14073 digits} & 	100.0 \\
			91 & 	4x3 &	1300.21  &	6  & 39597  & 	{14073 digits} & {7037 digits} &	{14073 digits} & 	100.0 \\
			92 &	2x6 &	0.15     &	2  &	13  &	94743346       & 10574 &	94842383&	99.9 \\
			93 &	2x6 &	0.13     &	2  &	14  &	98246088       & 10249 &	98364599&	99.9 \\
			94 &	2x6 &	0.98     &	2  &	17  &	489127502      & 	15828 &	493600387&	99.9 \\
			95 & 	2x6 &	0.59     &	5  &	155 & 	{55 digits}    &	{28 digits} &	{55 digits} & 	100.0 \\
			96 & 	2x6 &	12.96    &	3  & 19     & 	{1645 digits}  &	{823 digits} &	{1645 digits} & 	100.0 \\
			97 & 	2x6 &	131.73   &  3  & 16388  & 	{9864 digits}  &	{4933 digits} &	{9864 digits} & 	100.0 \\
			98 & 	2x6 &	0.51     &	2  & 13     & 	{9867 digits}  &	{4934 digits} &	{9867 digits} & 	100.0 \\
			99 &	7x2 &	0.16     &	3  & 7      & 	{90 digits}	   & {46 digits} &	{90 digits} & 100.0 \\
			101 &	5x2 &	0.05     &	2 &	11      &	11744416       & 	4097 &	11821234 &	99.4 \\
			102 &	5x2 &	0.06     &	2 &	11      &	11744368       & 	4097 &	11821220 &	99.4 \\
			103 &	5x2 &	0.06     &	2 &	11      &	11759266       & 	4096 &	11821190 &	99.5 \\
			104 &	5x2 &	0.05     &	2 &	11      &	11753274       & 	4096 &	11815076 &	99.5 \\
			105 &	5x2 &	0.05     &	2 &	11      &	11744416       & 	4097 &	11811040 &	99.5 \\
			106 &	5x2 &	0.06     &	3 &	12      &	11774496       & 	4097 &	11811040 &	99.7 \\
			107 &	5x2 &	0.06     &	2 &	11      &	11744368       & 	4097 &	11811026 &	99.5 \\
			108 &	5x2 &	0.06     &	2 &	11      &	11759266       & 	4096 &	11811010 &	99.6 \\
			109 &	5x2 &	0.04     &	2 &	11      &	11768486       & 	4097 &	11804940 &	99.7 \\
			110 &	5x2 &	0.04     &	2 &	11      &	11768446       & 	4097 &	11804926 &	99.7 \\
			111 &	5x2 &	0.05     &	1 &	11      &	11761838       & 	4097 &	11792724 &	99.7 \\
			112 &	5x2 &	0.05     &	1 &	11      &	11761838       & 	4097 &	11792696 &	99.7 \\
			113 &	5x2 &	0.03     &	1 &	11      &	11761838       & 	4097 &	11792682 &	99.7 \\
			114 &	5x2 &	2.92     &	11 & 32     &	2323796        & 	1471 &	2358064 &	99.0 \\
			115 &	5x2 &	3.02     &	11 & 32     &	2323796        & 	1471 &	2358064 &	99.0 \\
			116 & 	3x3 & 	0.23     & 	3 & 149     & 	{18 digits}    & 	374676383 & 	{18 digits} & 	100.0 \\
			117 &	3x3 &	0.57     &	1 &	11      &	15686426       & 	4096 &	15725629 &	99.8 \\
			118 &	3x3 &	0.06     &	1 &	11      &	15682242       & 	4098 &	15725661 &	99.7 \\
			119 &	3x3 &	0.06     &	1 &	11      &	15682242       & 	4098 &	15725659 &	99.7 \\
			120 &	3x3 &	47.42    &	1 &	21      &	452195755771728 & 	21264944 &	452196003014837 &	100.0 \\
			121 &	3x3 &	0.05     &	1 &	14      &	4144451519458  & 	2950149 &	4144465135614 &	100.0 \\
			122 &	3x3 &	10.12    &	1 &	10      &	51408293       & 	7205 &	51525774 &	99.8 \\
			123 &	3x3 &	0.95     &	1 &	16      &	1808265960     & 	43925 &	1808669046 &	100.0 \\
			124 &	3x3 &	0.93     &	1 &	16      &	1808265960     & 	43925 &	1808669046 &	100.0 \\
			125 &	3x3 &	0.66     &	1 &	16      &	1093020548     & 	34151 &	1093389035 &	100.0 \\
			126 &	3x3 &	0.16     &	2 &	12      &	47095060       & 	12290 &	47287015&	99.6 \\
			127-128 & 2x5 & Unproven & & & & & & \\						
			129     & 2x5 & 276.32	 & 3  & 563	    & {98 digits}	   & {49 digits}	& {98 digits} & 100.0 \\	
			130-137 & 2x5 & Unproven & - & - & - & - & - & - \\					
			138-146 & 2x6 & Overflow & - & - & - & - & - & - \\							
			147		& 2x6 & Unproven & - & - & - & - & - & - \\														
			148-151 & 2x6 & Overflow & - & - & - & - & - & - \\							
			152-153 & 2x6 & Unproven & - & - & - & - & - & - \\							
			154-157 & 2x6 & Overflow & - & - & - & - & - & - \\								
			158-164 & 3x4 & Overflow & - & - & - & - & - & - \\							
			165     & 3x4 &	69.5     &	4 &	11007      &	286557138 & 18046 &	347241183 &	82.5 \\
			166-172 & 3x4 & Overflow & - & - & - & - & - & - \\								
			173		& 3x4 &	Unproven & - & - & - & - & - & - \\							 						
			174-175 & 3x4 & Overflow & - & - & - & - & - & - \\							
			176     & 3x4 &	0.84     &	2 &	211      & 61568  & 286 & 83125  & 74.2 \\
			177     & 3x4 &	1.31     &	2 &	320      & 139879 & 427 & 187938 & 74.5 \\				
		\end{tabular}
	}
	\caption{Detailed results (2)}
	\label{table:detail2}
\end{table}

\begin{table}
	{\footnotesize
		\begin{tabular}{lllllllll}
			\textbf{No.} & \textbf{Dim.} & \textbf{Time} & \textbf{Otters} & \textbf{Ocelots} & \textbf{Ocelot Steps} & \textbf{Productivity} & \textbf{Activity} & \textbf{Predicted} \\
			178-181 & 4x3 & Overflow & - & - & - & - & - & - \\						
			182-184 & 4x3 & Unproven & - & - & - & - & - & - \\							
			185-190 & 4x3 & Overflow & - & - & - & - & - & - \\							
			191 & 	4x3 & 	7.67 & 	5 & 9 & {1236 digits} & 	{618 digits} & 	{1236 digits}	& 100.0 \\
			192-199 & 4x3 & Overflow & - & - & - & - & - & - \\							
			200-204 & 6x2 & Overflow & - & - & - & - & - & - \\						
			205-209 & 6x2 & Unproven & - & - & - & - & - & - \\						
			210	 	& 6x2 & Overflow & - & - & - & - & - & - \\												
			211   	& 6x2 & Unproven & - & - & - & - & - & - \\												
			212	    & 6x2 & Overflow & - & - & - & - & - & - \\													
			213-216 & 6x2 & Unproven & - & - & - & - & - & - \\	
			217-219 & 6x2 & Overflow & - & - & - & - & - & - \\	
			220-223 & 7x2 & Overflow & - & - & - & - & - & - \\							
			224	    & 7x2 & Unproven & - & - & - & - & - & - \\												
			225     & 7x2 & Overflow & - & - & - & - & - & - \\		
			226     & 7x2 & Unproven & - & - & - & - & - & - \\	
			227-239 & 7x2 & Overflow & - & - & - & - & - & - \\						
			240     & 3x5 & Unproven & - & - & - & - & - & - \\							
		\end{tabular}
	}
	\caption{Detailed results (3)}
	\label{table:detail3}
\end{table}

For the 130 termites,, all verified ocelot transitions arose from monoprogressive otter sequences, and 124 out of the 130 had verified ocelot transitions from only monoregressive ones. The remaining six (Dragons 17, 70, 77, 82, 93 and 116) had one ocelot transition each that arose from a sequence with two regressive variables. Two of the unproven cases (Dragons 131 and 134) also had multiregressive sequences, with the former discovering a pattern with three progressive and two regressive variables, but this was not verified. This suggests that there may be a slight mismatch between the identification of a potential pattern and the verification process, but we would rather a slightly over-eager pattern detection that an under-eager one. So it seems the vast majority of shelducks we have discovered so far involve only monoprogressive and monoregressive patterns.

Nested otters were more common, with 20 machines using at least one ocelot transition in a shape computation. While some of these were for unilinear shelducks, all 13 multilinear shape computations included at least one ocelot transition. It is also noticeable that all multilinear shelducks were of quadratic length, and that all of the 325 unilinear ones were either affine (280) or constant (45). Of the 280 of affine length, 10 were affine in the regressive variable (i.e.\ $a = 0$ in Case 2 of Theorem~\ref{theorem:hops}), and 270 were affine in the progressive variable (i.e.\ $b = 0$ in Case 2 of Theorem~\ref{theorem:hops}). 

To illustrate these concepts, consider the otters below found in Dragon 67. 

First we note the linear observant otter sequence below. 

\begin{center}
	\begin{tabular}{ll}
		\textbf{Hops} & \textbf{Configuration} \\
		61 & $\{a\}0\{r\}01^{9}01$ \\
		64 & $1\{a\}0\{r\}01^{8}01$ \\
		67 & $1^{2}\{a\}0\{r\}01^{7}01$ \\
	\end{tabular}
\end{center}

From this we find the otter pattern below. 

\begin{center}
	$1^{X}\{a\}0\{r\}01^{Y}01 \longmapsto 1^{X+1}\{a\}0\{r\}01^{Y-1}01$
\end{center} 

The shape computation corresponding to this pattern is then as follows. 

\begin{center}	
	\begin{tabular}{ll}
		\textbf{Hops} & \textbf{Configuration} \\
		$0$ & $1^{X}\{a\}0\{r\}01^{Y}01$ \\ % Hops: 0 Ones: 1 Pos: 0 Transition: init 
		$1$ & $1^{X+1}\{b\}0\{l\}1^{Y}01$ \\ % Hops: 1 Ones: 1 Pos: 1 Transition: complex(normal) 
		$2$ & $1^{X+1}0\{c\}1^{Y}\{l\}01$ \\ % Hops: 2 Ones: 1 Pos: 2 Transition: complex(normal) 
		$3$ & $1^{X+1}\{a\}0\{r\}01^{Y-1}01$ \\ % Hops: 3 Ones: 1 Pos: 1 Transition: complex(normal) 
	\end{tabular}
\end{center}

We thus have the Otter 1 transition is $1^{X}\{a\}0\{r\}01^{Y} \leadsto 1^{X+Y}\{a\}0\{r\}0$ in $3Y$ steps where $X \geq 0, Y \geq 1$. Next we observe the otter sequence below and the corresponding otter pattern. 

\begin{center}
	\begin{tabular}{ll}
		\textbf{Hops} & \textbf{Configuration} \\
		93  & $1^{10}\{e\}0\{r\}1^{2}$ \\
		135 & $1^{9}\{e\}0\{r\}1^{4}$ \\
		173 & $1^{8}\{e\}0\{r\}1^{6}$ \\
	\end{tabular}
\end{center}

\begin{center}
	$1^{Y}\{e\}0\{r\}1^{X} \longmapsto 1^{Y-1}\{e\}0\{r\}1^{X+2}$
\end{center} 

The shape computation corresponding to this pattern is then as follows. 

\begin{center}
	\begin{tabular}{ll}
		\textbf{Hops} & \textbf{Configuration} \\
		$0$    & $1^{Y}\{e\}0\{r\}1^{X}$ \\ % Hops: 0 Ones: 0 Pos: 0 Transition: init 
		$1$    & $\{a\}1^{Y}\{r\}1^{X+1}$ \\ % Hops: 1 Ones: 0 Pos: -1 Transition: complex(normal) 
		$2$    & $\{b\}1^{Y-1}\{r\}01^{X+1}$ \\ % Hops: 2 Ones: 0 Pos: -2 Transition: complex(normal) 
		$Y+1$  & $\{b\}0\{r\}1^{Y-1}01^{X+1}$ \\ % Hops: Y+1 Ones: 0 Pos: -Y1-1 Transition: expansion(1) 
		$Y+2$  & $\{c\}1^{Y-1}\{l\}01^{X+1}$ \\ % Hops: Y+2 Ones: 0 Pos: -Y1 Transition: complex(normal) 
		$Y+3$  & $\{a\}0\{r\}01^{Y-2}01^{X+1}$ \\ % Hops: Y+3 Ones: 0 Pos: -Y1-1 Transition: complex(normal) 
		$4Y-3$ & $1^{Y-2}\{a\}0\{r\}001^{X+1}$ \\ % Hops: 4*Y-3 Ones: 0 Pos: -3 Transition: ocelot(1) 
		$4Y-2$ & $1^{Y-1}\{b\}0^{2}\{l\}1^{X+1}$ \\ % Hops: 4*Y-2 Ones: 0 Pos: -2 Transition: complex(normal) 
		$4Y-1$ & $1^{Y-1}0\{c\}0\{l\}1^{X+1}$ \\ % Hops: 4*Y-1 Ones: 0 Pos: -1 Transition: complex(normal) 
		$4Y$   & $1^{Y-1}01\{d\}1^{X+1}\{l\}$ \\ % Hops: 4*Y Ones: 1 Pos: 0 Transition: complex(normal) 
		$4Y+1$ & $1^{Y-1}0\{f\}1\{r\}1^{X+1}$ \\ % Hops: 4*Y+1 Ones: 1 Pos: -1 Transition: complex(normal) 
		$4Y+2$ & $1^{Y-1}\{e\}0\{r\}1^{X+2}$ \\ % Hops: 4*Y+2 Ones: 0 Pos: -2 Transition: complex(normal) 
	\end{tabular}
\end{center}

Note that the length of this shape computation is $4Y+2$, so it is affine in the regressive variable. 

Hence the Otter 2 transition is is $1^{Y}\{e\}0\{r\}1^{X} \leadsto 1\{e\}0\{r\}1^{X+2Y-2}$ in $(Y-1)(2Y+6)$ steps where $X \geq 0, Y \geq 1$. Later we observe the sequence below and the corresponding otter pattern. 

Finally we observe the multilinear sequence and corresponding pattern below. 

\begin{center}
	\begin{tabular}{ll}
		\textbf{Hops} & \textbf{Configuration} \\
		$440$  & $1\{b\}1^{3} \{l\}(011)^{7}0101$ \\ 
		$508$  & $1\{b\}1^{12}\{l\}(011)^{6}0101$ \\ 
		$945$  & $1\{b\}1^{30}\{l\}(011)^{5}0101$ \\
		$3092$ & $1\{b\}1^{66}\{l\}(011)^{4}0101$ \\
	\end{tabular}
\end{center}

$1\{b\}1^{X}\{l\}(011)^{Y}0101 \longmapsto 1\{b\}1^{2X+6}\{l\}(011)^{Y-1}0101$

\begin{center}
	\begin{tabular}{ll}
		\textbf{Hops} & \textbf{Configuration} \\
		$0$           & $1\{b\}(1)^{X}\{l\}(011)^{Y}0101$ \\ % Hops:  0 Ones: 3 Pos: 0 Transition: init 
		$1$           & $\{b\}1\{r\}(1)^{X}(011)^{Y}0101$ \\ % Hops:  1 Ones: 3 Pos: -1 Transition: complex(normal) 
		$2$           & $\{b\}0\{r\}(1)^{X+1}(011)^{Y}0101$ \\ % Hops:  2 Ones: 2 Pos: -2 Transition: complex(acceleration) 
		$3$           & $\{c\}(1)^{X+1}\{l\}(011)^{Y}0101$ \\ % Hops:  3 Ones: 2 Pos: -1 Transition: complex(normal) 
		$4$           & $\{a\}0\{r\}0(1)^{X}(011)^{Y}0101$ \\ % Hops:  4 Ones: 2 Pos: -2 Transition: complex(normal) 
		$3X+4$        & $(1)^{X}\{a\}0\{r\}0(011)^{Y}0101$ ocelot(1) \\ % Hops:  3*X+4 Ones: 2 Pos: X-2 Transition: ocelot(1) 
		$3X+5$        & $(1)^{X+1}\{b\}0\{l\}(011)^{Y}0101$ \\ % Hops:  3*X+5 Ones: 2 Pos: X-1 Transition: complex(normal) 
		$3X+6$        & $(1)^{X+1}0\{c\}(011)^{Y}\{l\}0101$ \\ % Hops:  3*X+6 Ones: 2 Pos: X Transition: complex(normal) 
		$3X+9$        & $(1)^{X+1}\{e\}0\{r\}(1)^{3}(011)^{Y-1}0101$ \\ % Hops:  3*X+9 Ones: 5 Pos: X-1 Transition: complex(normal) 
		$11X+2X^2+9$  & $1\{e\}0\{r\}(1)^{2X+3}(011)^{Y-1}0101$ ocelot(2) \\ % Hops:  11*X+2*(X*X)+9 Ones: 3 Pos: -X-1 Transition: ocelot(2) 
		$11X+2X^2+10$ & $\{a\}1\{r\}(1)^{2X+4}(011)^{Y-1}0101$ \\ % Hops:  11*X+2*(X*X)+10 Ones: 3 Pos: -X-2 Transition: complex(normal) 
		$11X+2X^2+11$ & $\{b\}0\{r\}0(1)^{2X+4}(011)^{Y-1}0101$ \\ % Hops:  11*X+2*(X*X)+11 Ones: 2 Pos: -X-3 Transition: complex(normal) 
		$11X+2X^2+12$ & $\{c\}0\{l\}(1)^{2X+4}(011)^{Y-1}0101$ \\ % Hops:  11*X+2*(X*X)+12 Ones: 2 Pos: -X-2 Transition: complex(normal) 
		$11X+2X^2+13$ & $1\{d\}(1)^{2X+4}\{l\}(011)^{Y-1}0101$ \\ % Hops:  11*X+2*(X*X)+13 Ones: 3 Pos: -X-1 Transition: complex(normal) 
		$11X+2X^2+14$ & $\{f\}1\{r\}(1)^{2X+4}(011)^{Y-1}0101$ \\ % Hops:  11*X+2*(X*X)+14 Ones: 3 Pos: -X-2 Transition: complex(normal) 
		$11X+2X^2+15$ & $\{e\}0\{r\}(1)^{2X+5}(011)^{Y-1}0101$ \\ % Hops:  11*X+2*(X*X)+15 Ones: 2 Pos: -X-3 Transition: complex(normal) 
		$11X+2X^2+16$ & $\{a\}0\{r\}(1)^{2X+6}(011)^{Y-1}0101$ \\ % Hops:  11*X+2*(X*X)+16 Ones: 2 Pos: -X-4 Transition: complex(normal) 
		$11X+2X^2+17$ & $1\{b\}(1)^{2X+6}\{l\}(011)^{Y-1}0101$ \\ % Hops:  11*X+2*(X*X)+17 Ones: 3 Pos: -X-3 Transition: complex(normal) 
	\end{tabular}
\end{center}

We note that the final configuration for this transition is $1\{b\}1^{X2^Y + 6 \times 2^Y-6}{l}$, which is reached after \\ 
$((2^{2Y}-1)\times(2X^2+24X+72)+(2^Y-1) \times (-39X-234)+69Y)//3$  hops. Note also the dependence of the third (multilinear) transition on the two previous ones; it seems the order in which the otter patterns are found can have a profound influence, and clearly we wish to maximise the changes of finding multilinear transitions whenever possible. 

There are also several cases in which ocelot transitions are discovered, but used either 0 or 1 time only\footnote{We will refer to these as \textit{spurious spiders}}. This again suggests that more analysis is needed for the discovery and application of shape computations. 

%Flat otters. Constant camels?
%\textbf{\textit{Late otters. Tardy tarantulas?}}

We only found two machines amongst the 240 tested in which we found an issue with the way the modulus status was handled. In particular, only in Dragons 30 and 131 was it necessary to reformulate the modulus. So it seems this is not a major issue, at least for the cases we have tried. 

Naturally the 31 unproven cases above require further analysis, and we would of course like to reduce this number to 0 if possible. For some cases it is a matter of finding better methods for verifying otter patterns once found; for others (such as Dragons 5 and 73), it seems there configuration are generally less compact and predictable than many others, which may require some different techniques. Clearly further work is needed for these cases. 

A key result of interest is that for Dragon 72. This is still the machine of largest productivity and activity that we have shown to terminate. If we only use unilinear otter sequences, this finds four ocelot transitions and takes just under 1600 seconds to execute with around 30,000 applications of these transitions, resulting in a final configuration of size (18,268 digits) after a total of (36,535 digits) hops. However, allowing multilinear otter sequences generates these transitions plus a fifth one based on a multilinear sequence. Then the same productivity and activity is found in just over 3 seconds with a total of 12 applications of the ocelot transitions, including two with the multilinear transition resulting in a combined total of (36536 digits) hops (!!). This shows how the multilinear case can drastically improve performance. Similarly comments apply also to Dragons 71 and 88, 96 and 98, albeit with less spectacular success. 

The five ocelot transitions for Dragon 72 are given in Table~\ref{table:dragon72} and the applications in Table~\ref{table:dragon72a}. Note particularly the number of applications of each transition report in Table~\ref{table:dragon72a} and the resulting number of steps. 

\begin{table}
	{\footnotesize
		\begin{tabular}{llllll}
			& \textbf{Type} & \textbf{Initial} & \textbf{Final} &\textbf{ Constraints} & \textbf{Steps} \\
			1 & uni   & $(10)^{Y}(01)^{X}100\{b\}010\{l\}1(blank)$ & $(01)^{X+4Y}100\{b\}010\{l\}1$      & $X \geq 0, Y \geq 1$ & $Y(8X+16Y+46)$\\
			2 & uni   & $(blank)\{c\}(10)^{X}\{r\}01^{Y}$ & ${c}(10)^{X+Y-Y mod 2}\{r\}01^{Y mod 2}$     & $X \geq 0, Y \geq 2$ & $(Y//2)(4X+2(Y-Y mod 2)+2)$ \\
			3 & uni   & $(01)^{Y}0(01)^{X}1\{c\}01\{r\}0101(blank)$ & $0(01)^{X+4Y}1\{c\}01\{r\}0101$    & $X \geq 0, Y \geq 1$ & $Y(8X+16Y+46)$ \\
			4 & multi & $\{d\}(110)^{Y}\{r\}(10)^{X}1^{9}(blank)$ & $\{d\}(110)^{Y mod 3}\{r\}(10)^{Z_1}1^{9}$ & $X \geq 0, Y \geq 3$ & $Z_2$  \\
			5 & uni   & $(blank)1(01)^{X}0\{d\}1^{Y}\{l\}$ & $1(01)^{X+Y-Y mod 2}0\{d\}1^{Y mod 2}\{l\}$ & $X \geq 0, Y \geq 2$ & $(Y//2)(4X+2(Y-Y mod 2)+6)$ \\
		\end{tabular}
	}
	
	where $Z_1 = X \times 4^{Y//3}+(34\times 4^{Y//3}-34)//3$ \\
	and $Z_2 = ((4^{(2(Y//3)}-1)\times(432X^2+9792X+55488)+(4^{Y//3}-1)(-5760X+ -65280) -225(Y//3))//405$
	
	\caption{Dragon 72 ocelot transitions}
	\label{table:dragon72}
\end{table}

\begin{table}
	\centering
	\begin{tabular}{cccl}
		& \textbf{Type}  & \textbf{Applications} & \textbf{Steps Applied} \\
		1 & uni   & 3 & 1324 \\
		2 & uni   & 1 & 392 \\
		3 & uni   & 4 & 18642976712\\
		4 & multi & 2 & {36536 digits} \\
		5 & uni   & 2 & {18270 digits} \\
	\end{tabular}
	
	\caption{Dragon 72 ocelot transition applications}
	\label{table:dragon72a}
\end{table}

\section{Discussion} \label{section:discussion}

A clear limitation on our results is the memory size, and specifically the size of integers that can be manipulated. It is notable that of the 79 clattering clancies, 78 of them occur at an application of a multilinear ocelot transition, which implicitly shows that multilinear transitions have potential to be significantly more successful if larger integers can somehow be accommodated. The only exception to this observation involves comparing the expression below to 0; in the current implementation, we simply evaluate the expression and compare the resulting integer with 0. Clearly less simplistic methods will be required in future. 

$$2^{118842243771396506390315925502} + 5 \times 2^{118842243771396506390315925502} - 5$$

Note that writing $c = 2^{118842243771396506390315925502}$ means that the above expression can be rewritten as $6c -5$ and so it will follow that this number is positive if $c > \frac{5}{6}$, which is presumably can be determined without explicit evaluation of this very large integer. 

It is possible to increase space allocated to integers by SWI-Prolog, potentially by switching to Linux rather than Windows. This will increase our available integer capacity to something in the order of $2^{2^{64}}$ (and possibly further still using cloud computing rather than a desktop PC). However, there is no clear ``safe'' upper limit (or at least not clear to us now), and an explicit statement of a number thousands of digits long is both unwieldy and unreadable, and arguably not necessary for our purposes. An analogy can be made with formulae such as $\frac{\pi^2}{6}$ in that one can either be exact or explicit but not both, but an implicit and exact form may be more useful than an explicit one.  In particular, it seems that some kind of lazy evaluation of expressions can be developed in which numbers which exceed a certain bound are only evaluated in such lazy fashion, rather than explicitly calculated. We have made some initial steps in this direction, but it soon became clear that this would need not only some significant refactoring of the existing code, but also some further investigation of symbolic manipulation. 

As noted above, it seems intuitively obvious that the above expression is not zero; however, writing general methods for doing so is less obvious. For example, it seems relatively simple to argue that $2^{2^{60}} > 2^{2^{59}}$ by observing that $60 > 59$; however, it is more challenging for expressions such as the one below. 

$$(200 \times (2^{8589934592}-1)- 510 \times (2^{4294967296}-1)+734439407616)//3+307445734261178167080$$ 

We can simplify this to some extent, by first writing $2^{4294967296}$ as $c$ and then proceeding as follows. 

\begin{center}
	$(200(c^2-1)- 510(c-1)+734439407616)//3 + 307445734261178167080$ \\
	$ = (200c^2 - 200 - 510c + 510 + 734439407616)//3 + 307445734261178167080$ \\
	$ = (200c^2 - 510c + 734439407926)//3 + 307445734261178167080$ \\
	$ = (200c^2 + 734439407926)//3 -170c + 307445734261178167080$ \\
\end{center}

While this is more unwieldy then dealing with explicit integers, it seems that some kind of implicit evaluation and manipulation like this will be needed, given the number of clancies. 

A potential connection is to something like Knuth's up-arrow notation \cite{Knuth76}, and in particular to the relatively small ``two arrow'' case, in which $x \uparrow n$ is shorthand for $x^n$ and $x\uparrow \uparrow n$ is shorthand for $x\uparrow (x\uparrow (\ldots x\uparrow \ldots ))$ where there are $n$ occurrences of $x$. So for example $2\uparrow \uparrow 4$ is $2\uparrow (2 \uparrow (2\uparrow 2)) = 2\uparrow (2^4) = 2^{16} = 65,536$, and so $2 \uparrow \uparrow 5 = 2\uparrow (2\uparrow 4) = 2^{65,536}$.  Note that in the above example, $2^{32} = 4294967296$ and $2^{33} = 8589934592$, and so $c = 2^{2^{32}}$ and $c^2 = 2^{2^{33}}$. 

A further issue of note is that the order in which ocelot transitions are applied can be significant. 
In general, we use any applicable ocelot transitions as early as possible, and in the order in which they are generated. However, we have found some cases in which a little tweaking of this order can make a significant difference to the efficiency of the computation. For example, Dragon 191 will overflow if we apply the ocelot transitions found without restriction. However if we impose a requirement that ocelot transitions can only be used after 150 hops, then it executes to completion in just under 8 seconds. This particular property was found by hand inspection of the execution trace, and shows that further work on the order of the application of otters is required, as it may be that similar reordering or delayed application may provide significantly better performance. Clearly it is of central importance to find a systematic way in which to apply transitions found to the maximum possible effect. 

A related issue is the application of ocelot transitions within shape computations. Using the default order above (i.e.\ applying otters in the order in which they were generated) in some cases means some further shape computations are unable to be found. For example, in the attempted verification of otter 3 in Dragon 74, applying otter 2 in the shape computation means that the target configuration is ``overshot'', i.e.\ we proceed to a further stage without explicitly checking whether the target configuration has been reached. The relevant otter pattern and shape computation are below. 

\noindent
~\\
Otter 3 pattern: $\{a\}2^{Y}\{r\}1^{X} \longmapsto \{a\}2^{Y-3}\{r\}1^{X+5}$ \\ ~\\
Otter 1: $1^{X}\{c\}2\{l\}1^{Y} \leadsto 1^{X+Y}\{c\}2\{l\}$ \\
Otter 2: $2^{Y}1^{X}\{b\}0\{l\}(blank) \leadsto 2^{Y mod 3}1^{X+5(Y//3)}\{b\}0\{l\}$ \\ 

\begin{center}
	\begin{tabular}{ll}
		\textbf{Hops} & \textbf{Configuration} \\
		$0$     & $\{a\}2^{Y}\{r\}1^{X}$ \\ % Hops: 0 Ones: 0 Pos: 0 Transition: init 
		$1$     & $\{b\}2^{Y-1}\{r\}1^{X+1}$ \\ % Hops: 1 Ones: 0 Pos: -1 Transition: complex(normal) 
		$2$     & $\{c\}2^{Y-2}\{r\}21^{X+1}$ \\ % Hops: 2 Ones: 1 Pos: -2 Transition: complex(normal) 
		$3$     & $2^{Y-3}1\{c\}2\{l\}1^{X+1}$ \\ % Hops: 3 Ones: 2 Pos: -1 Transition: complex(normal) 
		$5X+8$  & $2^{Y-3}1^{X+2}\{c\}2\{l\}$ ocelot(1)  \\ % Hops: 5*X+8 Ones: 1 Pos: X Transition: ocelot(1) 
		$5X+9$  & $2^{Y-3}1^{X+3}\{c\}0\{l\}$ \\ % Hops: 5*X+9 Ones: 0 Pos: X+1 Transition: complex(acceleration) 
		$5X+10$ & $2^{Y-3}1^{X+3}3\{b\}0\{l\}$ \\ % Hops: 5*X+10 Ones: 1 Pos: X+2 Transition: complex(normal) 
		$5X+11$ & $2^{Y-3}1^{X+3}\{a\}3\{r\}$ \\ % Hops: 5*X+11 Ones: 1 Pos: X+1 Transition: complex(normal) 
		$5X+12$ & $2^{Y-3}1^{X+4}\{a\}0\{l\}$ \\ % Hops: 5*X+12 Ones: 0 Pos: X+2 Transition: complex(normal) 
		$5X+13$ & $2^{Y-3}1^{X+5}\{b\}0\{l\}$ \\ % Hops: 5*X+13 Ones: 0 Pos: X+3 Transition: complex(normal) 
		$5X+14$ & $2^{Y-3}\{a\}1^{X+5}\{r\}$ $\Leftarrow$ \textbf{NOT ocelot(2)!!} \\
		$6X+19$ & $\{a\}2^{Y-3}\{r\}1^{X+5}$ \\
	\end{tabular}
\end{center}

Note that if we apply $ocelot(2)$ at hop $5X+13$, we get the configuration $2^{(Y-3) mod 3} 1^{X+5+5(Y-3)//3} \{b\}0\{l\}$, which overshoots the target configuration of $\{a\}2^{Y-3}\{r\}1^{X+5}$ at hop $6X+19$.  Similar cases arise for Dragons 129, 154, 191, 209 and 237. In the above implementation we have provided manual exceptions for the relevant cases (i.e.\ in Dragon $M$, avoid using otter $K$ in the proof of otter $N$). However, this is hardly a satisfactory general solution, which would require more study of shape computations and how they can be found more effectively with more systematic methods. 

As mentioned in Section~\ref{sec:managing}, simplifying algebraic expressions is a significant part of execution, especially for the hops calculation, which of course is incremented by some expression at every step of a shape computation. As above, we make judicious use of Equation~\ref{equation:mod} and constraints on the input variables to reduce expressions such as $4X(X//2)$ to $2X(X - X \; mod \; 2)$ and, assuming $X \; mod \; 2 = 1$, to $2X(X-1)$. This may involve keeping unsimplified forms of terms such as $X(X//2)$ and $3X(X//2)$ before simplifying $X(X//2) + 3X(X//2)$ to $4X(X//2)$ and then as above. In cases such as Dragon 77 we find cases such as this that end up in ``neat'' quadratic formulae for the hops, such as $14X^2 + 36X + 17$. However, in others, such as in Dragons 221, 222, 237 and 238 we have found shape computations for which the hops calculation is as below. In the current implementation, we are unable to generate the corresponding ocelot transition due to the presence of $//$ in the final hops value, despite our successful search, which is frustrating. Clearly some further simplification techniques are needed, as well as possibly a means of dealing with terms including $//$ in the calculation of the hops for the ocelot transition from the hops in the shape computation. 

$$78X^2 + 39X(X+1)//2 + 3872X + 38188$$
$$2X^2 - X +11(X-1)//3 + 21$$
$$18X^2 + 126X + 9((X+1)//2)(X+1)//2 + 9(X//2)(X//2) + 25(X+1)//2 + 91X//2 + 390$$ 
$$18X^2 + 47X + 9((X+1)//2)(X+1)//2 + 9(X//2)(X//2) + 19(X//2) + 109(X+1)//2 + 429$$ 

It would seem using some computer algebra systems such as Mathematica \cite{mathematica} or Maple \cite{maple} may be an appropriate direction to follow for manipulations like these. 

\section{Conclusions and Further Work}

We have seen how shape computations can be used to express otter patterns found in computations, and which can then be used to infer ocelot transitions, which in turn can be applied to significantly reduce the execution time of machines of high productivity and activity. We can thus think of shape computations as a formal basis for such transitions, which provides verification of their correctness and conditions under which they can be applied. Clearly the science of shape computations is very much in its infancy, and one which will presumably repay further study. 

One clear direction for further work is to examine the unproven cases above. These include some for which otter patterns were found, but without a corresponding successful search for a shape computation. Naturally minimising such failures (possibly to 0) is highly desirable. There are also some cases which do not seem particularly compatible with observant otter patterns, in that they do not seem to be particularly ``compressible'', such as Dragons 5, 36 and 73. While we were able to get Dragons 5 and 73 to terminate, this was very much by sheer persistence and patience rather than any elegant or efficient analysis. These and others will most likely require some other pattern detection and analyses in order to ``unlock'' an efficient evaluation mechanism for them. % Comments about Dragon 37 and counting?

As discussed in Section~\ref{section:discussion}, manipulation and simplification of algebraic expressions is a crucial aspect of the reasoning required for evaluation of machines of high productivity and activity. In particular, techniques for simplifying formulae involving $//$ require some 
further investigation. For example, for an expression such as $(ax+b)//n$ when $a,b > n$, it is not hard to show that $(ax+b)//n = (a//n)x + b//n + ( (a\; mod\; n)x + (b\; mod\; n) )//n$. The trick is to know how to apply this in a simplification process along with some other relevant properties (such as Equation~\ref{equation:mod} above).  We also note that we can potentially exploit identities such as $n//2 + (n+1)//2 = n$ for any $n \geq 0$, and of course to explore similar potential simplifications for divisors other than 2. Some very early exploration of these ideas can be found in Section~\ref{sec:moduli}. 

A further area of study is to examine how computations may be divided into phases. In particular, it seems we need to run the machine naively for a smallish number of steps to determine patterns, from which we may then infer ocelot transitions. After generating a certain number of these, it is often the case that applications of existing transitions dominate the remaining computation, with relatively fewer transitions discovered as the computation proceeds. As we have noted for some examples, it seems ``tweaking'' or delaying the application of certain transitions can have a significant effect on the outcome. This will potentially require some analysis of whatever set of otter transitions have been generated in order to determine the most appropriate order in which these should be applied. Perhaps we cannot hope to reproduce the spectacular success of the single multilinear transition for Dragon 72 in every case, but it does seem that there is significantly more potential for such cases than we have been able to utilise so far. 

One step in this direction that we tried was to consider ``replaying'' the computation once all ocelot transitions were found, which may be thought of as knowing these in advance and seeing what difference that may make to the computation. some limited experimentation with our terminating termites seemed to produce very little difference in execution times, suggesting that finding ocelot transitions was much less of an issue then determining the most appropriate order in which they should be applied. One may liken our approach to playing ``transition golf'', i.e.\ finding the optional set of transitions and order in which they should be applied to find the final configuration in the smallest number of transition applications. This can be considered a quest for God's algorithm for dreadful dragons; it is possibly to define in our formalism transition of the form $\{a\}0\{l\} \leadsto (FinalState)$, and while it is utterly unrealistic to expect such a transition to be automatically discovered, it is intriguing to think how close we might get in a systematic manner to some minimal number of transitions required. 

A related observation is that all multilinear shelducks found so far have quadratic length, and typically involve at least one ocelot transition. This suggests that there is more to be analysed in the relationship between multilinear shelducks and the conditions that give rise to them. A further idea in this direction is to develop ocelot transitions directly from the machine definition, i.e,\ without observing execution. In other words \textit{analytical anacondas} would be found from the given set of transitions, and not, as in the observant otter, from particular shapes repeatedly occurring in computation. Part of our collection of the data for each ocelot transition found is the subset of machine transitions in it, meaning that we may be able to exploit this data to predict at least some ocelot transitions from the machine definitions alone. Such transition  identification may also be particularly helpful in the analysis of machines of unknown productivity and activity, or perhaps for explicitly constructing or recognising new dreadful dragons. 

Along these lines, it is also intriguing to think how shape computations could be used in other ways. For example, given a specific shape, it may be appropriate to start a shape computation with the shape alone, i.e.\ without a specific target pattern, and then seeing if there is any recurring pattern that can be found. This can be thought of as lifting the observation of patterns from ground computations to shape computations. 

A similar idea is to consider patterns and shape computations with only progressive variables. For example, if one can find a shape computation with initial configuration say $\{a\}0\{l\} (110)^X 111$ and target configuration $\{a\}0\{l\} (110)^{X+5} 111$ then we may safely conclude that the computation will never terminate (assuming of course that we can show that the computation starting with the blank input reaches at least one ground instance of the initial configuration such as $\{a\}0\{l\} (110)^{4} 111$). This may be too simplistic to capture all the subtle nuances of the various complex ``holdouts'' in the recent proof of the fifth busy beaver value \cite{BBChallenge26}, but nonetheless, it could be a useful way to formalise at least some non-termination proofs. 

\bibliographystyle{plain}
\bibliography{busybeaver}

@article{Aaronsen20,
author = {Scott Aaronsen},
title = {The Busy Beaver Frontier},
year = {2020},
issue_date = {August 2020},
publisher = {Association for Computing Machinery},
address = {New York, NY, USA},
volume = {51},
number = {3},
issn = {0163-5700},
url = {https://doi.org/10.1145/3427361.3427369},
doi = {10.1145/3427361.3427369},
journal = {SIGACT News},
month = sep,
pages = {32–54},
numpages = {23},
}

@misc{Quad,
	author = {Theoretical Computer Science on Stack Exchange}, 
	title = {Busy Beavers: quadratic relation of output size and running time?}, 
	howpublished = "\url{https://cstheory.stackexchange.com/questions/5532/busy-beavers-quadratic-relation-of-output-size-and-running-time}",
	year = {2011},
	note = "Accessed June, 2026"
}

@article{BBChallenge25,
author = {Justin Blanchard and Daniel Briggs and Konrad Deka and Nathan Fenner and Yannick Forster and Georgi Georgiev (Skelet) and Matthew L. House and Rachel Hunter and Iijil  and Maja Kadziołka and Pavel Kropitz and Shawn Ligocki and  mxdys and Mateusz Naściszewski and savask and Tristan Stérin and Chris Xu and Jason Yuen and  Théo Zimmermann},
title = {Determination of the fifth Busy Beaver value},
journal = {arXiv:2509.12337v1},
nourl = {"\url{https://doi.org/10.48550/arXiv.2509.12337}"},
year = 2025,
}

@inproceedings{BBChallenge26,
	author = {Blanchard, Justin and Briggs, Daniel and Deka, Konrad and Fenner, Nathan and Forster, Yannick and Georgiev (Skelet), Georgi and House, Matthew L. and K{\k a}dzio{\l}ka, Maja and Kropitz, Pavel and Ligocki, Shawn and mxdys and Na{\'s}ciszewski, Mateusz and St{\'e}rin, Tristan and Xu, Chris and Yuen, Jason and Zimmermann, Th{\'e}o},
	title = {Determination of the Fifth Busy Beaver Value},
	year = {2026},
	isbn = {9798400725364},
	publisher = {Association for Computing Machinery},
	address = {New York, NY, USA},
	url = {https://doi.org/10.1145/3798129.3806376},
	doi = {10.1145/3798129.3806376},
	booktitle = {Proceedings of the 58th Annual ACM Symposium on Theory of Computing},
	pages = {565–574},
	numpages = {10},
	location = {Salt Lake City, UT, USA},
	series = {STOC '26}
}

@misc{BBCpage,
author = {The Busy Beaver Challenge}, 
title = {Wiki Page}, 
howpublished = "\url{https://wiki.bbchallenge.org/wiki/Main_Page}",
note = "Accessed February, 2015"

}

@article{Brady83,
author = {Allen Brady}, 
title = {The Determination of the value of {R}ado's noncomputable function $\Sigma(k)$ for four-state {T}uring machines}, 
journal = {Mathematics of Computation},
volume = 40,
number = 162,
pages = {647-665}, 
year = 1983,
}

@article{Church36,
author = {Alonzo Church},
title = {An Unsolvable Problem of Elementary Number Theory},
journal = {American Journal of Mathematics},
volume = 58,
pages = {345-63},
year = 1936,
}

@inproceedings{Harland06,
author = {James Harland}, 
title = {The Busy Beaver, the Placid Platypus and Other Crazy Creatures}, 
booktitle = {Proceedings of Computing: the Australasian Theory Symposium (CATS'06)}, 
pages = {79-86}, 
address = {Hobart}, 
month = {January},
year = 2006
}

@inproceedings{Harland07,
author = {James Harland}, 
title = {Analysis of Busy Beaver Machines via Induction Proofs}, 
booktitle = {Proceedings of Computing: the Australasian Theory Symposium (CATS'07)},
pages = {71-78}, 
address = {Ballarat}, 
month = {January}, 
year = 2007,
}

@article{Harland16,
author = {James Harland}, 
title = {Busy {B}eaver {M}achines and the {O}bservant {O}tter {H}euristic (or How to Tame {D}readful {D}ragons)}, 
journal = {Theoretical Computer Science},
volume = {646},
pages = {61-85}, 
month = {September}, 
year = 2016,
note = {Available online at \url{http://www.sciencedirect.com/science/article/pii/S0304397516303504}.}, 
}

@article{Harland22,
author = {James Harland}, 
title = {Generating {C}andidate {B}usy {B}eaver {M}achines (or How to Build the {Z}any {Z}oo},
journal = {Theoretical Computer Science},
volume = {922},
pages = {368-394},
month = {June}, 
year = {2022},
note = {Available online at \url{doi:doi.org/10.1016/j.tcs.2022.04.040} },
}

@inproceedings{Holkner04,
author =  {Alex Holkner},  
title = {Acceleration Techniques for Busy Beaver Candidates}, 
noeditor = {Gad Abraham and Benjamin I.P. Rubenstein}, 
booktitle = {Proceedings of the Second Australian Undergraduate Students' Computing Conference},
pages = {75-80}, 
month = {December}, 
isbn = {0-975-71730-8},
year = 2004,
}

@article{Knuth76,
	author = {Donald E. Knuth },
	title = {Mathematics and Computer Science: Coping with Finiteness},
	journal = {Science},
	volume = {194},
	number = {4271},
	pages = {1235-1242},
	year = {1976},
	doi = {10.1126/science.194.4271.1235},
	URL = {https://www.science.org/doi/abs/10.1126/science.194.4271.1235},
	eprint = {https://www.science.org/doi/pdf/10.1126/science.194.4271.1235}
}

@inproceedings{LP07,
author = {Gregory Lafitte and Christophe Papazian}, 
title = {The fabric of small {T}uring machines}, 
booktitle = {Computation and Logic in the Real World, Proceedings of the Third Conference on Computability in Europe},
pages =  {219-227}, 
month = {June},
year = 2007,
}

@article{LR65,
author = {Shen Lin and Tibor Rado}, 
title = {Computer Studies of {T}uring Machine Problems}, 
journal = {Journal of the Association for Computing Machinery},
volume = 12,
number = 2,
pages = {196-212}, 
year = 1965
}

@article{MS90,
author = {Rona Machlin and Quentin Stout}, 
title = {The Complex Behavior of Simple Machines}, 
journal = {Physica D:Nonlinear Phenomena}, 
volume = 42,
pages = {85-98}, 
year = 1990
}

@misc{maple,
author = {MapleSoft},
title = {MapleSoft website},
howpublished = "\url{https://www.maplesoft.com/products/maple/}",
note = "Accessed October, 2026"
}

@misc{Marxen,
author = {Heiner Marxen}, 
title = {Busy Beaver web page}, 
howpublished = "\url{http://www.drb.insel.de/~heiner/BB/index.html}",
note = "Accessed October, 2016"
}

@article{HM90, 
author = {Heiner Marxen and J\"{u}rgen Buntrock}, 
title = {Attacking the Busy Beaver 5}, 
journal = {Bulletin of the EATCS}, 
volume = 40,
pages = {247-251}, 
month = {February},
year =  1990
}

@misc{mathematica,
author = {Wolfram},
title = {Wolfram Mathematica website},
howpublished = "\url{https://www.wolfram.com/mathematica/}",
note = "Accessed October, 2026"
}

@article{Michel93,
author =  {Pascal Michel}, 
title = {Busy Beaver Competition and {C}ollatz-Like Problems},
journal = {Archive for Mathematical Logic}, 
volume = 32,
number = 5,
pages = {351-367}, 
year = 1993
}

@article{Michel04,
author = {Pascal Michel}, 
title = {Small {T}uring machines and generalized busy beaver competition}, 
journal = {Theoretical Computer Science},
volume = 326,
number = {1-3},
pages = {45-56}, 
year = 2004,
}

@article{Michel13,
author = {Pascal Michel},
title = {Problems in Number Theory from Busy Beaver Competition},
journal = {arXiv:1311.1029},
nourl = {"\url{http://arxiv.org/abs/1311.1029}"},
year = 2013,
}

@article{Michel22,
author = {Pascal Michel},
title = {The Busy Beaver Competition: a historical survey},
journal = {arXiv:0906.3749v7},
nourl = {"\url{http://arxiv.org/abs/0906.3749v7}"},
year = 2022,
}

@article{Milner87,
author = {Robin Milner},
title = {Is Computing an Experimental Science?},
journal = {Journal of Information Technology},
volume = 2,
number = 2,
doi = {10.1177/026839628700200202},
howpublished = "\url{https://aisel.aisnet.org/jit/vol2/iss2/2}",
year = 1987,
}

@incollection{deMol,
author = {Liesbeth De Mol}, 
title = {Looking for Busy Beavers: A socio-philosophical study of a computer-assisted proof}, 
booktitle = {Foundations of the formal sciences VII}, 
noeditor = {K. Francois and  B. L\"{o}owe, and  T. M\"{u}ller and B. van Kerkhove}, 
publisher = {College Publications}, 
pages = {61-90}, 
year = 2011,
}

@book{Nahin26,
	title = {The {Delicate} {Art} of {Brute} {Force}},
	isbn = {978-0-691-26746-3},
	publisher = {Princeton University Press},
	author = {Nahin, Paul J.},
	year = {2026},
}

@book{Porges,
author = {Arthur Porges},
title = {The Devil and Simon Flagg and Other Fantastic Tales},
publisher = {Richard Simms Publications},
year = 2009,
}

@article{Rado62,
author = {Tibor Rado}, 
title = {On non-computable functions}, 
journal = {Bell System Technical Journal},
volume = 41,
pages = {877-884}, 
year = 1962,
}

@misc{swi,
author = {SWI Prolog},
title = {{SWI Prolog WWW Site}}, 
howpublished = "\url{http://www.swi-prolog.org}",
note = "Accessed October, 2016",
}

@book{Sudkamp,
author = {Thomas Sudkamp}, 
title = {Languages and Machines: An Introduction to the Theory of Computer Science (3rd ed.)}, 
publisher = {Addison-Wesley}, 
year = 2005,
}

@article{Turing36,
author = {Alan Turing},
title = {On Computable Numbers with an Application to the {Entscheidungsproblem}},
journal = {Proceedings of the London Mathematical Society},
volume = 42,
pages = {230-65},
year = {1936},
}

\appendix

\section*{Calculations}

\subsection*{Calculating Count}\label{sec:count}

Consider the following sequence. 

\begin{center}
\begin{tabular}{cc}
& \\
0 & $Y_0$ \\
1 & $Y_0 - l$ \\
2 & $Y_0 - 2 \times l$ \\
$\cdots$ & $\cdots$ \\
$k$ & $Y_0 - k \times l$ \\
& \\
\end{tabular}
\end{center}

for which we have the constraint that $Y \geq m$. Note that we must have that $m,l \geq 1$. 

So the final instance of the pattern will be when $Y_0 - (k \times l) < m$, i.e.\ when $Y_0 -m < k \times l$ and so $k > (Y_0 - m)//l$, i.e.\ $k \geq (Y_0 - m)//l + 1$. Hence if we let $Count$ be the number of iterations of the pattern, we have 

$$Count = (Y_0 - m)//l + 1 = (Y_0 - m + l)//l$$

For multiply regressive patterns, the same formula will apply for each regressive variable, and the overall count is then the minimum of each of the corresponding values.

\subsection*{Calculating Coefficients }\label{sec:coeffcalc}

For a multilinear progressive otter sequence, i.e. where $X_k = mX_{k-1} + n$, we note the following. 

\noindent
$X_1 = mX_0 + n$ \\
$X_2 = mX_1 + n = m(mX_0 + n) + n = m^2X_0 + n(m + 1)$ \\
$X_3 = mX_2 + n = m(m^2X_0 + mn + n) + n = m^3X_0 + n(m^2 + m+ 1)$ \\
$X_4 = mX_3 + n = m(m^3X_0 + n(m^2 + m + 1)) + n = m^4X_0 + n(m^3 + m^2 + m + 1)$ \\
\ldots \\
$X_k = mX_{k-1} + n = m^kX_0 + n \displaystyle\sum_{i=0}^{k-1} m^i$ \\

When $m = 1$ (i.e.\ the sequence is unilinear) this reduces to $X_k = X_0 + nk$. When $m > 1$, (i.e.\ the sequence is multilinear) we note that 

$$\displaystyle\sum_{i=0}^{k-1} m^i = m^{k-1} + m^{k-2} + \ldots + m + 1 = \frac{m^k - 1}{m-1}$$

So when $m = 1$ we have $X_k = X_0 + nk$ and when $m > 1$ we have $X_k = m^kX_0 + n\frac{m^k - 1}{m-1}$. 

\subsection*{Calculating Hops}\label{sec:hopscalc}

\subsubsection*{Unilinear sequences}

\begin{prop} \label{prop:linearaffine}
	Let $S_1, S_2, \ldots , S_k$ be a unilinear observant otter sequence for some minimal shape $S$ in an execution of a machine $M$ for which $S$ has affine length and which is executed $C$ times. Then the number of Hops in the corresponding ocelot transition is quadratic in $C$. 
\end{prop}

\begin{proof}
	
Assume we have a unilinear observant otter sequence for which the shape computation has affine length. 

Let the progressive variables be $X_1, \ldots, X_p$ and the regressive variables be $Y_1, \ldots, Y_q$. Let the relative difference for $X_i$ and $Y_j$ be $r_i$ and $s_j$ respectively. Because the otter is unilinear, at the $k$th iteration we have $X_ik = X_{i0} + kn_i$ and $Y_{jk} = Y_{j0} -k d_j$ for each $1 \leq i \leq p$, $1 \leq j \leq q$. 

We write the value of Hops after the $k$th iteration as $Hops(k)$, so that $Hops(0) = 0$ and $Hops(1) = \displaystyle\sum_{i=1}^{p} a_iX_{i0} + \sum_{j=1}^{q} b_jY_{j0} + c$. 
As the shape computation has affine length, $Hops(k+1)$ is 

$$\displaystyle \sum_{i=1}^{p} a_iX_{ik} + \sum_{j=1}^{q} b_jY_{jk} + c$$ 
$$= \displaystyle \sum_{i=1}^{p} a_i(X_{i0} + k n_i) + \sum_{j=1}^{q} b_j(Y_{j0} - k d_j) + c$$
$$= \displaystyle \sum_{i=1}^{p} a_iX_{i0} + k\sum_{i=1}^{p} a_in_i + \sum_{j=1}^{q} b_jY_{j0} - k \sum_{j=1}^{q} b_jd_j + c$$

Letting $\displaystyle A = \sum_{i=1}^{p} a_iX_{i0}, N = \sum_{i=1}^{p} a_in_i, B = \sum_{j=1}^{q} b_jY_{j0}$, and $\displaystyle D = \sum_{j=1}^{q} b_jd_j$, we can write this as

$$A + kN + B -kD + c = A+B+c + k(N-D)$$

Hence if the otter pattern is applied $C$ times, then the total number of steps is the sum of the hops at each of stages from $0$ up to $C-1$ is as below. 

$$\displaystyle \sum_{k=0}^{C-1} Hops(k+1)$$
$$\displaystyle \sum_{k=0}^{C-1} A+B+c + k(N-D)$$
$$\displaystyle = C(A+B+c) + (N-D)\sum_{k=0}^{C-1}k$$ 
$$= C(A+B+c) + C(C-1)/2 (N-D) = C(A+B+c + (C-1)/2(N-D))$$

Rewriting this in terms of the relative differences, the number of hops in the transition for a unilinear observant otter sequence for which the shape computation has affine length is as below. 

$$\displaystyle C \times \left( \sum_{i=1}^{p} a_iX_{i0} + \sum_{j=1}^{q} b_jY_{j0} + c  + \frac{C-1}{2}(\sum_{i=1}^{p} a_in_i - \sum_{j=1}^{q} b_jd_j) \right) $$
$$= \displaystyle C \times \left( Hops(1)  + \frac{C-1}{2}(\sum_{i=1}^{p} a_in_i - \sum_{j=1}^{q} b_jd_j) \right) $$

where $C$ is the number of iterations performed. 

Writing $R = \displaystyle\sum_{i=1}^{p} a_in_i - \sum_{j=1}^{q} b_jd_j$, we can also write this as below. 

$$= \displaystyle C \times \left( Hops(1)  + \frac{C-1}{2}R \right) $$
$$= \displaystyle C \times \left( Hops(1)  + \frac{R}{2}C - \frac{R}{2} \right) $$

This gives the number of Hops as quadratic in $C$, as claimed. $\qed$
\end{proof}

\subsubsection*{Multilinear sequences}

\begin{prop} \label{prop:multiaffine}
Let $S_1, S_2, \ldots , S_k$ be a multilinear observant otter sequence for some minimal shape $S$ in an execution of a machine $M$ for which $S$ has affine length and which is executed $C$ times. Then the number of Hops in the corresponding ocelot transition is exponential in $C$. 
\end{prop}

\begin{proof}

F%or a multilinear observant otter sequence for which the shape computation has affine length, we have the following. 

Let the progressive variables be $X_1, \ldots, X_p$ and the regressive variables be $Y_1, \ldots, Y_q$. Let the relative difference for $X_i$ and $Y_j$ be $(m_i,n_i)$ and $d_j$ respectively. Because the otter is multilinear, at the $k$th iteration we have 

$$X_{ik} = m_iX_{ik-1} + kn_i  = (m_i)^kX_{i0} + n_i\frac{(m_i)^k - 1}{m_i - 1} = (m_i)^k(X_{i0} +\frac{n_i}{m_i-1}) -\frac{n_i}{m_i-1}$$ 

and  $Y_{jk} = Y_{j0} -k d_j$, for each $1 \leq i \leq p$, $1 \leq j \leq q$. 

As the shape computation has affine length, the value of hops at the $(k+1)$ stage is 

$$\displaystyle \sum_{i=1}^{p} a_iX_{ik} + \sum_{j=1}^{q} b_jY_{jk} + c$$ 
$$= \displaystyle \sum_{i=1}^{p} a_i(m_i^kX_{i0} + n_i\frac{m_i^k - 1}{m_i - 1}) + \sum_{j=1}^{q} b_j(Y_{j0} - k d_j) + c$$
$$= \displaystyle \sum_{i=1}^{p} a_im_i^kX_{i0} + \sum_{i=1}^{p} a_in_i\frac{m_i^k - 1}{m_i - 1} + \sum_{j=1}^{q} b_jY_{j0} - k \sum_{j=1}^{q} b_jd_j + c$$

As above, letting $\displaystyle B = \sum_{j=1}^{q} b_jY_{j0}$, and $\displaystyle D = \sum_{j=1}^{q} b_jd_j$, we can write this as

$$\displaystyle \sum_{i=1}^{p} a_im_i^kX_{i0} + \sum_{i=1}^{p} a_in_i\frac{m_i^k - 1}{m_i - 1} + B - kD + c$$

Hence if the otter pattern is applied $C$ times, then the total number of steps is the sum of the hops at each of stages from $0$ up to $C-1$ is as below. 

$$\displaystyle \sum_{k=0}^{C-1} Hops(k+1)$$
$$\displaystyle \sum_{k=0}^{C-1} \left( \sum_{i=1}^{p} a_im_i^kX_{i0} + \sum_{i=1}^{p} a_in_i\frac{m_i^k - 1}{m_i - 1}) + B - kD + c \right)$$
$$= \displaystyle \sum_{k=0}^{C-1}\sum_{i=1}^{p}a_iX_{i0}m_i^k + \sum_{k=0}^{C-1} \sum_{i=1}^{p} \frac{a_in_i}{m_i -1}(m_i^k - 1) + C(B+c) - D\sum_{k=0}^{C-1}k$$
$$= \displaystyle \sum_{k=0}^{C-1}\sum_{i=1}^{p} a_iX_{i0}m_i^k + \sum_{k=0}^{C-1} \sum_{i=1}^{p} \frac{a_in_im_i^k}{m_i -1} -  \sum_{k=0}^{C-1} \sum_{i=1}^{p} \frac{a_in_i}{m_i -1} + C(B+c) - DC(C-1)/2$$
$$= \displaystyle \sum_{i=1}^{p}\sum_{k=0}^{C-1} a_iX_{i0}m_i^k +  \sum_{i=1}^{p} \sum_{k=0}^{C-1} \frac{a_in_im_i^k}{m_i -1} - C\sum_{i=1}^{p} \frac{a_in_i}{m_i -1} + C(B+c) - DC(C-1)/2$$
$$= \displaystyle \sum_{i=1}^{p}\sum_{k=0}^{C-1} (a_iX_{i0} + \frac{a_in_i}{m_i -1}) m_i^k   + C(B+c- \sum_{i=1}^{p} \frac{a_in_i}{m_i -1}) - DC(C-1)/2$$
$$= \displaystyle \sum_{i=1}^{p}  (a_iX_{i0} + \frac{a_in_i}{m_i -1}) \sum_{k=0}^{C-1} m_i^k  + C(B+c- \sum_{i=1}^{p} \frac{a_in_i}{m_i -1}) - DC(C-1)/2$$
$$= \displaystyle \sum_{i=1}^{p}  (a_iX_{i0} + \frac{a_in_i}{m_i -1}) \frac{m_i^C-1}{m_i -1}  + C(B+c- \sum_{i=1}^{p} \frac{a_in_i}{m_i -1}) - DC(C-1)/2$$
$$= \displaystyle \sum_{i=1}^{p}  \frac{a_iX_{i0}(m_i^C-1)}{m_i -1} + \sum_{i=1}^{p}\frac{a_in_i(m_i^C-1)}{(m_i -1)^2}  + C(B+c- \sum_{i=1}^{p} \frac{a_in_i}{m_i -1}) - DC(C-1)/2$$

Rewriting this in terms of the relative differences, the number of hops in the transition for a multilinear observant otter sequence for which the shape computation has affine length is as below. 

$$\displaystyle \sum_{i=1}^{p}  \frac{a_iX_{i0}(m_i^C-1)}{m_i -1} + \sum_{i=1}^{p}\frac{a_in_i(m_i^C-1)}{(m_i -1)^2} + C(\sum_{j=1}^{q} b_jY_{j0} + c -\sum_{i=1}^{p} \frac{a_in_i}{m_i -1}) - \frac{C(C-1)}{2}\sum_{j=1}^{q} b_jd_j$$
$\qed$
\end{proof}
 
 \begin{prop} \label{prop:multiquad}
 	Let $S_1, S_2, \ldots , S_k$ be a multilinear monoprogressive monoregressive observant otter sequence for some minimal shape $S$ in an execution of a machine $M$ for which $S$ has quadratic length over the progressive variable only, and which is executed $C$ times. Then the number of Hops in the corresponding ocelot transition is exponential in $C$. 
 \end{prop}
 
 \begin{proof}
 
For a multilinear monoprogressive monoregressive observant otter sequence for which the shape computation has quadratic length, we have the following.

Let the progressive variable be $X$ and the regressive variable be $Y$. Let the relative difference for $X$ and $Y$ be $(m,n)$ and $d$ respectively. Because the otter is multilinear, at the $k$th iteration we have 

$$\displaystyle X_{k} = mX_{k-1} + kn  = m^kX_{0} + n\frac{m^k - 1}{m - 1} = m^k(X_0 + \frac{n}{m-1}) - \frac{n}{m-1}$$

and  $Y_{k} = Y_{0} -k d$. 
 
% Let the shape computation have length $aX^2 + bXY + cY^2 + dX + eY + f$. 
Let the shape computation have length $aX^2 + bX + c$. 
So the value of $Hops(k+1)$ is

$$aX_k^2 + bX_k + c$$
$$\displaystyle= a(m^k(X_0 + \frac{n}{m-1}) - \frac{n}{m-1})^2 + b(m^k(X_0 + \frac{n}{m-1}) - \frac{n}{m-1}) + c$$
$$\displaystyle= am^{2k}(X_0 + \frac{n}{m-1})^2 - \frac{2an}{m-1}(X_0+\frac{n}{m-1})m^k +\frac{an^2}{(m-1)^2} + b((X_0 + \frac{n}{m-1})m^k - \frac{bn}{m-1} + c$$
$$\displaystyle= am^{2k}(X_0 + \frac{n}{m-1})^2 + \frac{b(m-1)- 2an}{m-1}(X_0+\frac{n}{m-1})m^k + \frac{an^2}{(m-1)^2} + c - \frac{bn}{m-1}$$
 
 Alternatively this is 
 
$$\displaystyle= a(m^kX_{0} + n\frac{m^k - 1}{m - 1})^2 + b(m^kX_{0} + n\frac{m^k - 1}{m - 1}) + c$$
$$\displaystyle= am^{2k}X_0^2 + \frac{2anX_0m^k(m^k-1)}{m-1} + \frac{an^2(m^k-1)^2}{(m-1)^2} + bm^kX_0 + \frac{bn(m^k-1)}{m-1}) + c$$
$$\displaystyle= am^{2k}X_0^2 + \frac{2anX_0m^k(m^k-1)}{m-1} + \frac{an^2(m^k-1)^2}{(m-1)^2} + bm^kX_0 + \frac{bn(m^k-1)}{m-1}) + c$$
$$\displaystyle= am^{2k}X_0^2 + \frac{2anX_0m^{2k} - 2anm^kX_0}{m-1} + \frac{an^2(m^{2k} - 2m^k +1)}{(m-1)^2} + bm^kX_0 + \frac{bnm^k- bn}{m-1} + c$$
$$\displaystyle= am^{2k}X_0^2 + \frac{2anX_0m^{2k}}{m-1} - \frac{2anm^kX_0}{m-1} + \frac{an^2m^{2k}}{(m-1)^2} - \frac{2an^2m^k}{(m-1)^2} + \frac{an^2}{(m-1)^2} + bm^kX_0 + \frac{bnm^k}{m-1}- \frac{bn}{m-1} + c$$
$$\displaystyle= m^{2k}\left(aX_0^2 +  \frac{2anX_0}{m-1} +  \frac{an^2}{(m-1)^2}  \right) 
+ m^k \left( bX_0 +  \frac{bn}{m-1}  - \frac{2anX_0}{m-1} - \frac{2an^2}{(m-1)^2} \right) 
+ \frac{an^2}{(m-1)^2} - \frac{bn}{m-1} + c$$ 
$$\displaystyle= m^{2k}\frac{aX_0^2(m-1)^2 + 2anX_0(m-1) +  an^2}{(m-1)^2}
 + m^k \frac{bX_0(m-1)^2 - 2anX_0(m-1) +  bn(m-1) - 2an^2}{(m-1)^2}$$
$$+ \frac{an^2 -bn(m-1) + c(m-1)^2}{(m-1)^2}$$ 
 
Hence if the otter pattern is applied $C$ times, then the total number of steps is the sum of the hops at each of stages from $0$ up to $C-1$ is as below. 

$$\displaystyle \sum_{k=0}^{C-1} Hops(k+1)$$
$$=\displaystyle \sum_{k=0}^{C-1} m^{2k}\frac{aX_0^2(m-1)^2 + (2anX_0)(m-1) +  an^2}{(m-1)^2}
+ m^k \frac{bX_0(m-1)^2 - 2anX_0(m-1) +  bn(m-1) - 2an^2}{(m-1)^2}$$
$$+ \frac{an^2 -bn(m-1) + c(m-1)^2}{(m-1)^2} $$
$$=\displaystyle \frac{aX_0^2(m-1)^2 + 2anX_0(m-1) +  an^2}{(m-1)^2} \sum_{k=0}^{C-1} m^{2k}
+ \frac{bX_0(m-1)^2 - 2anX_0(m-1) +  bn(m-1) - 2an^2}{(m-1)^2} \sum_{k=0}^{C-1} m^k $$
$$+ \frac{an^2 -bn(m-1) + c(m-1)^2}{(m-1)^2}\sum_{k=0}^{C-1} 1$$
$$=\displaystyle \frac{aX_0^2(m-1)^2 + 2anX_0(m-1) +  an^2}{(m-1)^2} \frac{m^{2C} - 1}{m^2-1}
+ \frac{bX_0(m-1)^2  - 2anX_0(m-1) +  bn(m-1)  - 2an^2}{(m-1)^2} \frac{m^C-1}{m-1} $$
$$+ C \times \frac{an^2 -bn(m-1) + c(m-1)^2}{(m-1)^2}$$
$$=\displaystyle \frac{  A + B + D }{(m-1)^3(m^2-1)}$$

where \\
$A = \left( aX_0^2(m-1)^2 + 2anX_0(m-1) +  an^2 \right)(m^{2C}-1)(m-1)$ \\
$B = \left( bX_0(m-1)^2  - 2anX_0(m-1) +  bn(m-1)  - 2an^2 \right)(m^C-1)(m^2-1)$ \\
$D = C \times \left( an^2 -bn(m-1) + c(m-1)^2\right)(m-1)(m^2-1) $ \\

$\qed$

\end{proof}

\subsection*{Moduli}\label{sec:moduli}

We note the following identity.

\begin{prop}
For all positive integers $m,n$, $n(m//n) + m \; mod \; n = m$. 
\end{prop}

\begin{proof}
The result is trivial when $m \leq n$, so we only need to show the case when $m > n$. 	

Let $p$ be such that $p \times n$ is the greatest multiple of $n$ that does not exceed $m$, i.e., $p \times n \leq m$ and $(p+1) \times n > m$. Hence there exists a non-negative integer $q$ such that $m = p \times n + q$ where $0 \leq q < n$. 

Hence $m // n = p$ and $m \; mod \; n = q$, and so $m = p \times n + q = n(m//n) + m \; mod \; n$ as required. $\qed$
\end{proof}

This means we can rewrite $n(m//n)$ as $m - m \; mod \; n$ if desired. This will be particularly helpful for simplifying expressions such as $2((X+1)//2)$ which we can write as $X+1 - (X+1) \; mod \; 2$. \\

%\textit{\textbf{Some more notes to be refined:}}

%\begin{prop}
%For any integer $n \geq 2$, $n//2 + (n+1)//2 = n$. 
%\end{prop}
%
%\begin{proof}
%Let $n = 2p+q$ where $q$ is either $0$ or $1$. 
%
%If $q = 0$, $n = 2p$, $n//2 = p$, $(n+1)//2 = p$ and so $n//2 + (n+1)//2 = 2p = n$. \\
%If $q = 1$, $n = 2p+1$, $n//2 = p$, $(n+1)//2 = p+1$ and so $n//2 + (n+1)//2 = 2p+1 = n$.
%$\qed$
%\end{proof}

We note also the following. 
%More generally consider the following. 

\begin{prop}
For any integers $n \geq m > 0$, $\displaystyle \sum_{i=0}^{m-1} (n+i)//m = n$
\end{prop}

\begin{proof}
Let $n = mp+q$ where $0 \leq q < m$. Then we have

$$\displaystyle \sum_{i=0}^{m-1} (n+i)//m = \sum_{i=0}^{m-1} (mp+q+i)//m = \sum_{i=0}^{m-1} p + (q+i)//m = \sum_{i=0}^{m-1} p + \sum_{i=0}^{m-1} (q+i)//m = mp + \sum_{i=0}^{m-1} (q+i)//m$$

Note that in the $m$ values in the sequence $q+i$, there are precisely $q$ values where $q+i \geq m$, and so for these $q$ values, $(q+i)//m = 1$ and for the rest $(q+i)//m = 0$. This means that $\displaystyle\sum_{i=0}^{m-1} (q+i)//m = q$, and so we have 

$$\displaystyle \sum_{i=0}^{m-1} (n+i)//m =  mp + \sum_{i=0}^{m-1} (q+i)//m = mp+q = n$$
$\qed$
\end{proof}

Note that it follows directly from the above result that $n//2 + (n+1)//2 = n$ for any $n \geq 2$. 

\section*{Dragons}

The 2-state machine mentioned in Section~\ref{sec:intro} is Dragon 48. 

Its productivity is 5,282,768,707,595,746,865,092,337,457,181,555,923,304,480,680,295,710,143,373,849,725,332,341, \\
654,013,512,661,461,020,020,766,119,592,641,932. 

Its activity is 16,744,587,130,771,702,585,056,729,777,499,619,894,814,779,221,058,942,033,129,756,857,605,021,935, \\
813,298,913,076,080,867,705,003,948,174,230,109,940,326,315,083,166,842,383,854,901,037,906,503,336,496,279,110,002, \\
007,373,200,041,049,688,400,411,330,516,770,862,100,706,147,243,958. 

\end{document}